\documentclass[aps,prd,nofootinbib,showkeys,preprint,floatfix]{revtex4}

\usepackage{amssymb}
\usepackage{amsmath}
\usepackage{amsfonts}
\usepackage{graphicx}
\usepackage{color}
\usepackage{xspace}
\usepackage{ulem}

\def\gsim{\raise0.3ex\hbox{$\;>$\kern-0.75em\raise-1.1ex\hbox{$\sim\;$}}}
\def\lsim{\raise0.3ex\hbox{$\;<$\kern-0.75em\raise-1.1ex\hbox{$\sim\;$}}}

\def\etsl{$E_T$ \hspace{-1.4em}/\;\: \hspace{0.4em}}

\newcommand {\SARAH} {{\tt SARAH}\xspace}
\newcommand {\SPheno} {{\tt SPheno}\xspace}

\newcommand{\FIG}[1]{fig.~\ref{#1}}
\newcommand{\TAB}[1]{table~\ref{#1}}

\newcommand{\AddrAHEP}{
  {\it AHEP Group, Instituto de F\'{\i}sica Corpuscular --
    C.S.I.C./Universitat de Val{\`e}ncia \\
    Edificio de Institutos de Paterna, Apartado 22085,
  E--46071 Val{\`e}ncia, Spain}}

\newcommand{\AddrWur}{%
Institut f\"ur Theoretische Physik und Astronomie, 
Universit\"at W\"urzburg\\
Am Hubland, 
97074 W\"urzburg}

\newcommand{\AddrBonn}{
{Bethe Center for Theoretical Physics \& Physikalisches Institut der 
Universit\"at Bonn, \\}
Nussallee 12, 53115 Bonn, Germany
}

\def\gsim{\raise0.3ex\hbox{$\;>$\kern-0.75em\raise-1.1ex\hbox{$\sim\;$}}}
\def\lsim{\raise0.3ex\hbox{$\;<$\kern-0.75em\raise-1.1ex\hbox{$\sim\;$}}}

\def\tb{\tan\beta}
\def\tbR{\tan\beta_R}

\newcommand{\Ds}{\Delta m^2_{\odot}}
\newcommand{\Da}{\Delta m^2_{\textsc{A}}}

\begin{document}

\preprint{IFIC/12-43, Bonn-TH-2012-11}  

\title{Phenomenology of the minimal supersymmetric $U(1)_{B-L}\times U(1)_R$ 
extension of the standard model}

\author{M. Hirsch} \email{mahirsch@ific.uv.es}\affiliation{\AddrAHEP}
\author{W. Porod}\email{porod@physik.uni-wuerzburg.de}\affiliation{\AddrWur \\ 
\AddrAHEP}
\author{L.~Reichert} \email{reichert@ific.uv.es}\affiliation{\AddrAHEP}
\author{F. Staub}\email{fnstaub@th.physik.uni-bonn.de}\affiliation{\AddrBonn}



\begin{abstract}
We discuss the minimal supersymmetric $U(1)_{B-L}\times U(1)_R$ extension of 
the standard model. Gauge couplings unify as in the MSSM, even if the 
scale of $U(1)_{B-L}\times U(1)_R$ breaking is as low as order TeV 
and the model can be embedded into an $SO(10)$ grand unified theory. 
The phenomenology of the model differs in some important aspects from 
the MSSM, leading potentially to rich phenomenology at the LHC. It 
predicts more light Higgs states and the mostly left CP-even Higgs 
has a mass reaching easily 125 GeV, with no constraints on the SUSY 
spectrum. Right sneutrinos can be the lightest supersymmetric 
particle, changing all dark matter constraints on SUSY parameter space.
The model has seven neutralinos and squark/gluino decay chains involve 
more complicated cascades than in the MSSM. We also discuss briefly 
low-energy and accelerator constraints on the model, where the most 
important limits come from recent $Z'$ searches at the LHC and upper 
limits on lepton flavour violation.

\end{abstract}

\maketitle

\tableofcontents

\section{Introduction}

Within the minimal supersymmetric extension of the standard model 
(MSSM) the gauge couplings unify nearly perfectly around an energy 
scale of approximately $m_G \simeq 2 \times 10^{16}$ GeV, if SUSY 
particles exist with masses of the order of ${\cal O}(1)$ TeV. 
Extending the MSSM with non-singlet superfields tends to destroy 
this attractive feature, unless (a) the additional fields come in 
complete $SU(5)$ multiplets or (b) the standard model gauge 
group is extended too. Here we study a model in which the SM group 
is enlarged to $SU(3)_c\times SU(2)_L\times U(1)_{B-L}\times U(1)_R$. 
It is a variant of the models first proposed in \cite{Malinsky:2005bi} 
and later discussed in more detail in \cite{DeRomeri:2011ie}. 

Our main motivation for studying this model can be summarized as: 
(i) It unifies, in the same way the MSSM does, even if the scale of 
$U(1)_{B-L}\times U(1)_R$ breaking is as low as the electro-weak 
scale; (ii) it can be easily embedded into an $SO(10)$ grand unified 
theory; (iii) it has the right ingredients to explain neutrino masses 
(and angles) by either an inverse \cite{Mohapatra:1986bd} or a linear 
\cite{Akhmedov:1995vm,Akhmedov:1995ip} seesaw; (iv) it 
allows for Higgs masses significantly larger than the MSSM without 
the need for a very heavy SUSY spectrum \cite{Hirsch:2011hg} and (v) 
it potentially leads to rich phenomenology at the LHC. 

With the data accumulated in 2011 both ATLAS \cite{ATLAS:2012ae} and 
CMS \cite{Chatrchyan:2012tx} have seen some indications for a Higgs 
boson with a mass of roughly $m_h \sim 125$ GeV. This result, perhaps 
unsurprisingly, has triggered an avalanche of papers studying the 
impact of such a relatively hefty Higgs on the supersymmetric parameter 
space \cite{
Hall:2011aa,Baer:2011ab,Feng:2011aa,Heinemeyer:2011aa,Arbey:2011ab,
Arbey:2011aa,Draper:2011aa,Moroi:2011aa,Carena:2011aa,Ellwanger:2011aa,
Buchmueller:2011ab,Akula:2011aa,Kadastik:2011aa,Cao:2011sn,
Arvanitaki:2011ck,Gozdz:2012xx,Gunion:2012zd,Ross:2011xv, Ross:2012nr,
FileviezPerez:2012iw,Karagiannakis:2012vk,King:2012is,Kang:2012tn,Chang:2012gp,
Aparicio:2012iw,Roszkowski:2012uf,Ellis:2012aa,Baer:2012uy,Desai:2012qy,
Cao:2012fz,Maiani:2012ij,Cheng:2012np,Christensen:2012ei,Vasquez:2012hn,
Ellwanger:2012ke}. The general consensus seems to be, that the MSSM can 
generate $m_h \sim 125$ GeV only if squarks and gluinos have masses 
in the multi-TeV range. While this is, of course, perfectly consistent 
with the lower bounds on SUSY masses obtained from \etsl searches 
at the LHC \cite{ATLAS-WEB,CMS-WEB}, such a heavy spectrum could make 
it quite difficult indeed for the LHC to find direct signals for SUSY.

There are, of course, several possibilities to circumvent this
conclusion. First of all, it is well-known that the loop corrections
to $h^0$ are dominated by the top quark-squark loops.  Thus, little or
no constraints on sleptons and on squarks of the first two generations
can in fact be derived from Higgs mass measurements, once the
assumption of universal boundary conditions for the soft SUSY
parameters is abandoned.  Second, in the next-to-minimal SSM (NMSSM)
the $h^0$ can be heavier than in the MSSM due to the presence of new
F-terms from the additional singlet Higgs
\cite{Ellwanger:2011aa,Ellwanger:2012ke}, especially in models with
non-universal boundary conditions for the (soft) Higgs mass terms
\cite{Gunion:2012zd} or in the generalized NMSSM \cite{Ross:2011xv,
Ross:2012nr}.  And, third, in models with an extended gauge group
additional $D$-terms contribute to the Higgs mass matrices, relaxing
the MSSM upper limit considerably
\cite{Haber:1986gz,Drees:1987tp,Cvetic:1997ky,Zhang:2008jm,Ma:2011ea}.
This latter possibility is the case we have studied in a previous
paper \cite{Hirsch:2011hg} using the minimal $U(1)_{B-L}\times U(1)_R$
models of \cite{Malinsky:2005bi}. Here, we extend the analysis of
\cite{Hirsch:2011hg}, including both Higgs and SUSY phenomenology.

Due to the extended gauge structure the model necessarily has more 
Higgses than the MSSM. Near D-flatness of the $U(1)_{B-L}\times U(1)_{R}$ breaking 
then results in one additional light Higgs, $h^0_{BLR}$ \cite{Hirsch:2011hg}. 
Mixing between the MSSM $h^0 \equiv h^0_L$ and $h^0_{BLR}$ enhances 
the mass of the mostly MSSM Higgs and, potentially, affects its decays. 
This is reminiscent to the situation in the NMSSM, where an additional 
light and mostly singlet Higgs state seems to be preferred 
\cite{Ellwanger:2011aa,Ellwanger:2012ke} if the signals found by ATLAS 
\cite{ATLAS:2012ae} and CMS \cite{Chatrchyan:2012tx} are indeed due to 
a 125 GeV Higgs. 

The MSSM-like $h^0_L$ in our model can have some exotic decays. For 
example, the $h^0_L$ will decay to two lighter Higgses, if kinematically 
possible, although this decay can never be dominant due to constraints 
coming from LEP. The model also includes right-handed 
neutrinos with electro-weak scale masses and there is a small but 
interesting part in parameter space where $m_{Z^0} \le m_{\nu_R} 
\le m_{h^0_L}$, where the Higgs decays to two neutrinos. These 
decays always lead to one light and one heavy neutrino, with the 
latter decaying promptly to either $W^{\pm}l^{\mp}$ or $Z^0\nu$. 
(Mostly right) sneutrinos can be lighter than the $h^0_L$, in which 
case the Higgs can have invisible decays.

The SUSY spectrum of the model is also richer than the MSSM: It 
has seven neutralinos and 
nine sneutrino states.
These additional sneutrinos can 
easily be the lightest supersymmetric particle (LSP) and thus 
change all the constraints on SUSY parameter space, usually derived 
from the requirement that the neutralino be a good dark matter 
candidate with the correct relic density 
\cite{Nakamura:2010zzi}\footnote{Also for the case of a neutralino LSP, 
constraints on the
SUSY parameter space from dark matter can change in case the right
Higgsino is light.}.  Even though the lightest sneutrino can also be
the LSP in the MSSM, direct detection experiments have ruled out this
possibility a long time ago \cite{Falk:1994es}.  In SUSY decays, within
the MSSM right squarks decay directly to the bino-like 
 neutralino,
leading to the standard missing momentum signature of supersymmetry.
Due to the extended gauge group, right squarks can decay also to
heavier neutralinos, leading to longer decay chains and potentially to
multiple lepton edges\footnote{Longer SUSY cascades from larger 
number of neutralino states have also been discussed in 
\cite{Ali:2009md,Belyaev:2012si}.}.
Decays of the heavier neutralinos also produce Higgses, both the 
$h^0_L$ and the $h^0_{BLR}$ appear, with ratios depending on the 
right higgsino content of the neutralinos in the decay chains.

The rest of this paper is organized as follows. In the next 
section we discuss the setup of the model, its particle content, 
superpotential and soft terms and the symmetry breaking. 
The phenomenologically most interesting mass matrices of the 
spectrum are given in section \ref{sec:masses} where we 
also discuss numerical results on the SUSY and Higgs mass 
eigenstates. Here, we focus on Higgs and slepton/sneutrino 
masses, which are the phenomenologically most interesting. 
In section \ref{sec:decays} we define some benchmark points for 
the model and discuss their phenomenologically most interesting decay 
chains. We then close with a short summary. In the appendix 
we give mass matrices not presented in the text, formulas for the
1-loop corrections in the Higgs sector and more information about 
the calculation of the RGEs, including anomalous dimensions as well
as the 1-loop $\beta$ functions for gauge couplings and gauginos.

\section{The model: $SU(3)_c \times SU(2)_L \times U(1)_{B-L}\times U(1)_R$}
\label{sec:model}

In this section we present the particle content of the model, 
its superpotential and discuss the symmetry breaking. 
We consider the simplest model based on the gauge 
group $SU(3)_{c}\times SU(2)_L\times U(1)_R \times U(1)_{B-L}$. 
We will call this the mBLR model below. 
As has been shown in \cite{Malinsky:2005bi} it can emerge 
as the low-energy limit of a certain class of $SO(10)$ GUTs broken 
along the ``minimal'' left-right symmetric chain
\begin{eqnarray}\label{chain}
SO(10) & \to &  SU(3)_c\times SU(2)_L\times SU(2)_R \times U(1)_{B-L} 
\\ \nonumber
& \to &
   SU(3)_c\times SU(2)_L\times U(1)_R \times U(1)_{B-L}.
\end{eqnarray}
The main virtue of this setting is that an MSSM-like gauge coupling 
unification is achieved with a sliding $U(1)_R \times U(1)_{B-L}$ 
breaking scale, i.e. this last stage can stretch down even to the 
electro-weak scale. Different from the previous works 
\cite{Malinsky:2005bi,DeRomeri:2011ie}, we assume that the first 
two breaking steps down to $U(1)_R \times U(1)_{B-L}$
happen {\em both} at (or sufficiently close to) the GUT scale. 
This assumption is used only for simplifying our setup, it does not 
lead to any interesting changes in phenomenology.

\subsection{Particle content, superpotential and soft terms}

\begin{table}
\begin{center} 
\begin{tabular}{|c|c|c|c|c|} 
\hline \hline 
\mbox{}\;\;\;\;\;\mbox{}& \; Superfield\;  & \; 
$SU(3)_c\times SU(2)_L\times U(1)_R\times U(1)_{B-L}$\;& \; Generations \; \\ 
\hline 
& \(\hat{Q}\)  
& \(({\bf 3},{\bf 2},0,+\frac{1}{6}) \) & 3 \\ 
&\(\hat{d^c}\) & 
\(({\bf \overline{3}},{\bf 1},+\frac{1}{2},-\frac{1}{6}) \)& 3 \\ 
&\(\hat{u^c}\) & 
\(({\bf \overline{3}},{\bf 1},-\frac{1}{2},-\frac{1}{6}) \)& 3 \\ 
&\(\hat{L}\)  & \(({\bf 1},{\bf 2},0,-\frac{1}{2}) \) & 3 \\ 
&\(\hat{e^c}\) & \(({\bf 1},{\bf 1},+\frac{1}{2},+\frac{1}{2}) \) & 3 \\ 
&\(\hat{\nu^c}\) & \(({\bf 1},{\bf 1},-\frac{1}{2},+\frac{1}{2}) \) & 3 \\ 
&\(\hat S\)& \(({\bf 1},{\bf 1},0,0) \) & 3 \\ 
\hline
&\(\hat{H}_u\)  
& \(({\bf 1},{\bf 2},+\frac{1}{2},0) \) & 1 \\ 
&\(\hat{H}_d\)  & \(({\bf 1},{\bf 2},-\frac{1}{2},0) \) & 1 \\ 
&\(\hat{\chi}_R\)  & \(({\bf 1},{\bf 1},+\frac{1}{2},-\frac{1}{2}) \) & 1 \\ 
&\(\hat{\bar{\chi}}_R\)  & \(({\bf 1},{\bf 1},-\frac{1}{2},+\frac{1}{2})\) 
& 1 \\ 
\hline \hline
\end{tabular} 
\end{center} 
\caption{\label{tab:fc}The Matter and Higgs sector field content of
the $U(1)_{R}\times U(1)_{B-L}$ model. Generation indices
have been suppressed. The $\hat S$ superfields are included to
generate neutrino masses via the inverse seesaw mechanism. Under
matter parity, the matter fields are odd while the Higgses are
even.}
\end{table}

The transformation properties of all matter and Higgs superfields
of the model are summarized in \TAB{tab:fc}. Apart from the 
MSSM fields, in the matter sector we have $\hat{\nu^c}$ and 
$\hat{S}$. The former are necessary in the extended gauge group 
for anomaly cancellation,\footnote{$\hat{\nu^c}$ is automatically 
part of the theory due to its $SO(10)$ origin.} while the fields 
$\hat{S}$ are included to explain neutrino masses by either an 
inverse \cite{Mohapatra:1986bd} or a linear \cite{Akhmedov:1995vm,
Akhmedov:1995ip} seesaw mechanism. Our Higgs sector, including 
the new fields $\hat{\chi}_R$ and $\hat{\bar{\chi}}_R$, 
is the minimal one for the breaking of $U(1)_{B-L}\times U(1)_R$ 
to $U(1)_{EM}$. 

The fields $\chi_{R}$ and $\bar{\chi}_{R}$ can be viewed as the
(electric charge neutral) remnants of $SU(2)_R$ doublets, which remain
light in the spectrum when the $SU(2)_R$ gauge factor is broken by the
vev of a $B-L$ neutral triplet down to the $U(1)_R$
\cite{Malinsky:2005bi}.  The presence of $\hat{\chi}_R$ and
$\hat{\bar{\chi}}_R$ makes it necessary to introduce an extra $Z_2^M$
matter parity, since otherwise $R$-parity is broken in a potentially
disastrous way, once these scalars acquire vacuum expectation values.
This $Z_2^M$ 
is not a particular feature of our setup; it is always needed in 
models where $U(1)_{B-L}$ is broken with doublets \cite{Martin:1992mq}. 
\footnote{In the normalization of \cite{Martin:1992mq} doublets have 
$U(1)_{B-L}=1$, i.e are ``odd'' under B-L.}

The relevant $R$-parity and $Z_2^M$ conserving superpotential is given 
by
\begin{equation}
W = W_{\rm MSSM} + W_{S}. 
\end{equation}
Here,
\begin{eqnarray}
W_{\rm MSSM} & = &  Y_u \hat{u^c}\hat{Q}\hat{H}_u 
  - Y_d \hat{d^c}\hat{Q}\hat{H}_d
    - Y_e \hat{e^c}\hat{L}\hat{H}_d
      +\mu\hat{H}_u\hat{H}_d \\ \nonumber
W_{S} & = & 
  Y_{\nu}\hat{\nu^c}\hat{L}\hat{H}_u
     +Y_s\hat{\nu^c}\hat{\chi}_R \hat{S} 
- \mu_{R}\hat{\bar{\chi}}_R\hat{\chi}_R
 + \mu_S \hat{S} \hat{S}.
\label{eq:superpot}
\end{eqnarray} 
where $Y_e$, $Y_d$ and $Y_u$ are the usual MSSM Yukawa couplings for
the charged leptons and the quarks. In addition there are the neutrino
Yukawa couplings $Y_\nu$ and $Y_s$; the latter mixes the $\nu^c$
fields with the $S$ fields giving rise to heavy SM-singlet 
pseudo-Dirac 
mass eigenstates. The term $\mu_R$ is completely analogous to the 
MSSM $\mu$ term. Note that the term $\mu_S$ is included to generate 
non-zero neutrino mass with an inverse seesaw mechanism. However, as 
always is done in inverse seesaw, we assume that $\mu_S$ is much smaller 
than all other dimensionful parameters of the model. Apart from 
neutrino masses themselves it will therefore not affect any of 
the mass matrices (or decays) of our interest.

Note that, besides the role it plays in neutrino physics,
the $Y_s$ coupling is relevant also for the Higgs phenomenology at the
loop level as it enters the mixing of $\chi_R$ and $\bar{\chi}_R$
Higgs fields with the $SU(2)_L$ Higgs doublets as well as the 
RGEs for $\chi_R$, see below. 

Following the notation and conventions of \cite{Allanach:2008qq} 
the soft SUSY breaking Lagrangian reads
\begin{eqnarray}
V_{soft} &  =&  \sum_{ij} m^2_{ij} \phi^*_i \phi_j+
\Big( \sum_a M_a \lambda_a \lambda_a +
 T_u \tilde{u}_R^* \tilde{Q} H_u 
- T_d \tilde{d}^*_R \tilde{Q} H_d
 + T_{\nu} \tilde{\nu}^*_R \tilde{L} H_u \nonumber \\
&&- T_e \tilde{e}^*_R \tilde{L} H_d
      +B_\mu H_u H_d- B_{\mu_{R}} \bar{\chi}_R \chi_R
     +T_s\tilde{\nu}^*_R \chi_R \tilde{S} 
     +B_{\mu_S} \tilde{S}\tilde{S}  \, + h.c. \Big) \,.
\label{eq:soft}
\end{eqnarray} 
The  first sum contains the scalar masses squared and the
second sum runs over all gauginos for the different gauge groups 
(called $\lambda_{BL}$, $\lambda_R$, $\lambda_L^i$  and $\lambda^\alpha_G$ 
in the following) and the second one contains the scalar masses squared. 
While 
$B_{\mu_S}$ is in principle a free parameter, a naive order 
of magnitude expectation for it is $B_{\mu_S} \sim \mu_S m_{SUSY}$. 
Thus, one expects that $B_{\mu_S}$ is much smaller than all other 
soft terms and can be safely neglected, see discussion of sneutrinos 
below.

To reduce the number of free parameters, in our numerical studies 
we will consider a scenario motivated by minimal supergravity. 
This means that we assume a GUT unification of all soft-breaking sfermion 
masses as well as a unification of all gaugino mass parameters
\begin{eqnarray}\label{eq:msugra}
m^2_0\delta_{ij} =  m_D^2 \delta_{ij} =  m_U^2 \delta_{ij} = m_Q^2 \delta_{ij}
= m_E^2 \delta_{ij} = m_L^2 \delta_{ij} = m_{\nu^c}^2 \delta_{ij} \\
\nonumber
 M_{1/2} =  M_{BL} = M_{R} = M_2 = M_3 
\end{eqnarray}
Also, for the trilinear soft-breaking coupling, the ordinary mSugra 
conditions are assumed
\begin{equation}\label{eq:msugra2}
 T_i = A_0 Y_i, \hspace{1cm} i = e,d,u,\nu,s \thickspace . 
\end{equation}
The GUT scale is chosen as the unification scale of $g_{BL}$, $g_R$
and $g_L$, while we allow $g_3$ to be slightly different, exactly as
in the MSSM.  A complete unification is assumed to happen due to GUT
threshold corrections.  For the remaining soft parameters in the Higgs
sector, $m^2_{H_d}$, $m^2_{H_u}$, $m^2_{\chi_R}$, $m^2_{{\bar\chi}_R}$
and $\mu, B_\mu, \mu_R$ and $B_{\mu_R}$, we have implemented two
different options. These are discussed in section
\ref{subsec:tadpoles}.

The presence of two Abelian groups gives rise to gauge kinetic mixing
\begin{equation}
\label{eq:offfieldstrength}
- \chi_{ab}  \hat{F}^{B-L, \mu \nu} \hat{F}^R_{\mu \nu} \,.
\end{equation}
This is allowed by gauge and Lorentz
invariance \cite{Holdom:1985ag}, as $\hat{F}^{B-L, \mu \nu}$ and
$\hat{F}^{R, \mu \nu}$ are gauge invariant, see
e.g. \cite{Babu:1997st}.  Even if $U(1)_R$ and $U(1)_{B-L}$ are
orthogonal in $SO(10)$ the kinetic mixing term will be induced during
the RGE running below the $SU(2)_R$ breaking scale because the light
fields remaining below the GUT scale can't be arranged in complete
$SO(10)$ multiplets: while all matter fields form three generations of
16-plets, $\hat{\chi}_R$ and $\hat{\bar{\chi}}_R$ induce off-diagonal
elements already in the 1-loop matrix of the anomalous dimensions
defined by $\gamma_{RBL} = \frac{1}{16 \pi^2} \mbox{Tr}Q_R
Q_{B-L}$. The matrix reads
\begin{equation}
\label{eq:AnaDim}
\gamma = \frac{1}{16 \pi^2} N 
\left( \begin{array}{cc} \frac{15}{2} & -\frac{1}{2} \\
 -\frac{1}{2} & \frac{9}{2} \end{array} \right) N.
\end{equation}
$N=\text{diag}(1,\sqrt{\frac{3}{2}})$ contains the GUT normalization
of the two Abelian gauge groups.  Our implementation follows the
description of \cite{Fonseca:2011vn}, where it is shown that terms of
the form as in eq.~(\ref{eq:offfieldstrength}) can be absorbed in the
covariant derivative by a re-definition of the gauge fields.
Therefore, we are going to work in the following with covariant
derivatives of the form
\begin{equation}\label{eq:deriv}
D_{\mu} = \partial_{\mu} - i Q^T_{\Phi} G A_{\mu} \,,
\end{equation}
where $Q^T_{\Phi}$ is a vector containing the charges of the 
field $\Phi$ with respect to the two Abelian gauge groups and 
$G$ is the gauge coupling matrix
\begin{equation}
 G =
 \left( 
  \begin{array}{cc}
   g_R      &  g_{RBL} \\
   g_{BLR}  &  g_{BL}
  \end{array}
 \right).
\end{equation}
$A_{\mu}$ contains the gauge bosons
$A_{\mu}=(A_{\mu}^R,A_{\mu}^{BL})^T$. Since the off-diagonal elements
in eq.~(\ref{eq:AnaDim}) are negative and roughly one order smaller
than the diagonal ones, it can be expected that the off-diagonal gauge
couplings at the SUSY scale are positive but also much smaller than
the diagonal ones. This is in some contrast to models in which kinetic
mixing arises due to the presence of $U(1)_Y \times U(1)_{B-L}$ \cite{
O'Leary:2011yq}.  In addition, a mixing term of the form
\begin{equation}
 M_{BL R} \lambda_{BL} \lambda_R 
\end{equation}
between the two gaugino $\lambda_{BL}$ and $\lambda_R$ will be present
\cite{Braam:2011xh}.  Since we have chosen the $SU(2)_R$ breaking
scale to be very close to the GUT scale we demand as additional
boundary conditions that the new parameters arising from kinetic
mixing vanish at the GUT scale, \textit{i.e.}
\begin{equation}
\label{eq:GUToffdiagonal}
g_{RBL} = g_{BLR} = 0 \, , \hspace{1cm} M_{BL R} = 0 \,.
\end{equation}
For more details on $U(1)$ mixing and its physical impact we refer the
interested reader also to recent papers
\cite{Chun:2010ve,Mambrini:2011dw, O'Leary:2011yq, Rizzo:2012rf}. Our
focus will be on the additional terms in the scalar mass matrices due
to the presence of non-diagonal couplings. 

\subsection{Tadpole equations and boundary conditions}
\label{subsec:tadpoles}

The $U(1)_{R}\times U(1)_{B-L}$ gauge symmetry is spontaneously
broken to the hypercharge $U(1)_{Y}$ by the vevs $v_{\chi_{R}}$ and
$v_{\bar\chi_{R}}$ of the scalar components of the $\hat\chi_R$ and
$\hat{\bar{\chi}}_R$ superfields while the $SU(2)_{L}\otimes 
U(1)_{Y}\to U(1)_{Q}$ is governed by the vevs $v_{d}$ and $v_{u}$ 
of the neutral scalar components of the $SU(2)_L$ Higgs doublets 
$H_d$ and $H_u$ up to gauge kinetic mixing effects. One can write
\begin{eqnarray}
\chi_R &=&
 \frac{1}{\sqrt{2}} \left( \sigma_R+ i \varphi_R+ v_{\chi_R}\right)
\,\,,\,\,
\bar{\chi}_R = \frac{1}{\sqrt{2}} \left( \bar{\sigma}_R 
+ i \bar{\varphi}_R + v_{\bar\chi_R}\right)\,,\\
H^0_d &=& \frac{1}{\sqrt{2}} \left( \sigma_d + i \varphi_d + v_d \right)
\,\,,\,\,\,\,\,\,\,
H^0_u = \frac{1}{\sqrt{2}} \left( \sigma_u + i \varphi_u + v_u \right)\,,
\end{eqnarray}
where the generic symbols $\sigma$ and $\varphi$ denote the CP-even
and CP-odd components of the relevant fields, respectively.

The minimum conditions for the four different vevs can be written at
tree-level as 
\begin{eqnarray}
&& t_d =  - B_{\mu} v_u 
      +  v_d 
  \left( m_{H_d}^2 + |\mu|^2 + 
  \frac{1}{8}  A_{LR,3} ( v_d^2 - v_u^2 ) + 
  \frac{1}{8}  A_{LR,2} ( v_{\bar{\chi}_R}^2 - v_{\chi_R}^2 ) \right)  \label{eq:tadd} \\
&& t_u = - B_{\mu} v_d 
     + v_u 
  \left( m_{H_u}^2 + |\mu|^2 - 
  \frac{1}{8}  A_{LR,3} ( v_d^2 - v_u^2 ) - 
  \frac{1}{8} A_{LR,2} ( v_{\bar{\chi}_R}^2 - v_{\chi_R}^2 ) \right)  \label{eq:tadu}\\
&& t_{\bar{\chi}_R}  = - B_{\mu_R} v_{\chi_R}
    +  v_{\bar{\chi}_R} 
  \left( m_{\bar{\chi}_R}^2 + |\mu_R|^2 + 
  \frac{1}{8}  A_{LR,1}  
 ( v_{\bar{\chi}_R}^2 - v_{\chi_R}^2 ) + 
  \frac{1}{8}A_{LR,2} (v_d^2 - v_u^2) \right)  \label{eq:tadRb} \\
&& t_{\chi_R}  =  - B_{\mu_R} v_{\bar{\chi}_R} 
     +  v_{\chi_R} 
  \left( m_{\chi_R}^2 + |\mu_R|^2 - 
  \frac{1}{8} A_{LR,1} 
( v_{\bar{\chi}_R}^2 - v_{\chi_R}^2 ) - 
  \frac{1}{8} A_{LR,2} (v_d^2 - v_u^2) \right)  \label{eq:tadR}
\end{eqnarray}

where we defined 
\begin{eqnarray}
 A_{LR,1} & = &  g_{BL}^2 + g_R^2 + g_{BLR}^2 + g_{RBL}^2 - 2 g_R g_{BLR} - 2 g_{BL} g_{RBL} \nonumber \, \\ 
 A_{LR,2} & = & g_R^2 + g_{RBL}^2 - g_R g_{BLR} - g_{BL} g_{RBL} \, \nonumber \\
 A_{LR,3} & = & g_L^2 + g_R^2 + g_{RBL}^2 \, .
\end{eqnarray}
For the vacuum expectation values we use the following 
parameterization:
\begin{eqnarray}\label{eq:deftbs}
 v_R^2 = v_{\chi_R}^2 + v_{\bar{\chi}_R}^2 ~, ~~
 v^2 = v_d^2 + v_u^2 \\ \nonumber
 \tbR = \frac{v_{\chi_R}}{v_{\bar{\chi}_R}} ~, ~~
 \tb = \frac{v_u}{v_d}.
\end{eqnarray}

The tadpole equations can analytically be solved for either (i) ($\mu,
B_\mu, \mu_R$, $B_{\mu_R}$) or (ii) ($\mu, B_\mu$, $m^2_{\chi_R}$,
$m^2_{{\bar\chi}_R}$) or (iii) ($m^2_{H_d}$, $m^2_{H_u}$,
$m^2_{\chi_R}$, $m^2_{{\bar\chi}_R}$).  Option (i) can be considered
the minimal version. We call this option CmBLR (constrained mBLR),
since it allows to define boundary conditions for all scalar soft
masses, $m^2_{H_d}=m^2_{H_u}=m_0^2$ and
$m^2_{\chi_R}=m^2_{{\bar\chi}_R}=m_0$ at $m_{GUT}$, reducing the
number of free parameters by four. This assumption, however, leads to
some important constraints on the parameter space, as we will discuss
next. Options (ii) and (iii) are more flexible. Option (ii) is similar
to the CMSSM with non-universal soft masses (NUHM)
\cite{Matalliotakis:1994ft,Polonsky:1994rz,Berezinsky:1995cj}, albeit
the non-universality is only in the $B-L$ sector. We will call this
the $\chi_R$mBLR (non-universal $\chi_R$ masses mBLR), and most of our
numerical results are based on this option. We mention option (iii)
for completeness, but we have not used it in our numerical studies.

As will be shown in section \ref{sec:MZP}, the mass of the
$Z'$-boson in the mBLR model is approximately given by
\begin{flalign}
 m_{Z'}^2 \simeq 
  \frac{1}{4}  A_{LR,1} v_R^2
\end{flalign}
We can use this expression and eqs (\ref{eq:tadd})-(\ref{eq:tadR}) to obtain 
an approximate relation between $m_{Z'}$ and $\mu_R$, $m^2_{\chi_R}$, 
$m^2_{{\bar\chi}_R}$ and $\tbR$. This leads to
\begin{equation}\label{eq:apprZP}
 m_{Z'}^2 \simeq 
          -2 (|\mu_R|^2 + m^2_{{\bar\chi}_R}) 
          + \frac{g_R^2}{4}  v^2 \cos(2 \beta)\frac{\tbR^2+1}{\tbR^2-1}
          + \Delta m^2_{\chi_R} \frac{2 \tbR^2}{\tbR^2 -1}
\end{equation}
where $\Delta m^2_{\chi_R} = m^2_{{\bar\chi}_R} - m^2_{\chi_R}$. 
We can roughly estimate $\Delta m^2_{\chi_R}$, if we make a 
mSugra-like assumption for the boundary conditions,  
$m^2_{{\bar\chi}_R} = m^2_{\chi_R}=m_0^2$ at the GUT scale. 
The running value of $\Delta m_{\chi_R}^2$ can then be found 
by a one-step integration of the RGEs at 1-loop level as:
\begin{equation}\label{eq:LLDelm}
 \Delta m^2_{\chi_R} \simeq \frac{1}{4\pi^2} \mbox{Tr}(Y_s Y^\dagger_s)
 (3 m^2_0 + A^2_0) \log\left(\frac{m_{GUT}}{M_{SUSY}}\right)
\end{equation}
with $T_s \simeq A_0 Y_s$. As eq.(\ref{eq:LLDelm}) shows, with these 
assumptions $\Delta m^2_{\chi_R}>0$ and the condition that $m_{Z'}$ of
eq.(\ref{eq:apprZP}) has to fulfill the experimental lower bound will
define an excluded area in the 3-dimensional parameter space 
[$\text{Tr}(Y_sY_s^{\dagger})$, $\tbR,m_{RGE}^2$], where $m_{RGE}^2=(3
m^2_0 + A^2_0)$. If we assume in addition that $Y_s$ is small enough 
to remain perturbative anywhere between the weak and
the GUT scale, a lower bound on $m_{RGE}^2$ as a function of $\tbR-1$
will result in the CmBLR.

\begin{figure}
 \begin{center}
  \hspace{-22mm}
  \begin{tabular}{cc}
   \resizebox{80mm}{!}{
\includegraphics{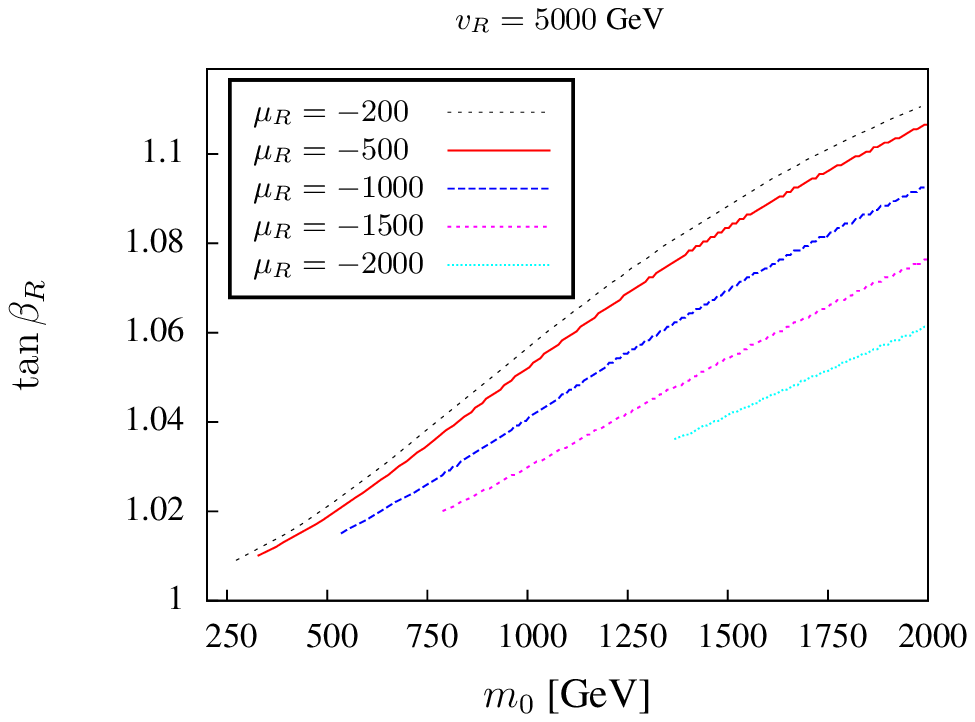}} &
   \resizebox{80mm}{!}{
\includegraphics{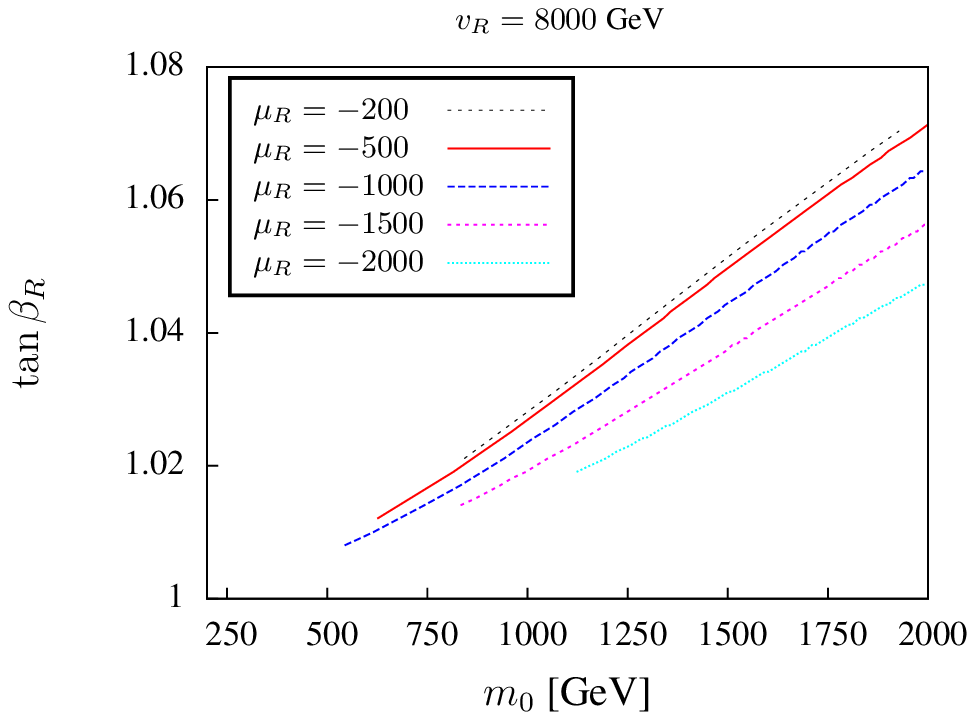}}  
  \end{tabular}
   \caption{\label{fig:m0vstbR}Constraints on the CmBLR parameter space from 
the condition of correct symmetry breaking, to the left: $v_R= 5$~TeV, 
to the right $v_R=8$~TeV. In both plots $M_{1/2} = 1000$~GeV,  
$\tan\beta = 10$ and $A_0 = 0$. Just above the lines for $\mu_R=-200$~GeV 
$|\mu_R|^2=0$, i.e. larger values of $\tbR$ do not lead to consistent 
solutions of the tadpole equations (for fixed $m_0$ and $A_0$). For 
detailed explanation see text. }
 \end{center}
\end{figure}

This can be understood in more details as follows. In the 
CmBLR $\Delta m^2_{\chi_R}\ge 0$, as shown by eq. (\ref{eq:LLDelm}) 
and the last term in eq. (\ref{eq:apprZP}) is positive only 
if $\tbR>1$. Since $\cos(2 \beta)<0$ the second term in eq. (\ref{eq:apprZP}) 
is positive only if $\tbR<1$. If $\Delta m^2_{\chi_R} \gsim | \frac{g_R^2}{4}  
v^2 \cos(2 \beta)|$, only solutions with $\tbR>1$ can be found. 
Since finally $|\mu_R|^2$ must be $|\mu_R|^2>0$ and $m^2_{\bar\chi_R}>0$ 
in the CmBLR we get the constraints on the parameter space shown in 
\FIG{fig:m0vstbR}. Here we show for two choices of $v_R$ 
contour lines of $\mu_R$ in the plane $(\tbR,m_0)$. Just above the 
lines for $\mu_R=-200$~GeV $|\mu_R|^2=0$, i.e. larger values of $\tbR$ 
do not lead to consistent solutions of the tadpole equations (for 
fixed $m_0$ and $A_0$). This restricts the model to values of 
$\tbR$ very close to $1$, as is clearly demonstrated in the 
figure. Note that for low values of $m_0$ the constraints on the 
viable region of $\tbR$ actually becomes stronger.
\footnote{$\tbR \simeq 1$ is also needed for a spectrum without 
tachyons since the additional D-terms can give large negative 
contributions to the sfermion masses, see below.}

No such constraint on $m_0$ and $A_0$ exists in the $\chi_R$mBLR, 
since here $\Delta m^2_{\chi_R}$ is a free parameter. However, 
if $(\Delta m^2_{\chi_R}/m^2_{\chi_R}) \ll 1$, values of $\tbR$ 
very close to $1$ are preferred by eq. (\ref{eq:apprZP}) in 
both, the CmBLR and the $\chi_R$mBLR.

\section{Masses}
\label{sec:masses}

In this section we give the most important mass matrices of the model 
at tree-level. In the numerical calculations we take also the 1-loop 
corrections \cite{Pierce:1996zz} into account, see appendix 
for more details. The numerical implementation of the model has been 
done using {\tt SPheno} \cite{Porod:2003um,Porod:2011nf}, for which the 
necessary subroutines and input files were generated using the 
package {\tt SARAH} \cite{Staub:2008uz,Staub:2009bi,Staub:2010jh}. The used model
files are included in the public version 3.1.0 of {\tt SARAH}. 

\subsection{Gauge bosons}
\label{sec:MZP}

In the basis ($W^0,B_{B-L},B_R$) 
the mass matrix for the neutral 
gauge bosons reads at tree-level
\begin{equation}\label{eq:ZMM}
 M_{VV}^2 = \frac{1}{4}
 \left(
 \begin{array}{ccc}
  g_L^2 v^2 & 
  -g_L g_{RBL} v^2 & 
  g_L g_R v^2 
  \\
  -g_L g_{RBL} v^2 & 
  g_{RBL}^2 v^2 + \tilde{g}_{BL}^2 v_R^2 &
  g_R g_{RBL} v^2 - \tilde{g}_{R} \tilde{g}_{BL} v_R^2
  \\
  -g_L g_R v^2 &
  g_R g_{RBL} v^2 - \tilde{g}_{R} \tilde{g}_{BL} v_R^2 &
  g_R^2 v^2 + \tilde{g}_{R}^2 v_R^2 
 \end{array}
 \right)
\end{equation}
where
\begin{equation}
 \tilde{g}_{BL} = (g_{BL} - g_{RBL}) ~,~~
 \tilde{g}_{R} = (g_R - g_{BLR}) \,.
\end{equation}
From eq. (\ref{eq:ZMM}) the masses of the photon, the $Z$ and 
the $Z'$ can be calculated analytically
\begin{equation}
 m_{\gamma} = 0 ~,~~
 m_{Z,Z'}^2 = \frac{1}{8} 
 \left(
  A v^2 + B v_R^2 \mp v_R^2 \sqrt{- 4 C \left( \frac{v^2}{v_R^2} \right) + 
  \left( A \left( \frac{v^2}{v_R^2} \right) + B \right)^2}
 \right)
 \label{eqn:massgammaZZp}
\end{equation}
with
\begin{eqnarray}
 A  &= & g_L^2 + g_R^2 + g_{RBL}^2
 \nonumber \\
 B  &=&  g_{BL}^2 + g_R^2 + g_{BLR}^2 + g_{RBL}^2 
         - 2 g_{BLR} g_R - 2 g_{RBL} g_{BL} 
 \nonumber \\
 C  &=&  g_L^2 (g_R - g_{BLR})^2 + g_{BL}^2 (g_L^2 + g_R^2) - 
  2 g_{BL} (g_L^2 + g_{BLR} g_R) g_{RBL} + (g_{BLR}^2 + g_L^2) g_{RBL}^2
\,. \nonumber \\
\end{eqnarray}
Expanding eq. \ref{eqn:massgammaZZp} in powers of $v^2/v_R^2$, we 
find up to first order
\begin{equation}
 m_Z^2 = \frac{C v^2}{4 B} ~,~~
 m_{Z'}^2 = \frac{(A B - C) v^2 + B^2 v_R^2}{4 B}\,. 
\end{equation}
In the limit $g_{BLR} = 0$ and $g_{RBL} = 0$ we then get
\begin{equation}
 m_Z^2 = \frac{(g_{BL}^2 g_L^2 +  g_{BL}^2 g_R^2 + g_L^2 g_R^2) v^2}
{4 (g_{BL}^2 + g_R^2)}
 ~,~~
 m_{Z'}^2 = \frac{g_R^4 v^2}{4 (g_{BL}^2 + g_R^2)} 
          + \frac{1}{4} (g_{BL}^2 + g_R^2) v_R^2\,. 
\end{equation}
ATLAS has recently published updated lower limits on $Z'$ searches 
\cite{ATLAS-CONF-2012-007}. Our $Z'$ corresponds to the $Z_\chi$ in
the notation of \cite{Erler:2011ud}, i.e. \cite{ATLAS-CONF-2012-007} 
gives a lower limit of $Z' \gsim 1.8$~TeV, which corresponds to 
roughly $v_R \gsim 5$~TeV for our choice of couplings\footnote{The
 condition that the gauge couplings reproduce correctly 
the standard model hypercharge, plus the assumption of unification 
lead to values of roughly $g_{BL}\sim 0.57$, $g_R \sim 0.45$, 
$g_{BLR}\sim 0.014$ and $g_{RBL} \sim 0.012$ at the SUSY scale.}, 
see, however, the discussion in section \ref{subsec:zp}.

\subsection{Higgs bosons}

\subsubsection{Pseudoscalar Higgs bosons}

At the tree level we find that in the $(\varphi_d,\varphi_u,\bar{\varphi}_R,
\varphi_R)$ basis the pseudoscalar sector has a block-diagonal form 
and reads in  Landau gauge

\begin{eqnarray}
M^2_{AA} = 
\left( \begin{array}{cc}
M^2_{AA,L} & 0 \\ 0 & M^2_{AA,R}
\end{array} \right)
\end{eqnarray}
with

\begin{eqnarray}
M^2_{AA,L} = B_\mu \left( \begin{array}{cc}
\tan \beta & 1 \\ 1 & \cot \beta
\end{array} \right) \,\,, \,\,
M^2_{AA,R} = B_{\mu_R} \left( \begin{array}{cc}
\tan \beta_R & 1 \\ 1 & \cot \beta_R
\end{array} \right) \,.
\end{eqnarray} 
From these four states two are Goldstone bosons which become
the longitudinal parts of the massive neutral vector bosons $Z$ and a
$Z'$. In the physical spectrum there are two pseudoscalars $A^0$ and $A^0_R$
with masses
\begin{equation}
 m_A^2 = B_{\mu} (\tan\beta + 1/\tan\beta) \,,\qquad
 m_{A_R}^2 = B_{\mu_R} (\tan\beta_R + 1/\tan\beta_R)\,.
\end{equation}

\subsubsection{Scalar Higgs bosons}

The tree-level CP-even Higgs mass matrix  in the $(\sigma_d,\sigma_u,
\bar{\sigma}_R,\sigma_R)$ basis reads 
\begin{eqnarray}\label{HHmat}
M_{hh}^2 &=&
\left(
\begin{array}{cc}
m_{LL}^2 & m_{LR}^2 \\
m_{LR}^{2,T} & m_{RR}^2
\end{array}
\right)\;,
\label{eq:Hmass}
\end{eqnarray}
where
\begin{eqnarray}\label{HHmat2}
m_{LL}^2 &=&
\left(
\begin{array}{cc}
(g_Z^2 + \frac{1}{4} g_{RBL}^2) v^2 c^2_{\beta} + m_A^2 s^2_{\beta } 
& -\frac{1}{2} 
\left( m_A^2 + (g_Z^2 + \frac{1}{4} g_{RBL}^2) v^2\right) s_{2 \beta} \\ 
 -\frac{1}{2} 
\left( m_A^2 + (g_Z^2 + \frac{1}{4} g_{RBL}^2) v^2\right) s_{2 \beta} 
& (g_Z^2 + \frac{1}{4} g_{RBL}^2) v^2 s^2_{\beta} + m_A^2 c^2_{\beta }
\end{array}
\right) 
\;,\\ 
m_{LR}^2 &=&
\frac{1}{4} \left(
\begin{array}{cc}
   (\tilde{g}_R g_R - \tilde{g}_{BL} g_{RBL}) v v_R c_{\beta} c_{\beta_R} 
&- (\tilde{g}_R g_R - \tilde{g}_{BL} g_{RBL}) v v_R c_{\beta} s_{\beta_R} \\ 
 - (\tilde{g}_R g_R - \tilde{g}_{BL} g_{RBL}) v v_R s_{\beta} c_{\beta_R} 
&  (\tilde{g}_R g_R - \tilde{g}_{BL} g_{RBL}) v v_R s_{\beta} s_{\beta_R} 
\end{array}
\right)\;, \\ 
m_{RR}^2 &=&
\left(
\begin{array}{cc}
 \tilde{g}_{Z_R}^2 v_R^2 c^2_{\beta_R} +m_{A_R}^2 s^2_{\beta_R} 
& -\frac{1}{2} \left( m_{A_R}^2 + \tilde{g}_{Z_R}^2 v_R^2\right) s_{2 \beta_R} \\
  -\frac{1}{2} \left( m_{A_R}^2 + \tilde{g}_{Z_R}^2 v_R^2\right) s_{2 \beta_R}
& \tilde{g}_{Z_R}^2 v_R^2 s^2_{\beta_R} + m_{A_R}^2 c^2_{\beta_R} 
\end{array}
\right)\;,  
\end{eqnarray}
$s_x = \sin(x)$, $c_x=\cos(x)$ ($x=\beta, \beta_R, 2 \beta, 2
\beta_R$), $g_Z^2 = (g_L^2 + g_R^2)/4$, $\tilde{g}_{Z_R}^2 = 
(\tilde{g}_{BL}^2 + \tilde{g}_R^2)/4$.
The matrix $m_{LL}^2$ contains the standard MSSM doublet mass matrix.
To see this explicitly one has to integrate out the additional Higgs 
fields in the $v_R \to \infty$ limit which yields a shift in the gauge 
couplings such that the MSSM limit is achieved. $m_{RR}^2$ corresponds 
to the $U(1)_{R}\times U(1)_{B-L}$ Higgs bosons and $m_{LR}^2$ provides 
the essential mixing between the two sectors. 

Note that it is straightforward to show that the determinant of the 
mass matrix eq.~(\ref{HHmat}) goes to zero, whenever one of the 
parameters ($(\tan\beta-1),m_A,(\tbR-1),m_{A_R}$) goes to zero. One can 
also calculate analytically that in the limit of $v_R\to \infty$ the 
lightest eigenvalue of eq.~(\ref{HHmat}) obeys the MSSM tree-level 
limit for $h^0$. For finite $v_R$ corrections to $m_{LL}^2$ 
appear, of the order of $g_R^2 v^3/v_R$, which lead to a shift 
in the lightest eigenvalue. Thus the MSSM tree-level upper bound of 
$m_{h^0}^{\rm tree} \le m_{Z^0}$ for the lightest Higgs can be 
violated.

\subsubsection{Numerical examples}

\begin{figure}
 \begin{center}
  \hspace{-22mm}
  \begin{tabular}{cc}
   \resizebox{83mm}{!}{\includegraphics{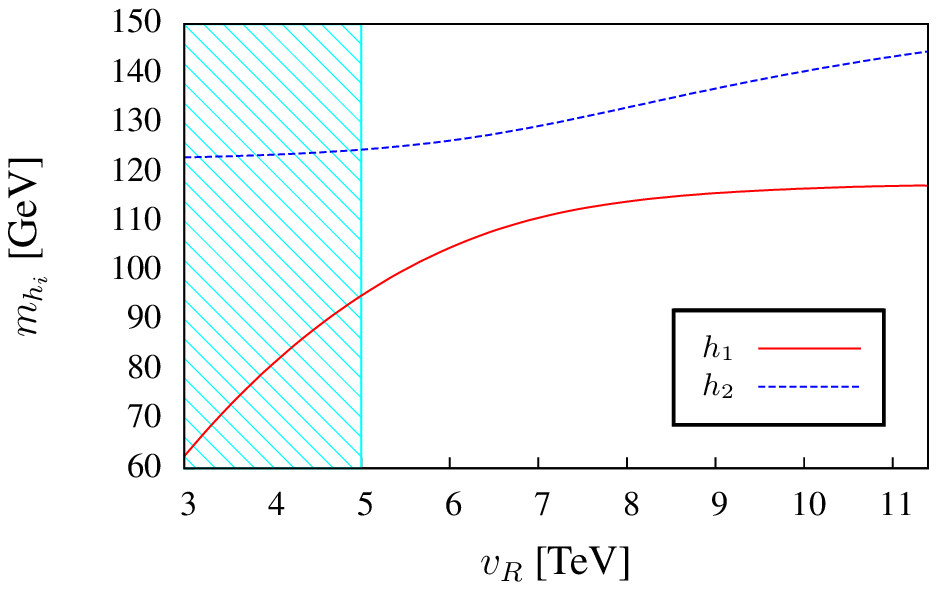}} &
   \resizebox{80mm}{!}{\includegraphics{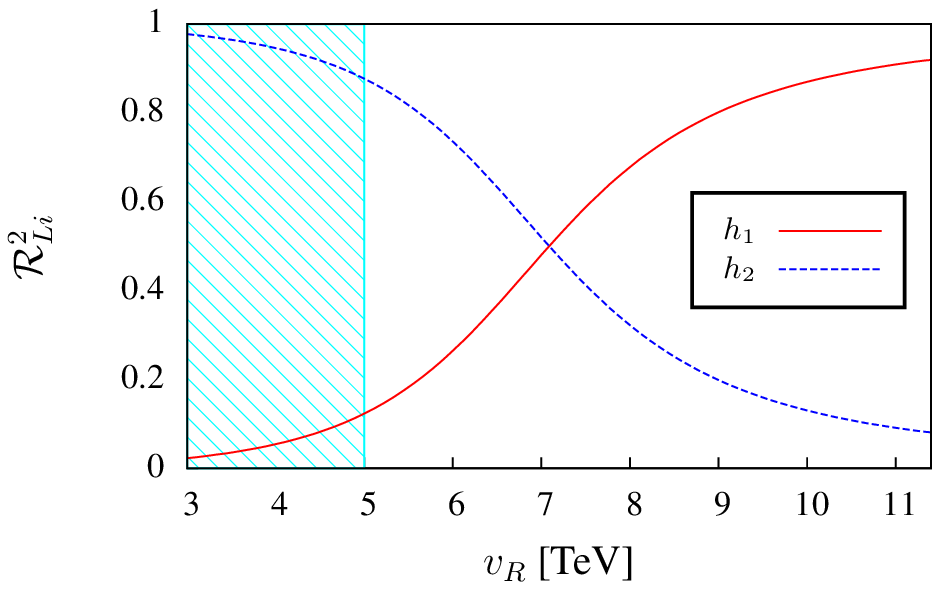}}  
  \end{tabular}
   \caption{\label{fig:hhvsvR}Example plot for the masses (left) and
   ``leftness'' (right) of two lightest eigenvalues of the CP-even
   Higgs sector as a function of $v_R$ for fixed choices of the other
   parameters: $m_0 = 250$~GeV, $M_{1/2} = 800$~GeV, $\tan\beta = 10$,
   $A_0 = 0$, $\tan\beta_R = 0.94$, $\mu_R = -800$~GeV, $m_{A_R} =
   2350$~GeV.}
 \end{center}
\end{figure}

In \FIG{fig:hhvsvR} we show the two lightest Higgs boson masses 
(to the left) together with
\begin{equation}
{\cal R}^2_{Li} \equiv R^2_{i1} + R^2_{i2}
\end{equation} 
(to the right) as a function of $v_R$. Here $i=1,2$ labels the light
Higgs scalars in the model and $R_{ij}$ is the rotation matrix which 
diagonalizes the CP-even Higgs sector.  Note that the quantity ${\cal
R}^2_{Li}$, which reaches one in the MSSM limit, is a rough measure of
how much the corresponding Higgs with index $i$ resembles an MSSM
Higgs boson. We will call this leftness. 
Roughly speaking, the smaller this quantities is, the
smaller is the $i$-th Higgs coupling to the $Z$- and $W$-bosons,
implying a reduced production cross sections at LEP, Tevatron and the
LHC.

Since the MSSM Higgs and the two additional Higgs $\chi_R$ and
$\bar{\chi}_R$ are both charged under $U(1)_R$ the two lightest
Higgs states mix due to additional D-terms in the CP-even Higgs matrix
(see eq. (\ref{HHmat2})). Thus, both the masses and the mixing of the
two lightest Higgs $h_1$ and $h_2$ depend strongly on $v_R$, see
\FIG{fig:hhvsvR}.  In this example, up to approximately $v_R$=6~TeV
$h_1$ is mostly the singlet Higgs whereas $h_2$ is the MSSM-like
Higgs. For larger $v_R$ a level-crossing occurs and the situation is
reversed. Note that, although $m_0$, $M_{1/2}$ and $A_0$ have rather
moderate values in this example, $h_2$ has a mass of the order of
$m_{h_2} \simeq 125$~GeV for $v_R \simeq 5-6$~TeV, i.e. the D-terms
have shifted the MSSM-like Higgs mass into the region preferred by
ATLAS and CMS.

\begin{figure}
 \begin{center}
  \hspace{-22mm}
  \begin{tabular}{cc}
   \resizebox{83mm}{!}{\includegraphics{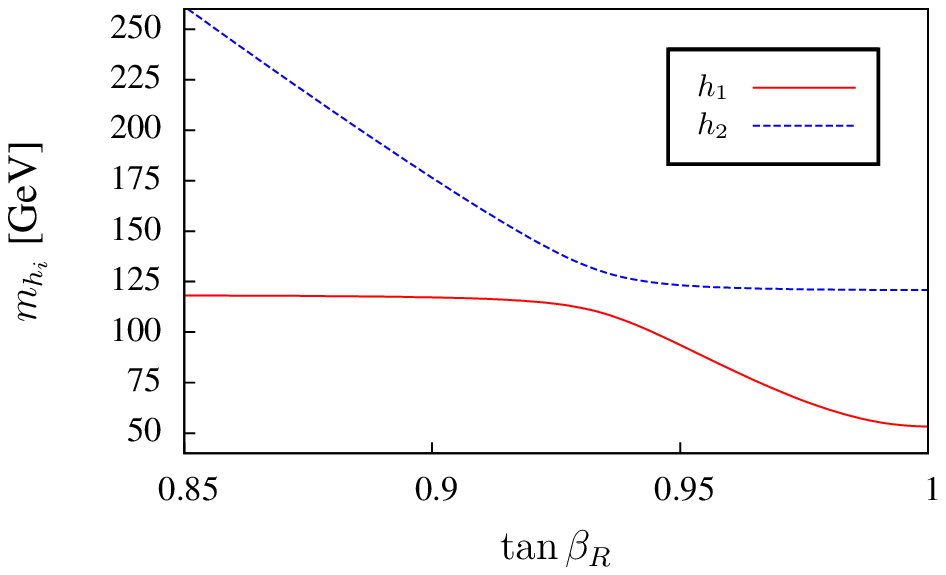}} &
   \resizebox{80mm}{!}{\includegraphics{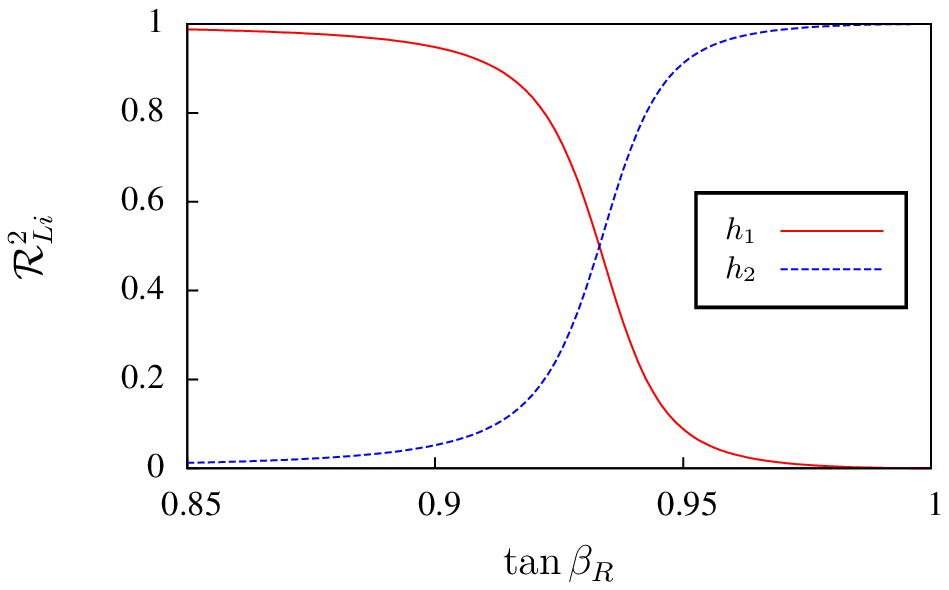}}  
  \end{tabular}
   \caption{\label{fig:hhvstbR}Example plot for the masses (left) and
   ``leftness'' (right) of two lightest eigenvalues of the CP-even
   Higgs sector as a function of $\tbR$ for fixed choices of the other
   parameters: $m_0 = 250$~GeV, $M_{1/2} = 800$~GeV, 
    $\tan\beta = 10$, $A_0 = 0$, $v_R = 6000$~GeV, $\mu_R = -800$~GeV, 
    $m_{A_R} = 2350$~GeV}
 \end{center}
\end{figure}

\begin{figure}
 \begin{center}
  \hspace{-22mm}
  \begin{tabular}{cc}
   \resizebox{83mm}{!}{\includegraphics{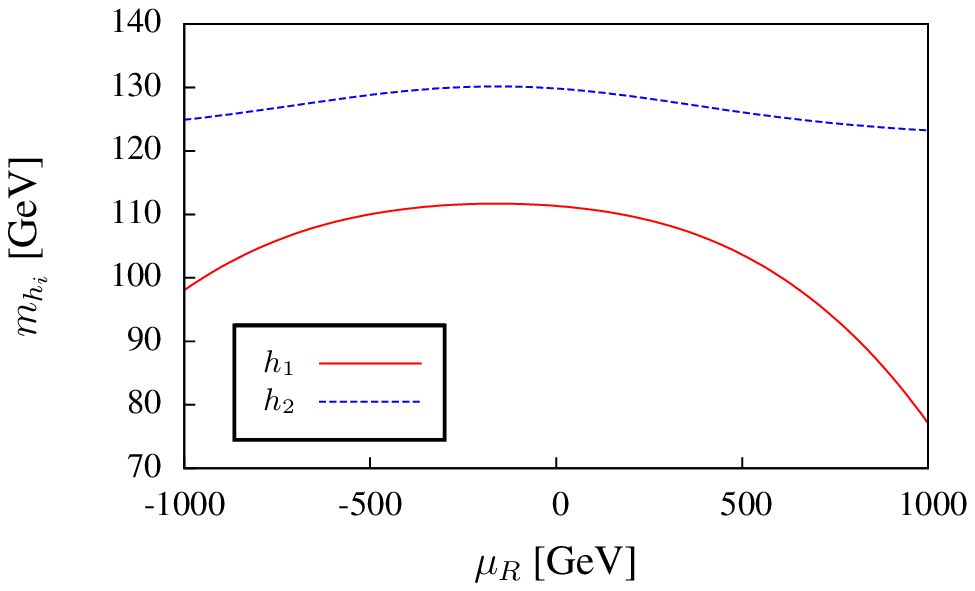}} &
   \resizebox{80mm}{!}{\includegraphics{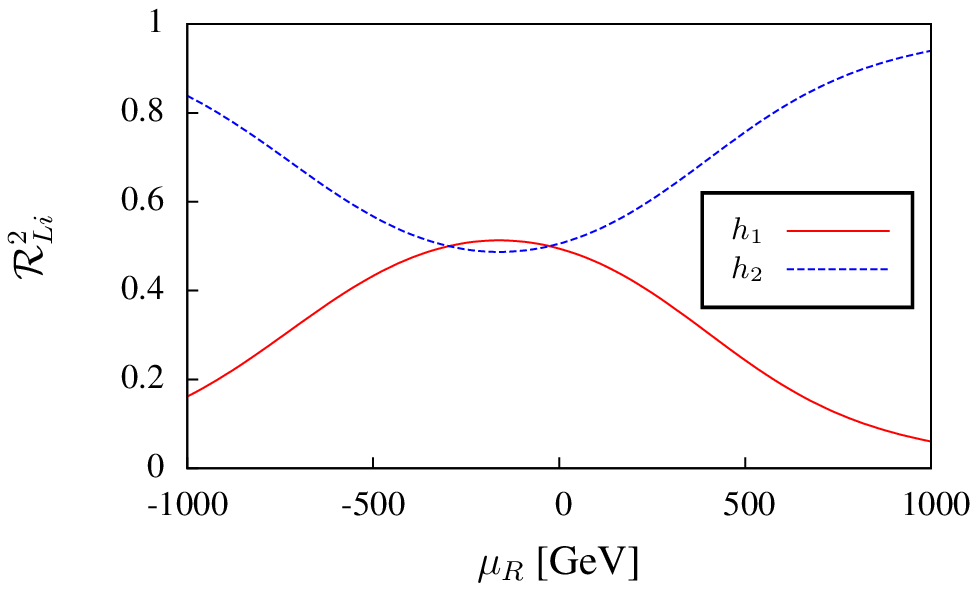}}  
  \end{tabular}
   \caption{\label{fig:hhvsmuR}Example plot for the masses (left) and
   ``leftness'' (right) of two lightest eigenvalues of the CP-even
   Higgs sector as a function of $\mu_R$ for fixed choices of the other
   parameters: $m_0 = 250$~GeV, $M_{1/2} = 800$~GeV, 
    $\tan\beta = 10$, $A_0 = 0$, $v_R = 6000$~GeV, $\tan\beta_R = 0.94$, 
    $m_{A_R} = 2350$~GeV}
 \end{center}
\end{figure}

In \FIG{fig:hhvstbR} masses and ``leftness'' of the two lightest
Higgs eigenstates are plotted against $\tbR$. For $\tbR$ close to
$\tbR=1$ one gets a very light singlet Higgs as expected, see
discussion above.  As is the case when varying $v_R$ a level-crossing
appears also when $\tbR$ is changed. In this figure the $\chi_R$mBLR
version was used, thus we can put $\tbR<1$.  Note, however, that the
masses of $h_1$ and $h_2$ show a behavior which is symmetric with
respect to $|\tbR-1|$. Again the figure demonstrates that the 
MSSM limit of the lightest Higgs mass can be violated at the expense 
of a reduced coupling of the MSSM-like state to SM gauge bosons.

1-loop corrections play in general an important role, not only for the
MSSM-like Higgs but also in the singlet sector. This can be seen in
\FIG{fig:hhvsmuR}, where we plot masses and mixings of $h_1$ and
$h_2$ versus the parameter $\mu_R$. Increasing or decreasing $\mu_R$,
respectively, changes the mass of the mostly-singlet Higgs by
considerable factors. In fact, for larger values of $|\mu_R|$ one can
get easily a negative mass squared for $h_1$, which is of course
forbidden phenomenologically. The importance of $\mu_R$ stems from
1-loop contributions to the Higgs mass matrix with a higgsino-right in
the loop. Loop corrections for the mostly-singlet Higgs are, in fact, 
even more important numerically than for the MSSM-like Higgs and 
many points which are allowed at tree-level lead to tachyonic states, 
once 1-loop corrections are taken into account.

Finally we note, that in the plots in this section we have not 
shown the regions excluded by LEP or the LHC searches, since we 
were interested only in showing the parameter dependencies of our 
numerical results. In the study points of the next section, however, 
we have taken care that our points survive all known experimental 
constraints.

\subsection{Neutrinos}
The mBLR model contains beside the usual three left-handed neutrinos six
additional states which are singlets with respect to the SM group.
The corresponding mass matrix is in the basis $(\nu_L, \nu^c,S)$ given by 
\begin{equation} 
\label{eq:neutrinoMM}
m_{\nu} = \left( 
\begin{array}{ccc}
0 &\frac{1}{\sqrt{2}} v_u Y_{\nu}^{T}  &0\\ 
\frac{1}{\sqrt{2}} v_u Y_\nu  &0 &\frac{1}{\sqrt{2}} v_{\chi_R} Y_s \\ 
0 &\frac{1}{\sqrt{2}} v_{\chi_R} Y_{s}^{T}  & \mu_S\end{array} 
\right)  \,.
\end{equation}
This matrix is diagonalized by \(U^\nu\):
\begin{equation} 
U^{\nu,*} m_{\nu} U^{\nu,\dagger} = m^{dia}_{\nu}.
\end{equation}
Eigenvalues for the three light (and mostly left-handed) neutrinos 
can be found in the seesaw approximation as:
\begin{equation} 
\label{eq:meff}
m_{\nu}^{\rm eff} = - \frac{v_u^2}{v_R^2} Y_{\nu}^{T}Y_s^{-1} \mu_S 
(Y_s^T)^{-1}Y_{\nu} \,.
\end{equation}
Neutrino data implies that either $Y_{\nu}$ and/or $\mu_S$ is small and 
in inverse seesaw the smallness of neutrino mass is attributed to the 
smallness of the latter. As we will
discuss in section \ref{sec:lepton_LFV}, the bounds on rare lepton decays
imply that the off-diagonal terms of $Y_s$ and $Y_\nu$ have to be small
compared to their diagonal entries, unless their diagonal values are 
small too. 

The smallness of $\mu_S$ implies that the six heavy states form three 
``quasi-Dirac'' pairs. For vanishing off-diagonal entries in
$Y_s$ and $Y_\nu$ a good estimate of the masses of the
heavy states is:
\begin{equation}
m_{\nu_h,ii} \simeq 
\pm \sqrt{|Y_{\nu,ii}|^2 v^2_u + |Y_{s,ii}|^2 v^2_{\chi_R} } \, .
\end{equation}

\subsection{Sparticles}

\subsubsection{Neutralinos}

The mass matrix of the neutralinos reads in the basis $(\lambda_{BL}, \lambda_L^0, \tilde{h}_d^0, 
\tilde{h}_u^0, \lambda_R, \tilde{\bar{\chi}}_R, \tilde{\chi}_R)$:
\begin{flalign}
 M_{\tilde{\chi}^0} = 
 \left(
 \begin{array}{ccccccc}
  M_{BL} & 0 & - \frac{1}{2} g_{RBL} v_d & \frac{1}{2} g_{RBL} v_u & \frac{M_{BLR}}{2} & 
  \frac{1}{2} v_{\bar{\chi}_R} \tilde{g}_{BL} & - \frac{1}{2} v_{\chi_R} \tilde{g}_{BL} \\
  0 & M_2 & \frac{1}{2} g_L v_d & - \frac{1}{2} g_L v_u & 0 & 0 & 0 \\
  - \frac{1}{2} g_{RBL} v_d & \frac{1}{2} g_L v_d & 0 & -\mu & - \frac{1}{2} g_R v_d & 0 & 0 \\
  \frac{1}{2} g_{RBL} v_u & - \frac{1}{2} g_L v_u & -\mu & 0 & \frac{1}{2} g_R v_u & 0 & 0 \\
  \frac{M_{BLR}}{2} & 0 & - \frac{1}{2} g_{R} v_d & \frac{1}{2} g_{R} v_u & M_{R} & 
  - \frac{1}{2} v_{\bar{\chi}_R} \tilde{g}_{R} & \frac{1}{2} v_{\chi_R} \tilde{g}_{R} \\
  \frac{1}{2} v_{\bar{\chi}_R} \tilde{g}_{BL} & 0 & 0 & 0 & 
  - \frac{1}{2} v_{\bar{\chi}_R} \tilde{g}_{R} & 0 & - \mu_R \\
  - \frac{1}{2} v_{\chi_R} \tilde{g}_{BL} & 0 & 0 & 0 & 
  \frac{1}{2} v_{\chi_R} \tilde{g}_{R} & -  \mu_R & 0  \\
 \end{array}
 \right) \,.
\end{flalign}
The eigenvalues of this matrix are not completely arbitrary. Since
$U(1)_{B-L}\times U(1)_R$ is broken in such a way as to produce
correctly the SM group $U(1)_Y$ in the limit of $v<<v_R$ the matrix
contains one state which corresponds to the MSSM bino, ${\tilde B}$,
which is a superposition of $\lambda_{BL}$ and $\lambda_{R}$. In
addition the matrix contains an orthogonal state, which we will call
${\tilde B}_{\perp}$ in the following.

For CMSSM like boundary conditions, $M_{BL}=M_2=M_R=M_{1/2}$,  
the bino is usually the lightest of the three gaugino like 
states, with the ${\tilde W}$ being approximately twice as heavy. 
The ${\tilde B}_{\perp}$ is very often mixed with one of the right 
higgsinos, and, since $M_{BL}$ at low energies is much smaller 
than $v_R$ this mixing is often important. In addition, there is 
the standard quasi-Dirac pair of ``left'' higgsinos, plus two more 
states which are mostly right higgsinos. Of the latter one is usually 
rather heavy, while the other can be light, if $\mu_R$ is small.

\begin{figure}
 \begin{center}
  \hspace{-22mm}
  \begin{tabular}{cc}
   \resizebox{83mm}{!}{\includegraphics{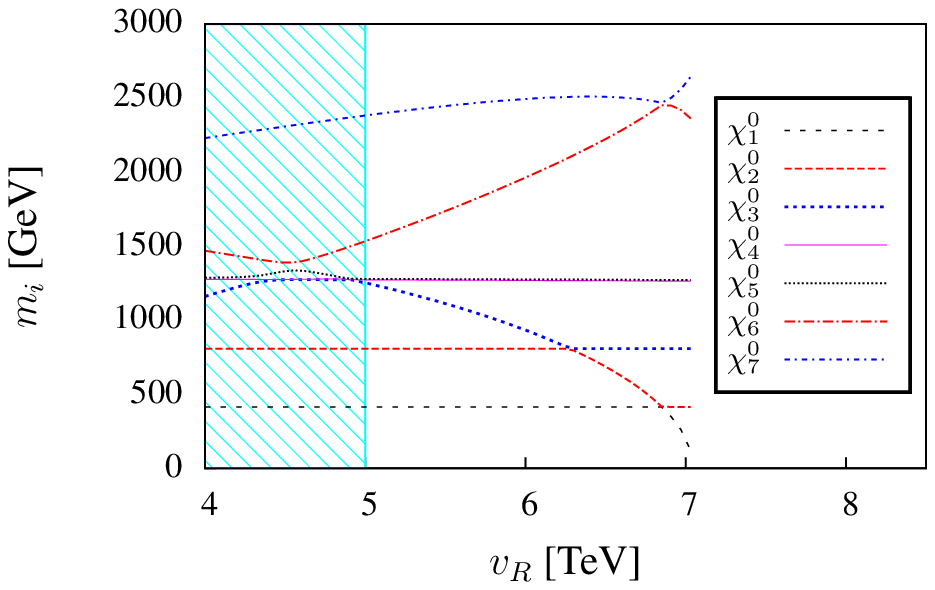}} &
   \resizebox{80mm}{!}{\includegraphics{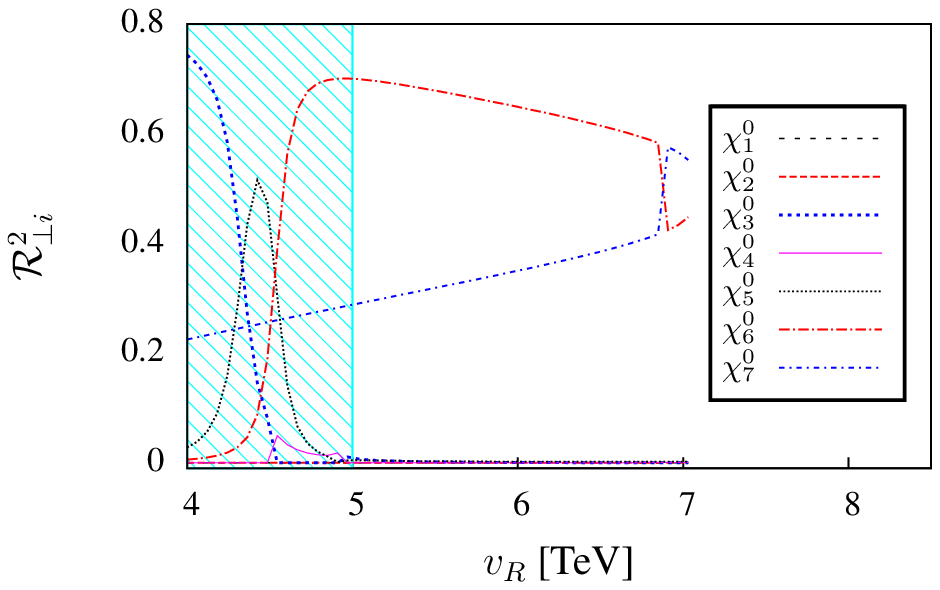}}  
  \end{tabular}
   \caption{\label{fig:ChivsvRGUT} Neutralino masses (left) and 
   ${\cal R}^2_{{\perp i}}$ (right) versus $v_R$ for otherwise 
   fixed choice of parameters: $m_0 = 1000$~GeV, $M_{1/2} = 1000$~GeV, 
    $\tan\beta = 10$, $A_0 = -600$, $\tan\beta_R = 1.04$. This plot 
   uses the CmBLR version of the model.}
 \end{center}
\end{figure}

\begin{figure}
 \begin{center}
  \hspace{-22mm}
  \begin{tabular}{cc}
   \resizebox{83mm}{!}{\includegraphics{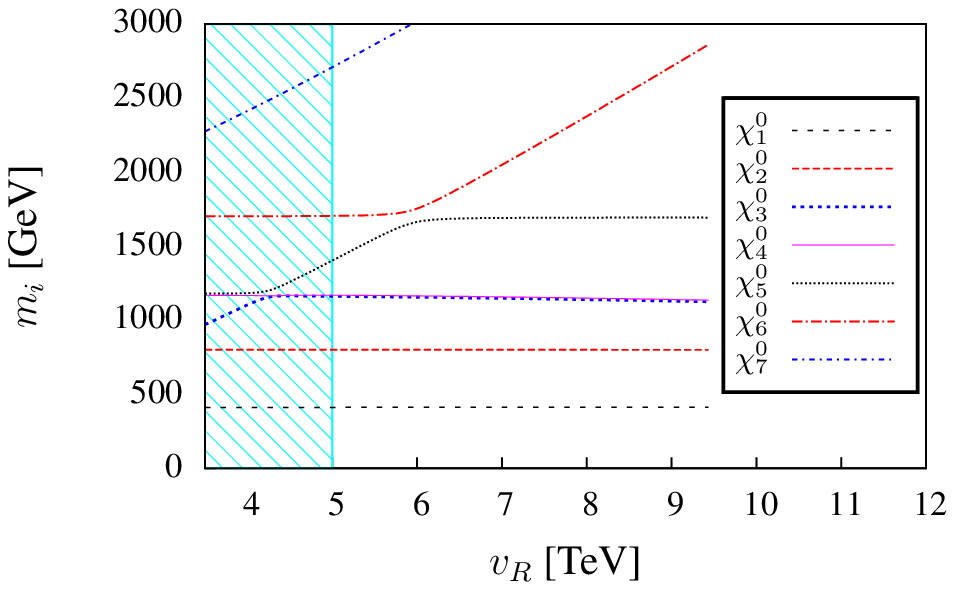}} &
   \resizebox{80mm}{!}{\includegraphics{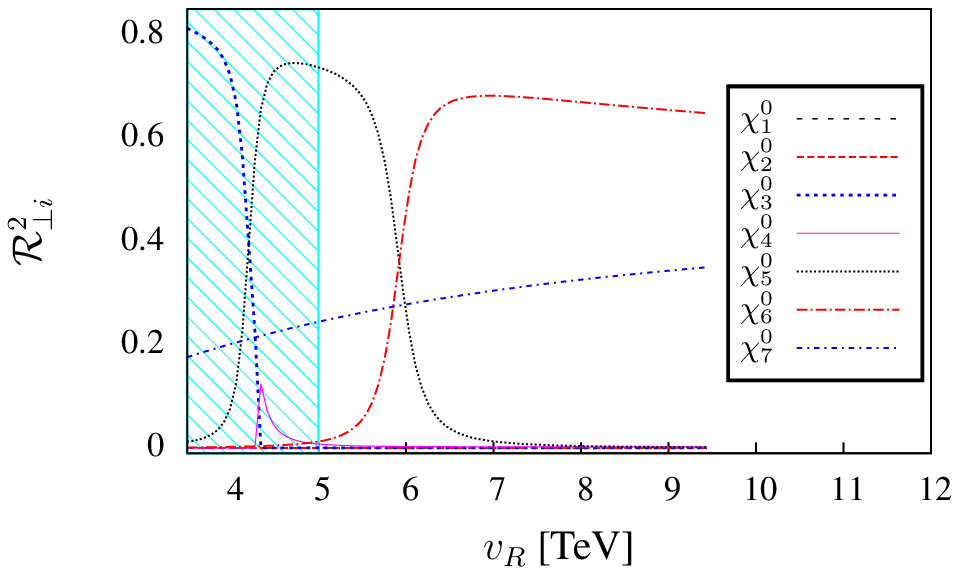}}  
  \end{tabular}
   \caption{\label{fig:ChivsvR} Neutralino masses (left) and ${\cal
   R}^2_{{\perp i}}$ (right) versus $v_R$ for otherwise fixed choice
   of parameters: $m_0 = 630$~GeV, $M_{1/2} = 1000$~GeV, $\tan\beta =
   10$, $A_0 = 0$, $\tan\beta_R = 1.05$, $\mu_R = -1700$~GeV, $m_{A_R}
   = 4800$~GeV. This plot 
   uses the $\chi_R$mBLR version of the model.}
 \end{center}
\end{figure}

In \FIG{fig:ChivsvRGUT} the neutralino masses and ${\cal
R}^2_{{\perp i}}$ for CmBLR are plotted against $v_R$ for some
arbitrary choice of other parameters. As discussed, there are in total
seven eigenstates. Of special interest is the ${\tilde B}_{\perp}$, so
in the plot on the right we show the percentage of ${\tilde
B}_{\perp}$ (${\cal R}^2_{{\perp i}}$) in the corresponding mass
eigenstate. Here ${\cal R}^2_{{\perp i}}=1$ means that the $i$-th
neutralino is a pure ${\tilde B}_{\perp}$. As one can see in
\FIG{fig:ChivsvR} the masses and mixing of the three new states
depend strongly on $v_R$. For small $v_R$ all three states mix to each
other.  Increasing $v_R$ leads to a decoupling of the lighter
higgsino-right from the ${\tilde B}_{\perp}$ which decreases in mass
since $\mu_R$ becomes smaller for large $v_R$ while the masses of the
two remaining states get large.  Since the MSSM Neutralinos mix very
little with the new states, there are four eigenvalues which show
almost no dependence on the parameters $v_R$ and $\mu_R$.

In \FIG{fig:ChivsvR} the neutralino masses and ${\cal
R}^2_{{\perp i}}$ are plotted against $v_R$ for the case of
$\chi_R$mBLR. In this calculation, $\mu_R$ and $m_{A_R}$ can take
fixed values while $v_R$ is varied freely. Two of the three new
Neutralino states are a mixture of the higgsino-right and ${\tilde
B}_{\perp}$ and therefore depend on $v_R$. Since the lighter
higgsino-right hardly mix to the ${\tilde B}_{\perp}$ it has a
constant mass at $m_{\tilde{h}_R} \simeq |\mu_R| = 1700$~GeV in this
example. The lighter of the two new states that show dependence on
$v_R$ is mostly a ${\tilde B}_{\perp}$, whereas the one with larger
mass is mostly a higgsino-right. The smaller $v_R$ the smaller the
mixing between these two states and thus the larger the coupling of
the mostly ${\tilde B}_{\perp}$-state to the MSSM particles.  This
will be important when we discuss LHC phenomenology in
section \ref{sec:susycascades}.

\begin{figure}
 \begin{center}
  \hspace{-22mm}
  \begin{tabular}{cc}
   \resizebox{83mm}{!}{\includegraphics{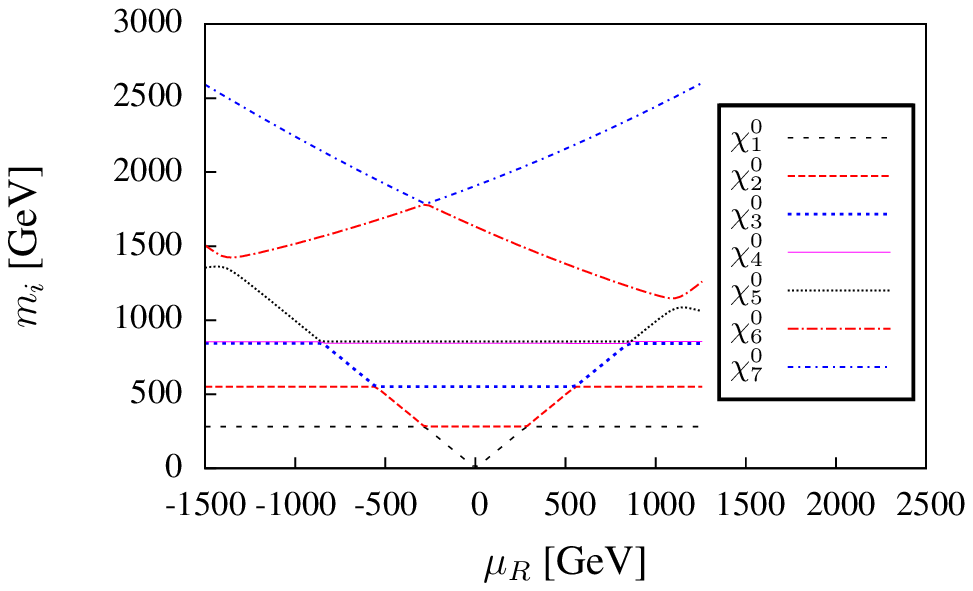}} &
   \resizebox{80mm}{!}{\includegraphics{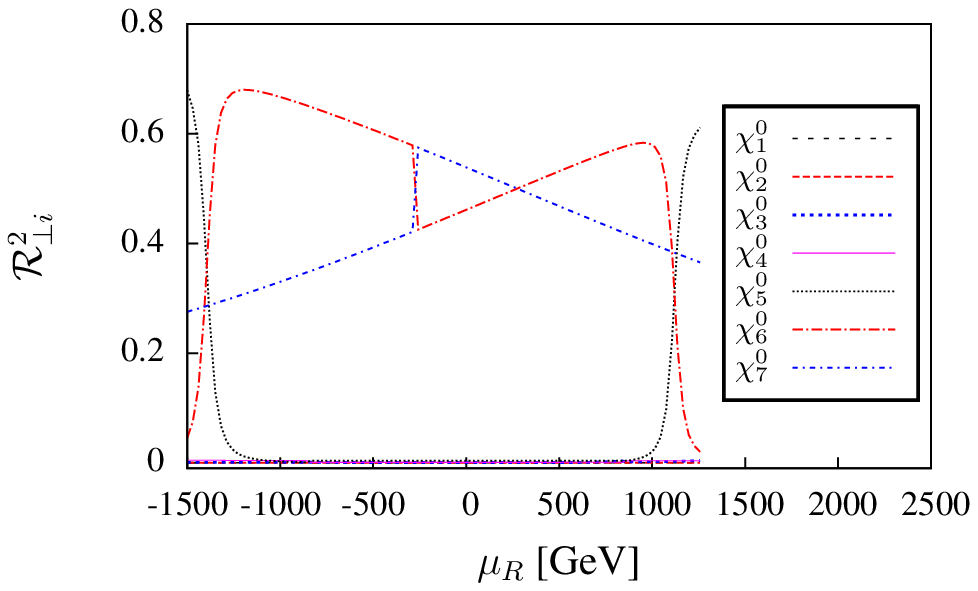}}  
  \end{tabular}
   \caption{\label{fig:ChivsmuR} Neutralino masses (left) and ${\cal
   R}^2_{{\perp i}}$ (right) versus $\mu_R$ for otherwise fixed choice
   of parameters: $m_0 = 400$~GeV, $M_{1/2} = 700$~GeV, $\tan\beta =
   10$, $A_0 = 0$, $v_R = 5000$~GeV, $\tan\beta_R = 1.05$, $m_{A_R}
   =3000$~GeV.}
 \end{center}
\end{figure}

The dependence of the neutralino masses and ${\cal R}^2_{{\perp i}}$
on $\mu_R$ is shown in \FIG{fig:ChivsmuR}. Since the
higgsino-right and the ${\tilde B}_{\perp}$ mix, all three states show
a dependence on $\mu_R$. The state which hardly mixes to the ${\tilde
B}_{\perp}$ decreases in mass for small $|\mu_R|$. So we can easily
have a higgsino-right as LSP choosing $\mu_R$ close to zero. The state
which is mostly the ${\tilde B}_{\perp}$ gets a smaller mass for large
$|\mu_R|$, while the one which is mostly a higgsino-right increases in
mass.

\subsubsection{Sleptons and sneutrinos}

In models in which lepton number is broken, the scalar neutrinos 
split into a real and an imaginary part with slightly different 
masses \cite{Hirsch:1997vz}. Since we assume that the smallness 
of neutrino masses is due to the smallness of the parameter 
$\mu_S$ (and, therefore, $B_{\mu_S}$ is supposed to be small too), 
this splitting between sneutrino mass eigenstates is too small to be 
of any relevance, except neutrino masses themselves. 

Neglecting $\mu_S$ and $B_{\mu_S}$ the sneutrino mass matrix 
is given by
\begin{flalign}
 M_{\tilde{\nu}}^2 =
 \left(
 \begin{array}{ccc}
  m_{LL,\tilde{\nu}}^2 & 
  \frac{1}{\sqrt{2}} Y_{\nu}^{\dagger} v_u (A_0 - \cot{\beta} \mu) &
  \frac{1}{2} v_u v_{\chi_R} Y_{\nu}^{\dagger} Y_{s} \\
  \frac{1}{\sqrt{2}} Y_{\nu} v_u (A_0 - \cot{\beta} \mu^{\ast}) & 
  m_{RR,\tilde{\nu}}^2 &
  \frac{1}{\sqrt{2}} Y_{s} v_{\chi_R} (A_0 - \cot{\beta_R} \mu_R^{\ast}) \\
  \frac{1}{2} v_u v_{\chi_R} Y_{s}^{\dagger} Y_{\nu} &
  \frac{1}{\sqrt{2}} Y_{s}^{\dagger} v_{\chi_R} (A_0 - \cot{\beta_R} \mu_R) &
  m_S^2 + \frac{v_{\chi_R}^2}{2} Y_{s}^{\dagger} Y_s
 \end{array}
 \right)
\end{flalign}
where
\begin{flalign}
 m_{LL,\tilde{\nu}}^2 & =
 m_L^2 + \frac{v_u^2}{2} Y_{\nu}^{\dagger} Y_{\nu} ~- 
 \nonumber \\
 & \frac{1}{8} \left(
 (g_{BL}^2 + g_{BLR}^2 - g_{BL} g_{RBL}) (v_{\bar{\chi}_R}^2 - v_{\chi_R}^2) + 
 (g_L^2 + g_R^2 + g_{BL} g_{RBL}) (v_d^2 - v_u^2)
 \right) {\bf 1} 
 \nonumber \\
 m_{RR,\tilde{\nu}}^2 & =
 m_{\nu}^2 + \frac{v_u^2}{2} Y_{\nu} Y_{\nu}^{\dagger} ~+ 
 \frac{v_{\chi_R}^2}{2} Y_{s}^{\dagger} Y_s  + 
 \nonumber \\
 & \frac{1}{8} \left(
 (g_{BL}^2 + g_R^2 + g_{BLR}^2  + g_{RBL}^2 -2 g_{BL} g_{RBL} - 2 g_R g_{BLR}) 
(v_{\bar{\chi}_R}^2 - v_{\chi_R}^2) + 
 \right.
 \nonumber \\
 & ~~~~ 
\left. (g_R^2 + g_{RBL}^2 - g_{BL} g_{RBL} - g_R g_{BLR}) (v_d^2 - v_u^2)
 \right) {\bf 1} 
\end{flalign}
For charged sleptons one gets:
\begin{flalign}
 M_{\tilde{l}}^2 =
 \left(
 \begin{array}{cc}
  m_{LL,\tilde{l}}^2 & \frac{1}{\sqrt{2}} Y_l^{\dagger} v_d (A_0 - \tb \mu) \\
  \frac{1}{\sqrt{2}} Y_l v_d (A_0 - \tb \mu^{\ast}) & m_{RR,\tilde{l}}^2
 \end{array}
 \right)
\end{flalign}
where
\begin{flalign}
 m_{LL,\tilde{l}}^2 & =
 m_L^2 + \frac{v_d^2}{2} Y_l^{\dagger} Y_l - 
   \frac{1}{8} \Big(
 (g_{BL}^2 + g_{BLR}^2 - g_{BL} g_{RBL} - g_R g_{BLR}) 
(v_{\bar{\chi}_R}^2 - v_{\chi_R}^2) \nonumber \\ & \hspace{1.0cm} - 
 (g_L^2 - g_{BL} g_{RBL} - g_R g_{BLR}) (v_d^2 - v_u^2) 
 \Big) {\bf 1} 
 \nonumber \\
 m_{RR,\tilde{l}}^2 & =
 m_E^2 + \frac{v_d^2}{2} Y_l Y_l^{\dagger} + 
  \frac{1}{8} \Big( (g_{BL}^2 - g_R^2 + g_{BLR}^2 - g_{RBL}^2) 
 (v_{\bar{\chi}_R}^2 - v_{\chi_R}^2)   \nonumber \\ & \hspace{1.0cm} -
 (g_R^2 + g_{RBL}^2 + g_{BL} g_{RBL} + g_R g_{BLR}) (v_d^2 - v_u^2) 
 \Big) {\bf 1} 
\end{flalign}

\begin{figure}
 \begin{center}
  \hspace{-22mm}
  \begin{tabular}{cc}
   \resizebox{80mm}{!}{\includegraphics{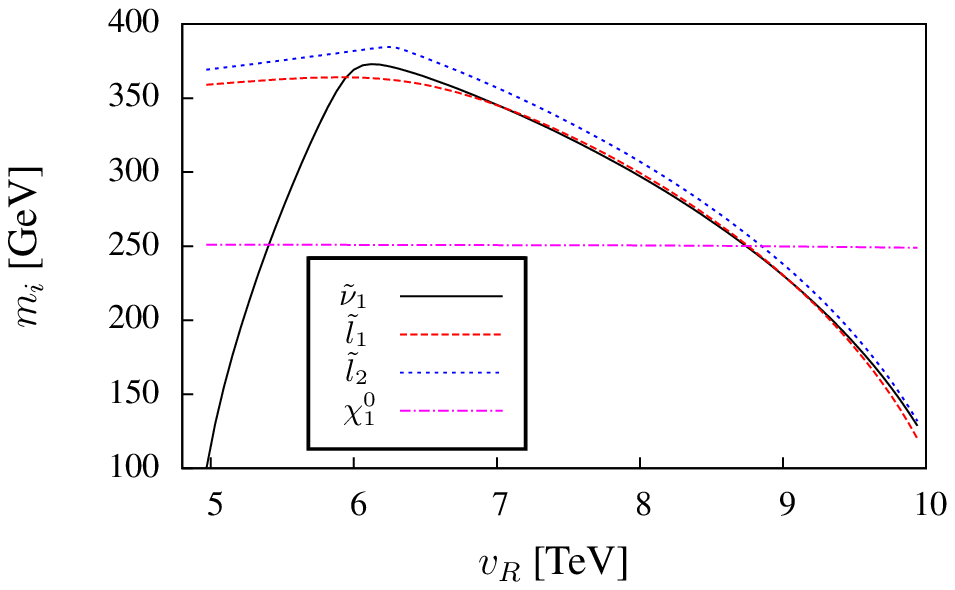}}    &
   \resizebox{80mm}{!}{\includegraphics{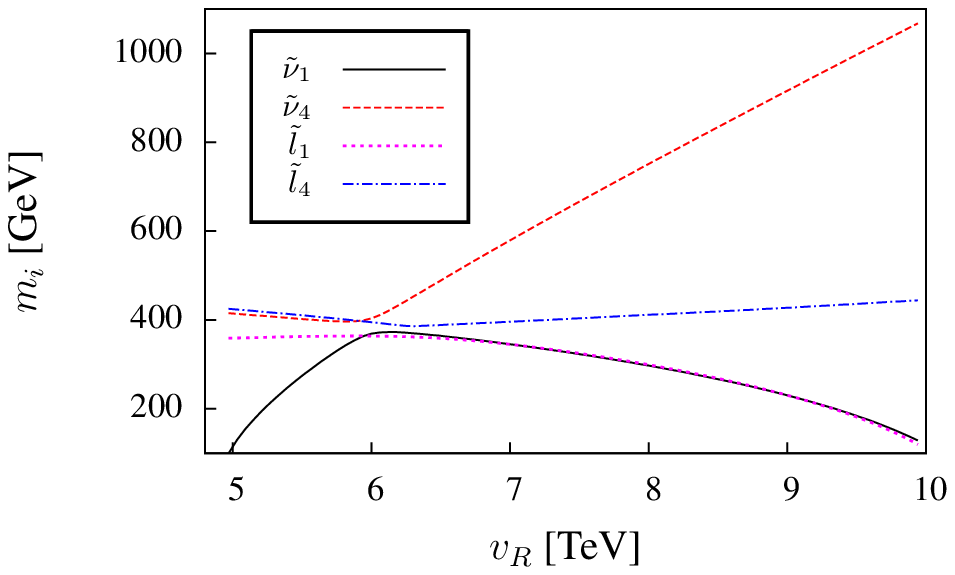}} \\
   \resizebox{80mm}{!}{\includegraphics{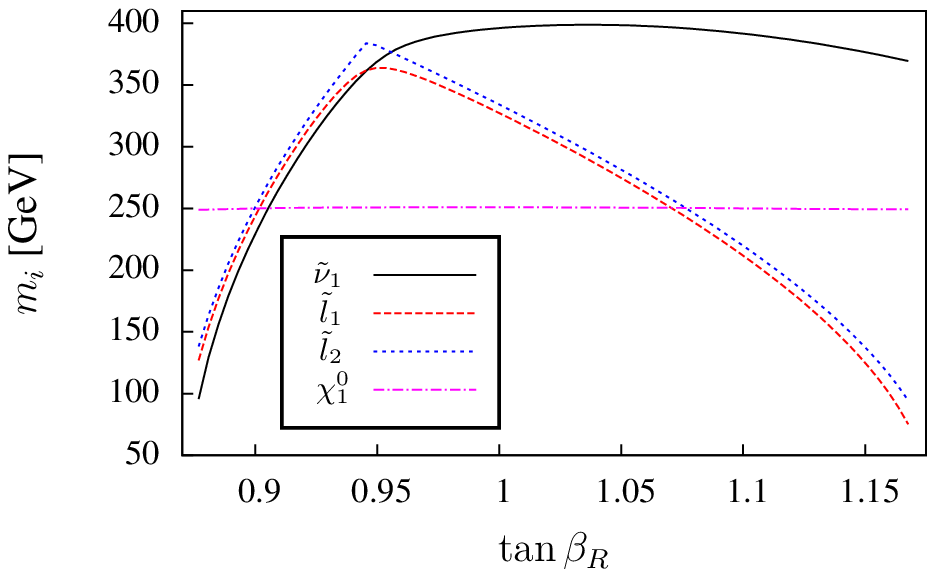}}    &
   \resizebox{80mm}{!}{\includegraphics{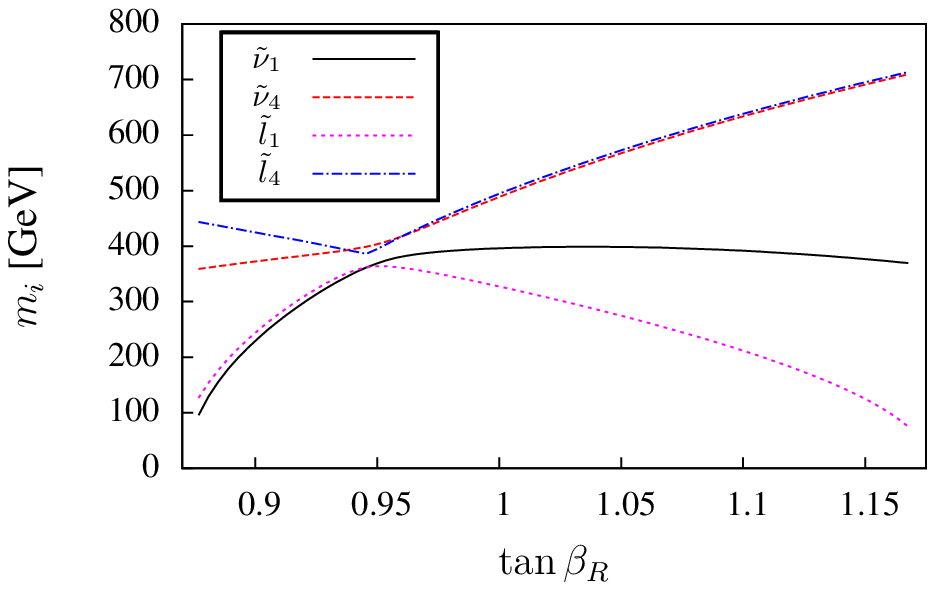}} \\
   \resizebox{80mm}{!}{\includegraphics{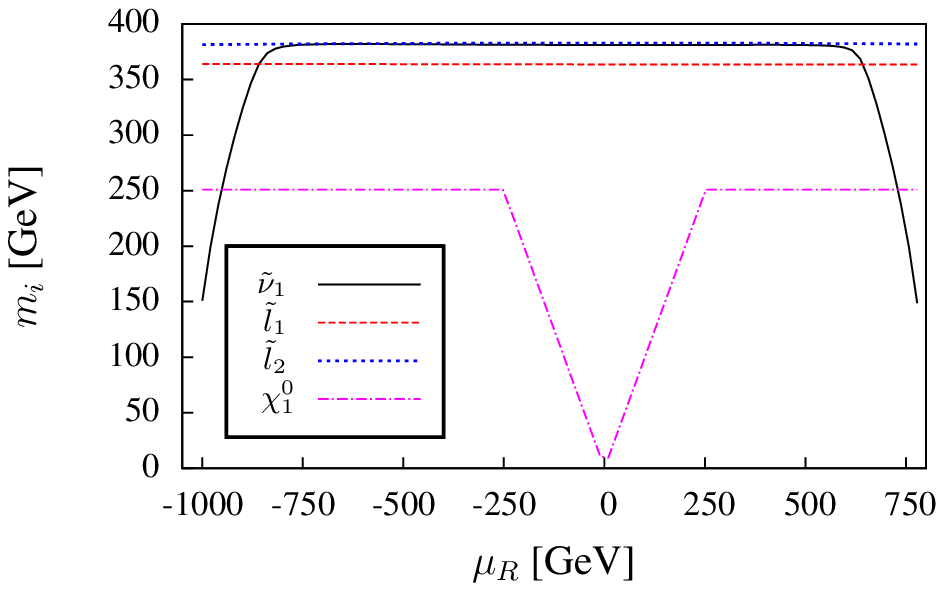}}    &
   \resizebox{80mm}{!}{\includegraphics{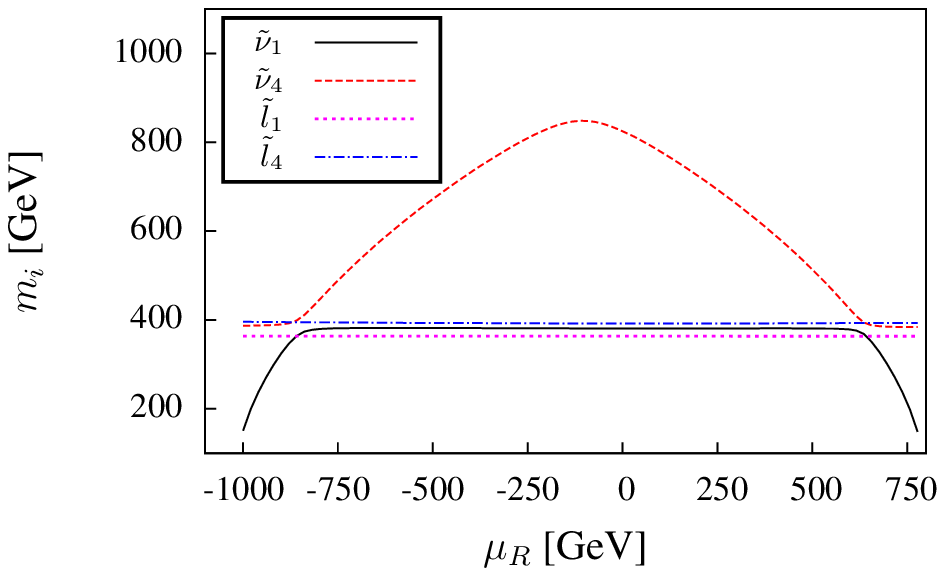}}
  \end{tabular}
\caption{\label{fig:massSvSl} Lightest slepton (and neutralino) masses
as function of $v_R$, $\tbR$ and $\mu_R$ for a fixed but arbitrary
choice of other parameters: $m_0 = 220$~GeV, $M_{1/2} = 630$~GeV,
$\tan\beta = 10$, $A_0 = 0$, $\tan\beta_R = 0.95$, $v_R = 6000$,
$\mu_R = -850$~GeV, $m_{A_R} = 2200$~GeV, $Y_{s,ii}=0.3$. 
Plots on the left show a
zoom into the light mass region, such that mass differences between
the lightest sneutrino and the lightest charged slepton are resolved,
figures to the right show the overall dependence, for a discussion see
text.}
\end{center}
\end{figure}

In \FIG{fig:massSvSl} 
sneutrino and slepton masses are plotted
against $v_R$, $\tbR$ and $\mu_R$. The figures on the left show a zoom
into the region of the lightest states, whereas the figures on the
right show a larger range of masses for a better understanding of the
overall behavior. To see which particle is the LSP, while varying
$v_R$, $\tbR$ and $\mu_R$, we included in all plots on the left the
mass of the lightest neutralino state. This state is always a Bino,
except for the plot against $\mu_R$. Here the LSP becomes a
higgsino-right for $|\mu_R|<250$~GeV. The plots show that the masses
depend strongly on the choice of $v_R$ and $\tbR$. In the case of
charged sleptons the dependence on $v_R$ and $\tbR$ comes only from
additional D-terms at the tree level.  This is different for
sneutrinos. Here we can have an interplay between new D-terms and
terms coming from the coupling $Y_s$ which both depend on $v_R$ and
$\tbR$.  The additional D-terms force left sparticle to become light
for $\tbR<1$ while for $\tbR>1$ right sparticle masses decrease. Up to
$v_R=6$~TeV $\tilde{\nu}_1$ is a right handed sneutrino and therefore
the mass increase for increasing $v_R$. For $v_R>6$~TeV the mass of
$\tilde{\nu}_1$ drops down again since here it is mainly a left handed
sneutrino. Thus, increasing $v_R$ leads to a level-crossing in the
mass spectrum of left and right handed sneutrinos. The same holds
for the sleptons. In the plot against $v_R$ the mass of the right
sneutrino decreases much faster for $v_R<6$~TeV than the mass of the
right sleptons. This is due to the off-diagonal terms proportional to
$Y_s$, which contain also $\mu_R$ in the sneutrino mass matrix. These
terms mix the scalar component of $\hat{S}$ to $\tilde{\nu}_R$.  Thus
for low values of $v_R$ in this example the LSP is neutral, which is
allowed, whereas for larger values of $v_R$ (with left sleptons being
light) there are parts of the parameter space, where the lightest
slepton is charged, which is phenomenologically forbidden. Whether 
in the left sector charged or neutral states are lighter, depends 
heavily on the choice of parameters. 

Varying $\tbR$ the right slepton masses
decrease faster than the right sneutrino masses for $\tbR>0.95$
due the additional  sneutrino mixing.
Since the sneutrino and slepton masses depend strongly on the choice
of $v_R$, $\tbR$ and $\mu_R$ one obtains limits on combination on 
these parameters. On the
one hand, one has to avoid tachyonic states 
and on the other hand one has to take care not to get charged
sleptons as LSP. The combination of
both conditions forces us to choose $\tbR$ 
close to one and gives us an upper limit on $v_R$ and $|\mu_R|$ 
as function of $|\tbR-1|$. 

\section{Constraints, sample spectra and decays}
\label{sec:decays}

In this section we discuss several interesting phenomenological
aspects which potentially allow the BLR model to be discriminated 
from the MSSM at the LHC and exemplify the most important features 
for a few study points. We include a discussion of the direct 
production of new states and characteristic changes in the cascade 
decays of supersymmetric particles.  

For brevity we will call these benchmark points BLRSP1- BLRSP5, 
the corresponding input parameters are listed in 
\TAB{tab:input}. All of these points have been calculated with the
$\chi_R$mBLR version of the model. However, note that for BLRSP5 
the input is chosen to be consistent with the CmBLR variant.

A few comments on the input parameters and the resulting mass spectra
are in order, before we discuss the phenomenology in detail. As shown
below, the bounds on rare lepton flavour violating decays require
$Y_{\nu}$ and $Y_s$ to be essentially flavour-diagonal, unless these
couplings are very small.  Therefore we have chosen $Y_{\nu}$ and
$Y_s$ diagonal as starting point implying that all points satisfy
trivially the LFV constraints. A correct explanation for the neutrino
angles then requires flavour violating entries in the parameter
$\mu_S$, which we do not give in \TAB{tab:input}, since they
are irrelevant for collider phenomenology.

\begin{table}
\begin{center} 
\begin{tabular}{|c|c|c|c|c|c|} 
\hline 
\hline
  & BLRSP1 & BLRSP2 & BLRSP3 & BLRSP4 & BLRSP5 \\
\hline
\multicolumn{6}{|c|}{cMSSM} \\
\hline
$m_0$ [GeV]            & 470  &  1000 &  120 & 165  &  500  \\
$M_{1/2}$ [GeV]        & 700  &  1000 &  780 & 700  &  850  \\
$\tan\beta$            &  20  &    10 &   10 &  10  &   10  \\
$A_0$                  &   0  & -3000 & -300 &   0  & -600  \\
\hline
\multicolumn{6}{|c|}{Extended gauge sector} \\
\hline
$v_R$ [GeV]         &  4700  & 6000  &  6000  &  5400 &  5000 \\
$\tan\beta_R$       &  1.05  & 1.025 &  0.85  &  1.06 &  1.023 \\
$\mu_R$ [GeV]       & -1650  & -780  & -1270  &   260 &  (-905) \\
$m_{A_R}$ [GeV]     &  4800  & 7600  &   800  &  2350 &  (1482) \\
\hline
\multicolumn{6}{|c|}{Yukawas} \\
\hline
$Y_{\nu,11}$             & 0.04   & 0.1   & 0.1  & 0.1  & 0.1  \\
$Y_{\nu,22}$             & 0.04   & 0.1   & 0.1  & 0.1  & 0.1  \\
$Y_{\nu,33}$             & 0.04   & 0.1   & 0.1  & 0.1  & 0.1  \\
$Y_{s,11}$               & 0.04   & 0.042 & 0.3  & 0.3  & 0.3  \\
$Y_{s,22}$               & 0.05   & 0.042 & 0.3  & 0.3  & 0.3  \\
$Y_{s,33}$               & 0.05   & 0.042 & 0.3  & 0.3  & 0.3  \\
\hline 
\hline
\end{tabular} 
\end{center} 
\caption{\label{tab:input} Parameters of the various study points. In
BLRSP1-BLRSP4 $\mu_R$ and $m_{A_R}$ are input whereas in BLRSP5 the
constrained version of the model has been used and, thus, these two
parameters are output. For a discussion of these points see text.}
\end{table}

The input values of \TAB{tab:input} lead to the mass spectrum
shown in \TAB{tab:spectra}. We give the masses and in brackets
the particle character. In case of mixed states the two largest
components, for example (${\tilde W},{\tilde h}_L$), are given 
where
the first entry accounts for the larger contribution. If the ordering
in the composition changes like in the case of $m_{\tilde{u}_{5,6}}$ we use 
squared brackets. Therefore we have $(\tilde{c}_L,\tilde{u}_L)$ 
for $m_{\tilde{u}_{5}}$ and $(\tilde{u}_L,\tilde{c}_L)$ for 
$m_{\tilde{u}_{6}}$.
In all cases input parameters have been chosen such, that the squark and
gluino masses are outside the region currently excluded by pure CMSSM
searches at ATLAS \cite{ATLAS-WEB} and CMS \cite{CMS-WEB}. Since (a) we
expect the missing momentum signal to be smaller in these points than
in a true CMSSM spectrum and (b) our squark spectra are less
degenerate than the CMSSM case, we believe this is a conservative
choice. Two of the points have a sneutrino LSP (BLRSP1 and BLRSP3),
while three points have a neutralino LSP (for BLRSP2 and BLRSP5 mostly
a bino, for BLRSP4 a state which is mostly a ${\tilde h}_R$).

\begin{table}
\begin{center} 
\begin{tabular}{|c|c|c|c|c|c|} 
\hline 
\hline
  & BLRSP1 & BLRSP2 & BLRSP3 & BLRSP4 & BLRSP5 \\
\hline
\multicolumn{6}{|c|}{Sneutrinos and Sleptons} \\
\hline
$m_{\tilde{\nu}_{1}}$ [GeV]       & 102.3  $(\tilde{\nu}_R)$                   & 
                                    797.0  $(\tilde{\nu}_R)$                   &   
                                    91.6   $(\tilde{\nu}_R,\tilde{\nu}_L)$     & 
                                    542.3  $(\tilde{\nu}_R,\tilde{\nu}_L)$     & 
                                    753.4  $(\tilde{\nu}_R,\tilde{\nu}_L)$    \\
$m_{\tilde{\nu}_{2}}$ [GeV]       & 102.3  $(\tilde{\nu}_R)$                   & 
                                    797.0  $(\tilde{\nu}_R)$                   & 
                                    92.6   $(\tilde{\nu}_R,\tilde{\nu}_L)$     & 
                                    542.3  $(\tilde{\nu}_R,\tilde{\nu}_L)$     & 
                                    753.9  $(\tilde{\nu}_R,\tilde{\nu}_L)$    \\
$m_{\tilde{\nu}_{3}}$ [GeV]       & 203.0  $(\tilde{\nu}_R)$                   & 
                                    797.0  $(\tilde{\nu}_R)$                   &   
                                    92.6   $(\tilde{\nu}_R,\tilde{\nu}_L)$     & 
                                    542.3  $(\tilde{\nu}_R,\tilde{\nu}_L)$     & 
                                    753.9  $(\tilde{\nu}_R,\tilde{\nu}_L)$    \\
$m_{\tilde{\nu}_{4}}$ [GeV]       & 573.8  $(\tilde{\nu}_R)$                   & 
                                    1120.1 $(\tilde{\nu}_R,\tilde{\nu}_L)$     &  
                                    253.4  $(\tilde{\nu}_L,\tilde{\nu}_R)$     & 
                                    585.4  $(\tilde{\nu}_L,\tilde{\nu}_R)$     & 
                                    785.5  $(\tilde{\nu}_L,\tilde{\nu}_R)$    \\
$m_{\tilde{\nu}_{5,6}}$ [GeV]     & 604.4  $(\tilde{\nu}_R)$                   & 
                                    1120.3 $(\tilde{\nu}_R,\tilde{\nu}_L)$     &  
                                    258.2  $(\tilde{\nu}_L,\tilde{\nu}_R)$     & 
                                    586.7  $(\tilde{\nu}_L,\tilde{\nu}_R)$     & 
                                    789.0  $(\tilde{\nu}_L,\tilde{\nu}_R)$    \\
$m_{\tilde{\nu}_{7}}$ [GeV]       & 725.2  $(\tilde{\nu}_L)$                   & 
                                    1220.0 $(\tilde{\nu}_L,\tilde{\nu}_R)$     & 
                                    1374.0 $(\tilde{\nu}_L,\tilde{\nu}_R)$     & 
                                    953.4  $(\tilde{\nu}_R)$                   & 
                                    950.1  $(\tilde{\nu}_R)$                  \\
$m_{\tilde{\nu}_{8,9}}$ [GeV]     & 734.1  $(\tilde{\nu}_L)$                   & 
                                    1236.6 $(\tilde{\nu}_L,\tilde{\nu}_R)$     & 
                                    1374.0 $(\tilde{\nu}_R)$                   & 
                                    953.4  $(\tilde{\nu}_R)$                   & 
                                    950.1  $(\tilde{\nu}_R)$                  \\
$m_{\tilde{e}_1}$ [GeV]           & 484.1  $(\tilde{\tau}_R)$                  & 
                                    1013.9 $(\tilde{\tau}_R)$                  &  
                                    254.7  $(\tilde{\tau}_L,\tilde{\tau}_R)$   & 
                                    263.0  $(\tilde{\tau}_R)$                  & 
                                    580.4  $(\tilde{\tau}_R)$                 \\
$m_{\tilde{e}_{2,3}}$ [GeV]       & 512.7  $(\tilde{\mu}_R)/(\tilde{e}_R)$     & 
                                    1055.3 $(\tilde{\mu}_R)/(\tilde{e}_R)$     &  
                                    265.6  $(\tilde{\mu}_L)/(\tilde{e}_L)$     & 
                                    270.5  $(\tilde{\mu}_R)/(\tilde{e}_R)$     & 
                                    592.3  $(\tilde{\mu}_R)/(\tilde{e}_R)$    \\
$m_{\tilde{e}_{4}}$ [GeV]         & 732.1  $(\tilde{\tau}_L)$                  & 
                                    1222.4 $(\tilde{\tau}_L)$                  &  
                                    447.7  $(\tilde{\tau}_R,\tilde{\tau}_L)$   & 
                                    591.6  $(\tilde{\tau}_L$                   & 
                                    788.0  $(\tilde{\tau}_L$                  \\
$m_{\tilde{e}_{5,6}}$ [GeV]       & 738.8  $(\tilde{\mu}_L)/(\tilde{e}_L)$     & 
                                    1237.9 $(\tilde{\mu}_L)/(\tilde{e}_L)$     &  
                                    450.6  $(\tilde{\mu}_R)/(\tilde{e}_R)$     & 
                                    592.2  $(\tilde{\mu}_L)/(\tilde{e}_L)$     & 
                                    790.9  $(\tilde{\mu}_L)/(\tilde{e}_L)$    \\
\hline
\multicolumn{6}{|c|}{Squarks} \\
\hline
$m_{\tilde{u}_1}$ [GeV]         & 1144.0 $(\tilde{t}_R,\tilde{t}_L)$     & 
                                  1185.4 $(\tilde{t}_R,\tilde{t}_L)$     & 
                                  1247.0 $(\tilde{t}_R,\tilde{t}_L)$     & 
                                  1111.3 $(\tilde{t}_R,\tilde{t}_L)$     & 
                                  1316.0 $(\tilde{t}_R,\tilde{t}_L)$    \\
$m_{\tilde{u}_2}$ [GeV]         & 1392.1 $(\tilde{t}_L,\tilde{t}_R)$     & 
                                  1851.9 $(\tilde{t}_L,\tilde{t}_R)$     & 
                                  1526.9 $(\tilde{t}_L,\tilde{t}_R)$     & 
                                  1361.4 $(\tilde{t}_L,\tilde{t}_R)$     & 
                                  1643.2 $(\tilde{t}_L,\tilde{t}_R)$    \\
$m_{\tilde{u}_{3,4}}$ [GeV]     & 1456.0 $(\tilde{c}_R)/(\tilde{u}_R)$   & 
                                  2154.7 $(\tilde{c}_R)/(\tilde{u}_R)$   & 
                                  1565.9 $(\tilde{c}_R)/(\tilde{u}_R)$   & 
                                  1392.4 $(\tilde{c}_R)/(\tilde{u}_R)$   & 
                                  1728.0 $(\tilde{c}_R)/(\tilde{u}_R)$  \\
$m_{\tilde{u}_{5,6}}$ [GeV]     & 1509.0 $[\tilde{c}_L,\tilde{u}_L]$     & 
                                  2227.3 $[\tilde{c}_L,\tilde{u}_L]$     & 
                                  1634.0 $[\tilde{c}_L,\tilde{u}_L]$     & 
                                  1448.8 $[\tilde{c}_L,\tilde{u}_L]$     & 
                                  1795.8 $[\tilde{c}_L,\tilde{u}_L]$    \\
$m_{\tilde{d}_1}$ [GeV]         & 1359.2 $(\tilde{b}_L,\tilde{b}_R)$     & 
                                  1819.2 $(\tilde{b}_L)$                 & 
                                  1409.8 $(\tilde{b}_R,\tilde{b}_L)$     & 
                                  1326.3 $(\tilde{b}_L)$                 & 
                                  1611.8 $(\tilde{b}_L)$                \\ 
$m_{\tilde{d}_2}$ [GeV]         & 1464.0 $(\tilde{b}_R,\tilde{b}_L)$     & 
                                  2148.1 $(\tilde{b}_R)$                 & 
                                  1462.3 $(\tilde{s}_R)$                 & 
                                  1420.1 $(\tilde{b}_R)$                 & 
                                  1724.5 $(\tilde{b}_R)$                \\ 
$m_{\tilde{d}_3}$ [GeV]         & 1489.8 $(\tilde{s}_R)$                 & 
                                  2175.9 $(\tilde{s}_R)$                 & 
                                  1462.3 $(\tilde{d}_R)$                 & 
                                  1426.2 $(\tilde{s}_R)$                 & 
                                  1734.8 $(\tilde{s}_R)$                \\ 
$m_{\tilde{d}_4}$ [GeV]         & 1489.8 $(\tilde{d}_R)$                 & 
                                  2175.9 $(\tilde{d}_R)$                 & 
                                  1496.2 $(\tilde{b}_L,\tilde{b}_R)$     & 
                                  1426.2 $(\tilde{d}_R)$                 & 
                                  1734.8 $(\tilde{d}_R)$                \\ 
$m_{\tilde{d}_{5,6}}$ [GeV]     & 1509.0 $[\tilde{s}_L,\tilde{d}_L]$     & 
                                  2228.9 $[\tilde{s}_L,\tilde{d}_L]$     & 
                                  1635.9 $[\tilde{s}_L,\tilde{d}_L]$     & 
                                  1450.9 $[\tilde{s}_L,\tilde{d}_L]$     & 
                                  1795.8 $[\tilde{s}_L,\tilde{d}_L]$    \\ 
\hline
\multicolumn{6}{|c|}{Higgs (1-loop/2-loop)} \\
\hline
$m_{h_1}$ [GeV]             & 59.1/59.6   &  119.2/125.4  & 92.7/93.1     & 100.8/102.6 & 18.8/18.8   \\
$m_{h_2}$ [GeV]             & 119.0/124.1 &  139.7/140.4  & 114.5/120.1   & 121.0/124.8 & 115.7/121.8 \\ 
\hline
${\cal R}_{L1}^2$           & 0.05/0.04   &  0.90/0.83  & 0.07/0.04     & 0.33/0.22 &  0.001/0.001 \\
${\cal R}_{L2}^2$           & 0.95/0.96   &  0.10/0.17  & 0.93/0.96     & 0.67/0.78 &  0.999/0.999 \\
\hline
\multicolumn{6}{|c|}{Neutralinos} \\
\hline
$m_{\chi_1^0}$ [GeV]            & 282.2  $(\tilde{B})$                       & 
                                  416.7  $(\tilde{B})$                       & 
                                  312.9  $(\tilde{B})$                       & 
                                  258.5  $(\tilde{h}_R)$                     & 
                                  346.6  $(\tilde{B})$                      \\ 
$m_{\chi_2^0}$ [GeV]            & 552.3  $(\tilde{W},\tilde{h}_L)$           & 
                                  780.0  $(\tilde{h}_R)$                     & 
                                  615.3  $(\tilde{W},\tilde{h}_L)$           & 
                                  279.7  $(\tilde{B})$                       & 
                                  679.5  $(\tilde{W},\tilde{h}_L)$          \\ 
$m_{\chi_3^0}$ [GeV]            & 828.9  $(\tilde{h}_L)$                     & 
                                  817.5  $(\tilde{W})$                       & 
                                  1086.6 $(\tilde{h}_L)$                     & 
                                  549.0  $(\tilde{W},\tilde{h}_L)$           & 
                                  902.7  $(\tilde{h}_R)$                    \\ 
$m_{\chi_4^0}$ [GeV]            & 838.9  $(\tilde{h}_L,\tilde{W})$           & 
                                  1865.5 $(\tilde{h}_L)$                     & 
                                  1092.8 $(\tilde{h}_L,\tilde{W})$           & 
                                  844.9  $(\tilde{h}_L)$                     & 
                                  1133.1 $(\tilde{h}_L)$                    \\ 
$m_{\chi_5^0}$ [GeV]            & 1230.4 $({\tilde B}_{\perp},\tilde{h}_R)$  & 
                                  1865.7 $(\tilde{h}_L)$                     & 
                                  1232.2 $(\tilde{h}_R,{\tilde B}_{\perp})$  & 
                                  856.8  $(\tilde{h}_L,\tilde{W})$           & 
                                  1139.4 $(\tilde{h}_L,\tilde{W})$          \\ 
$m_{\chi_6^0}$ [GeV]            & 1650.9 $(\tilde{h}_R)$                     & 
                                  2017.6 $({\tilde B}_{\perp},\tilde{h}_R)$  & 
                                  1811.3 $({\tilde B}_{\perp},\tilde{h}_R)$  & 
                                  1639.0 $({\tilde B}_{\perp},\tilde{h}_R)$  & 
                                  1489.8 $({\tilde B}_{\perp},\tilde{h}_R)$ \\ 
$m_{\chi_7^0}$ [GeV]            & 2608.3 $(\tilde{h}_R,{\tilde B}_{\perp})$  & 
                                  2392.3 $(\tilde{h}_R,{\tilde B}_{\perp})$  & 
                                  2741.4 $(\tilde{h}_R,{\tilde B}_{\perp})$  & 
                                  2174.6 $(\tilde{h}_R,{\tilde B}_{\perp})$  & 
                                  2056.5 $(\tilde{h}_R,{\tilde B}_{\perp})$ \\ 
\hline 
\hline
\end{tabular} 
\end{center} 
\caption{\label{tab:spectra} Spectra of our study points, for discussion see text.
$(\tilde{\nu}_R)$ is a nearly maximal mixture of the right sneutrinos
and the $S$-fields.}
\end{table}

Note that the ordering of sfermion mass eigenstates does in many cases
not follow the standard CMSSM patterns: $m_{{\tilde\tau}_1} \le
m_{{\tilde\mu}_R} \simeq m_{{\tilde e}_R} < m_{{\tilde\mu}_L} \simeq
m_{{\tilde e}_L} \le m_{{\tilde\tau}_2}$ and $m_{{\tilde t}_1} \le
m_{{\tilde c}_R} \simeq m_{{\tilde u}_R} < m_{{\tilde c}_L} \simeq
m_{{\tilde u}_L} \le m_{{\tilde t}_2}$ (similar for sdowns). These
patterns are distorted in the study points due to the unconventional
D-terms of the model and this feature gets enhanced for larger
$|\tbR-1|$ and/or larger values of $v_R$. We note also that for
sneutrinos and charged sleptons many states are quite degenerate. For
example ${\tilde\mu}_R$ and ${\tilde e}_R$ have practically the same
mass in all points. While these degeneracies are always true in CMSSM
spectra, in our case this is not necessarily so, but simply
reflects the fact that both $Y_{\nu}$ and $Y_s$ have been chosen
generation independent in all points, except BLRSP1. As this point
shows, even a rather moderate generation dependent value of $Y_s$ can
lead to large mass splittings in the sneutrino sector. A generation 
dependent value of $Y_{\nu}$ would not only split sneutrino masses 
but also charged slepton masses.

\subsection{Lepton decays and LFV}
\label{sec:lepton_LFV}
To explain the measured neutrino mixing angles by the mass matrix
given in eq.~(\ref{eq:neutrinoMM}) the Yukawa couplings $Y_\nu$, $Y_s$
and/or the the bilinear term $\mu_S$ have to contain off-diagonal
elements. In case that the neutrino mixing is explained by
the form of $Y_\nu$ or $Y_s$ also 
lepton flavor violation in the charged lepton
sector will be induced. On the one hand the contributions of $Y_\nu$
to the RGE evaluation branching ratios of $Y_e$ and to the
soft-breaking terms in the lepton sector open decay channels like
$l_i \to l_j \gamma$ and $l_i \to 3 l_j$ similarly to high-scale
seesaw type I--III \cite{Hirsch:2008dy,Hirsch:2008gh,Esteves:2010ff}.
On the other hand, in inverse seesaw the entries of $Y_\nu$ can be
potential large and the dominant contributions to LFV decays can come
from diagrams which are proportional to $(Y_\nu Y_\nu^\dagger)_{ij}$.
For a long time it has been assumed that the most stringent bounds on
$(Y_\nu Y_\nu^\dagger)_{ij}$ come from the radiative decay $\mu \to e
\gamma$  while the photonic contributions to $\mu \to 3 e$ 
are always smaller and therefore $\text{Br}(\mu \to e \gamma) >
\text{Br}(\mu \to 3 e)$ must hold. However, recently it has been
pointed out that in presence of new Yukawa couplings like $Y_\nu$ the
$Z$-penguin contributions to $\text{Br}(\mu \to 3 e)$ can dominate
\cite{Hirsch:2012ax}. These are less suppressed then the photonic
contributions by a factor $\left(\frac{M_{SUSY}}{M_Z}\right)^4$. As
consequence, the experimental limits on the decay into two charged
electrons can be much more constraining than the radiative decay 
in case of a heavy SUSY spectrum. We
can see this by parametrizing the neutrino Yukawa coupling as
\begin{equation}
Y_\nu = f \left( \begin{array}{ccc}
 0 &   0 &  0 \\
 a &   a &  -a \\
 b &  1 &  1 \\
\end{array}  \right) \,,
\end{equation}
with
\begin{equation}
a = \left( {\Ds}/{\Da} \right)^{\frac{1}{4}} \sim 0.4\; ~~\text{and}~~~ b = 0.23.
\end{equation}
Here, $\Ds$ and $\Da$ are the mass differences measured in solar and athmospheric neutrino oscillations. 
The value of $b$ accommodates for $\sin^2\theta_{13}=0.026$.

\begin{figure}
 \begin{center}
 \hspace{-22mm}
  \begin{tabular}{cc}
   \resizebox{100mm}{!}{\includegraphics{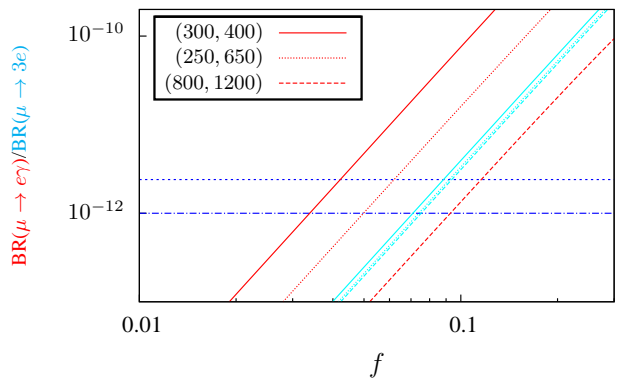}}
  \end{tabular}
  \caption{\label{fig:LowEnergyConstraints1} Branching ratios
  of lepton flavour violating processes as a function of $f$ for
  $\tan\beta = 10$, $A_0 = 0$, 
   $v_R = 5000$~GeV, $\tan\beta_R = 1.05$, $\mu_R = -500$~GeV, $m_{A_R} = 1000$~GeV and three $(m_0,M_{1/2})$ combinations.}
 \end{center}
\end{figure}

$\text{Br}(\mu \to e \gamma)$ and $\text{Br}(\mu \to 3 e)$ as function of $f$  are depicted in
\FIG{fig:LowEnergyConstraints1}, and we show also the most recent, experimental limits of \cite{Adam:2011ch,Nakamura:2010zzi}
\begin{equation}
\text{Br}(\mu \to e \gamma) < 2.4\cdot 10^{-12} \,, \hspace{1cm} \text{Br}(\mu \to 3 e) < 1.0 \cdot 10^{-12}
\end{equation}

\begin{figure}
 \begin{center}
 \hspace{-22mm}
  \begin{tabular}{c}
   \resizebox{90mm}{!}{\includegraphics{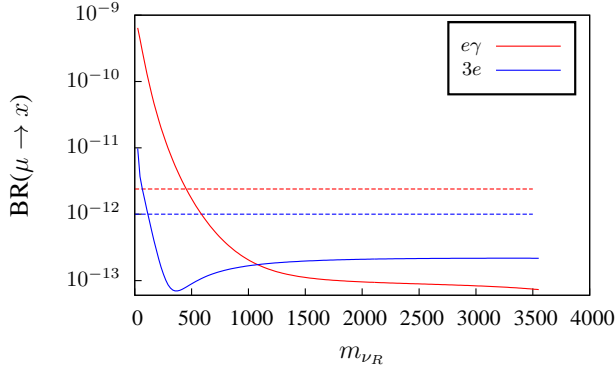}}
  \end{tabular}
   \caption{\label{fig:LowEnergyConstraints2} Branching ratios
  of lepton flavour violating processes as a function of
  $m_{\nu_R}$ for $m_0=800$~GeV, $M_{1/2}=1200$~GeV, $\tan\beta = 10$, $A_0 = 0$, 
   $v_R = 10$ TeV, $\tan\beta_R = 1.05$, $\mu_R = -500$~GeV, $m_{A_R} = 1000$~GeV. The dashed red line is the upper limit for 
    $\mu \rightarrow e \gamma$ and the dashed blue line for $\tau \rightarrow \mu \gamma$.}
 \end{center}
\end{figure}

For the plot we chose three different points for ($m_0$, $M_{1/2}$).
Since $\text{Br}(\mu \to 3 e)$ hardly depends on SUSY masses for $m_{SUSY} \gg m_Z$ all 
three lines lie very close together in contrast to $\text{Br}(\mu \to e \gamma)$. For light
SUSY spectra $\text{Br}(\mu \to e \gamma)$ is dominant whereas the heavier the SUSY particles
the more important gets Br($\mu \to 3 e$). As shown in BLRSP1 and BLRSP2 we
can have light right-handed neutrinos such that contributions
from the $W$-graph to LFV can not be neglected anymore. In 
\FIG{fig:LowEnergyConstraints2} 
$\text{Br}(\mu \to e \gamma)$ and $\text{Br}(\mu \to 3 e)$ 
are plotted as a function of $m_{\nu_R}$. For masses below 300~GeV
contributions from right-handed neutrinos start to dominate. 
The minimum in $\text{Br}(\mu \to 3 e)$ comes from a cancelation between the 
right-handed neutrino and the corresponding SUSY graph.
In the limit of large $m_{\nu_R}$ $\text{Br}(\mu \to e \gamma)$ and $\text{Br}(\mu \to 3 e)$ converge to the value coming from SUSY contributions.

However, it is possible to circumvent  these 
bounds by assuming that $Y_\nu$ and $Y_s$ are diagonal and the entire 
neutrino mixing is explained by $\mu_S$. Of course, this 
will not only reduce $\text{Br}(\mu \to 3 e)$ but also $\text{Br}(\mu \to e \gamma)$.

\subsection{Higgs physics, direct production}

In all study points of \TAB{tab:input} there is one Higgs boson
with mass between 120 and 125~GeV. In addition, there is a second
state with masses varying between 19 and 140~GeV. In BLRSP1, BLRSP3
and BLRSP5 the mass eigenstate $h_2$ is SM-like, with $R_{L2}^2>0.9$.
In BLRSP2 it is $h_1$, which has a large content of $H_d$ and $H_u$
and BLRSP4 is a case where $h_1$ and $h_2$ have large mixing. Since we
have often a mass eigenstate below the LEP limit of 115~GeV for a
standard model Higgs boson, we have checked the consistency of these
eigenstates with data using {\tt HiggsBounds 3.4.0beta} 
\cite{Bechtle:2008jh,Bechtle:2011sb}. All points are allowed by
accelerator constraints, but sometimes very close to existing bounds,
especially BLRSP4 and also BLRSP2.
As an indication for the theoretical uncertainties in the mass
calculation we give the masses using the complete 1-loop formulas
and the ones adding the dominant 2-loop corrections to
the MSSM sector \cite{Degrassi:2001yf,Brignole:2001jy,Brignole:2002bz,%
Dedes:2002dy,Dedes:2003km,Allanach:2004rh}.

In BLRSP1 and BLRSP5 $h_1$ is so light that the decay $h_2 \to h_1 h_1$
is kinematically allowed. However, the mixing between both sectors is
so small that for BLRSP1 the corresponding branching ratio is about
1 per-cent whereas for BLRSP5 it is a few per-mile. The smallness of 
this decay is a direct consequence of the bounds imposed by LEP and 
the decay $h_2 \to h_1 h_1$ can never be dominant in the BLR model.
The $h_2$ can also decay into a combination of heavy and light neutrinos
with a branching ratio of a few per-cent, as for example in case of
BLRSP1 leading to the final states
\begin{eqnarray}
h_2 &\to & \nu_i \nu_k \to \nu_i l^\pm W^\mp \\ 
h_2 &\to & \nu_i \nu_k \to \nu_i \nu_j Z  
\end{eqnarray}
with $i,j=1,2,3$ and $k=4,\dots,9$. These final states can also be
obtained via intermediate states containing an off-shell vector boson,
e.g.~$W W^*$ and $Z Z^*$. However, their existence implies that ratio
of quark versus lepton final states will not correspond to the
branching ratios of the vector bosons. Note, however, that for 
hadronic W-boson decays the invariant mass of $jj$+lepton system 
would show a peak at the heavy neutrino mass, which allows to 
identify this signature, in principle. Apart from these decays, 
the $h_2$ can also decay to two scalar neutrinos and, if kinematically 
allowed, this decay can become dominant, leading to a (nearly) 
invisible Higgs boson.

The hint for a 125~GeV Higgs boson \cite{ATLAS:2012ae,Chatrchyan:2012tx}, 
see also \cite{ATLAS-WEB} and \cite{CMS-WEB}, indicates a slightly 
larger than expected branching ratio into the two-photon final state. 
In \cite{Ellwanger:2011aa} it was shown that the NMSSM can, in principle, 
explain such an enhanced di-photon rate, due to a possible mixing 
of the singlet and the Higgs, which reduces the coupling of the Higgs 
to bottom quarks, thus reducing the total width, without affecting 
the production cross section. In the case of the 
BLR model, such a construction is not possible, since our singlets are 
charged under $U(1)_R$ and the mixing between SM and BLR sectors is 
controlled by $\tbR$. Since we have to choose $\tbR$ close to 
one, the singlets mix to the up and down components of the Higgs 
equally. Therefore a reduction of $h \to b \bar{b}$ causes a reduction 
of the coupling for gluon fusion as well. Thus, a sizeable 
enhancement of Br($h \to \gamma \gamma$) by reducing simply the total 
width is not possible in the BLR model. Currently the discrepancy of the 
data with expectations is only at the level of about 1 $\sigma$ c.l. 
However, should future data show indeed an enhanced rate for the 
$\gamma\gamma$ final state, this would be hard to
 explain in the BLR model.

In the four points (BLRSP1, BLRSP3-BLRSP5) $h_1$ has approximately the
same branching ratios for the decays into SM-fermions as a SM Higgs
boson of the same mass. However, the corresponding widths are
suppressed by the mixing with the usual MSSM sector which reduces the
width by a factor between $10^2$ and $10^4$. At the LHC the main
production of this particle is via SUSY cascade decays, e.g.\ it
appears in the decays of $\tilde \nu_{4,5,6}$ (BLRSP1, BLRSP3),
$\tilde \chi^0_3$ (BLRSP4) or in the decays of the heavy neutrinos
which are produced via the $Z'$ (BLRSP1, BLRSP4) as discussed in
section \ref{tab:ZpBRs}.  However, in case of BLRSP5 LHC will miss
$h_1$ as it only appears in the decays of the heavy Higgs bosons which
have masses in the TeV range.

Study point BLRSP2 differs from the others as here $h_1$ is the
MSSM-like Higgs boson and $h_2$ has a mass of 140~GeV which could
explain the slight excess in this region observed by ATLAS and CMS in
the early data \cite{ATLAStalk,CMStalk}.  In this region of the parameter
space the Higgs at 125~GeV is made as in the MSSM, implying
a rather heavy SUSY spectrum.  This is due to the fact that a 
140~GeV Higgs with reduced couplings can only be the $h_{BLR}$, i.e. 
this points exist to the right of the level-crossing region shown 
in \FIG{fig:hhvsvR}. Due to the choice of a rather small $Y_s$ 
in this point the heavy neutrinos masses are below the mass
of $h_2$. This leads to non-standard decays into the heavy
neutrinos which dominantly decay to a lepton and a W-boson.

\subsection{$Z'$ physics }
\label{subsec:zp}

As already mentioned in sect.~\ref{sec:MZP}, our $Z'$ corresponds
essentially to the $Z_\chi$ in the notation of \cite{Erler:2011ud}.
In previous studies usually two assumptions have been made in the
construction of mass bounds: (i) the $Z'$ decays only into the known
SM particles \cite{Etzion2012} and (ii) the effects of gauge kinetic
mixing are neglected. Both assumptions are not truly valid in the BLR
model. As shown in \TAB{tab:ZpBRs} we find in all our points
that the heavy neutrinos appear as final states beside the
SM-fermions. Moreover, in all but BLRSP5 also supersymmetric particles
appear as decay products, in particular sneutrinos and sleptons.  On
the other hand, gauge kinetic effects are in this model less important
and were only important if one could measure the branching with a
precision of 1 per-cent or better.

\begin{table}
\caption{Branching ratios of the dominant $Z'$ decay modes. Here
 we have summed over the generations in case of the charged fermions
 and sfermions. For the neutrinos we have splitted this sum into a 
 sum over the light (heavy) states denoted by $\nu_l$ ($\nu_h$).}
\begin{tabular}{lccccc} \hline
final state & BLRSP1 & BLRSP2 & BLRSP3 & BLRSP4 & BLRSP5 \\ \hline
$BR(dd)$                    & 0.31 & 0.35 & 0.35 & 0.37 &  0.43 \\ 
$BR(uu)$                    & 0.06 & 0.07 & 0.07 & 0.07 & 0.08 \\ 
$BR(ll)$                    & 0.12 & 0.14 & 0.14 & 0.14 & 0.16 \\ 
$BR(\nu_l \nu_l)$           & 0.10 & 0.11 & 0.12 & 0.12 & 0.12 \\ 
$BR(\nu_h \nu_h)$           & 0.27 & 0.30 & 0.13 & 0.11 & 0.13  \\ 
$BR(\tilde \nu \tilde \nu)$ & 0.05 &  --- & 0.05 & 0.03 &  --- \\ 
$BR(\tilde l \tilde l)$     &  --- &  --- & 0.05 & 0.03 &  --- \\ 
$BR(\tilde \chi^+_2 \tilde \chi^-_2)$ &  --- & --- &  --- & 0.02 &  --- \\ 
$BR(\tilde \chi^0_4 \tilde \chi^0_5)$ &  --- & --- &  --- & 0.02 &  --- \\ 
\hline
\end{tabular}
\label{tab:ZpBRs}
\end{table}

The $Z'$ couples to leptons and quarks as follows
\begin{equation}
Z'_\mu \bar{f} \gamma^\mu (c^f_L P_L + c^f_R P_R) f 
\end{equation}
The different coefficients are given in \TAB{tab:coeffZR}.

\begin{table}
\begin{tabular}{|c|c|c|}
\hline
  & $c_L$ & $c_R$ \\
\hline
$d$   & $ -\frac{i}{6}\left(-3 g_L Z^{13} + g_{BL} Z^{23} + g_{BLR} Z^{33}\right) $ & $ -\frac{i}{6}\left((g_{BL} - 3 g_{RBL}) Z^{23} + (g_{BLR} - 3 g_R) Z^{33} \right)$ \\
\hline
$u$   & $ -\frac{i}{6}\left(3 g_L Z^{13} + g_{BL} Z^{23} + g_{BLR} Z^{33}\right)  $ & $ -\frac{i}{6}\left((g_{BL} + 3 g_{RBL}) Z^{23} + (g_{BLR} + 3 g_R) Z^{33} \right)  $ \\
\hline
$l$   & $\frac{i}{2}\left(g_L Z^{13} + g_{BL} Z^{23} + g_{BLR} Z^{33}\right) $ & $ \frac{i}{2}\left((g_{BL} + g_{RBL}) Z^{23} + (g_{BLR} + g_R) Z^{33} \right)$ \\
\hline
$\nu$ & $ \begin{array}{c}  \frac{i}{2}\Big[\sum_{x=1}^3 Z_\nu^{j 3+x,*} Z_\nu^{i 3+x} \big( (-g_{BL}+g_{RBL}) Z^{23} \\ + (g_R - g_{BLR}) Z^{33}\big)+ \\
   \sum_{x=1}^3 Z_\nu^{j x,*} Z_\nu^{i x} \left(g_{BL} Z^{23} - g_L Z^{13} + g_{BLR} Z^{33} \right)  \Big] \end{array} $ &
  $\begin{array}{c} -\frac{i}{2}\Big[\sum_{x=1}^3 Z_\nu^{i 3+x,*} Z_\nu^{j 3+x} \big( (-g_{BL}+g_{RBL}) Z^{23} \\ + (g_R - g_{BLR}) Z^{33}\big) +  \\
   \sum_{x=1}^3 Z_\nu^{i x,*} Z_\nu^{j x} \left(g_{BL} Z^{23} - g_L Z^{13} + g_{BLR} Z^{33} \right)  \Big] \end{array} $ \\
\hline
\end{tabular}
\caption{Coefficients $c^f_L$ and $c^f_R$ for the coupling between $Z_R$ and two leptons or quarks.  Here, $Z$ is the rotation matrix diagonalizing the neutral gauge boson
 mass matrix and $Z_\nu$ is the neutrino mixing matrix. }
\label{tab:coeffZR}
\end{table}

Note, that in
the couplings to the $u$-quarks a partial cancellation occurs in
contrast to the ones to $d$-quarks, which get enhanced.  Moreover, the same
feature appears in the vertex $\tilde q$-$q$-$\tilde{B}_{\perp}$ which
leads to some interesting consequences discussed in section
\ref{sec:susycascades}.

We find that the decays into the heavy neutrino states are always
possible and have a sizable branching ratio provided Tr$(|Y_s|) \lsim
1$. In \TAB{tab:ZpBRs} we summarize the most important final
states of the $Z'$ for the different scenarios. As can be seen the
heavy neutrino final states have always a sizable branching ratios with up to
about 30 per-cent when summing over the generations. But even for
rather heavy neutrinos as in BLRSP5 own finds for this channel a 15
per-cent branching ratio. In several cases also channels into SUSY
particles are open, in particular in scenarios with sneutrino LSPs. In
case of supersymmetric particles the final states containing sleptons
or sneutrinos have the largest branching ratios. Channels into
neutralinos or charginos are suppressed. They proceed either via the
mixing with the $Z$ which is rather small or via the projection of the
higgsino-right onto the corresponding neutralino state.

\begin{figure}
 \begin{center}
 \hspace{-22mm}
  \begin{tabular}{c}
   \resizebox{80mm}{!}{\includegraphics{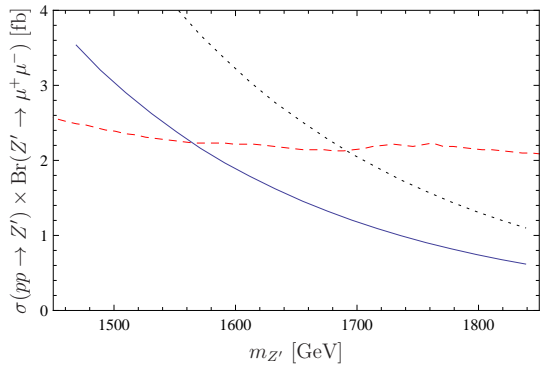}}
  \end{tabular}
   \caption{\label{fig:ppMuMu} Cross section of $p p \to Z' \to \mu^+
  \mu^-$ near the $Z'$ peak as function of $m_{Z'}$ taking into account
  a K-factor of 1.3 \cite{Fuks:2007gk}. For the black,
  dotted line the $Z'$ width has been calculated allowing only SM
  final states, while the blue solid includes also right-handed
  neutrinos and SUSY states.  The red line shows the ATLAS exclusion
  limit \cite{ATLAS-CONF-2012-007}.  We have used as input BLRSP1 and
  varied $v_R = [4.1,5.1]$~TeV.}
 \end{center}
\end{figure}

The appearance of additional final states leads to a reduction of the
event numbers in the most sensitive search channels, i.e. reducing
cross section times branching ratio, and, thus, the bounds obtained by
the LHC collaborations
\cite{ATLAS-CONF-2012-007,Collaboration:2011dca,Chatrchyan:2011wq} are
less constraining in the BLR model. This is depicted in
\FIG{fig:ppMuMu} where we show the production cross section
$\sigma(p p \to \mu^+ \mu^-)$ arround the $Z'$ resonance. 
\footnote{For the calculation
of the cross section we used {\tt WHIZARD} \cite{Kilian:2007gr,Moretti:2001zz} 
and implemented the
model using the {\tt SUSY Toolbox} \cite{Staub:2011dp}.}
In case that
the width of the $Z'$ is calculated using only SM final states the
cross section is increased roughly by a factor 1.6 in comparison to
the case where also right handed neutrinos and SUSY particle contribute
to the width of $Z'$. With this choice of parameters, the main effect is
due to R-neutrinos. We attribute the remaining difference to the official 
ATLAS result to slightly different values in the couplings and slightly
different branching ratios of the final states. Our results agree also
with the ones of ref.~\cite{Erler:2011ud}.
 We conclude that, although in our benchmark points
we take always $m_{Z'}> 1.8$~TeV, a significantly lower mass is
possible consistent with data.

\subsection{Heavy neutrinos}
\label{sec:heavyneutrinos}
As discussed above, the heavy neutrino states can be produced via the
$Z'$ with a considerable branching ratio of about 30 per-cent when
summing over all heavy neutrinos. Moreover, see below, they can also
be produced in the cascade decays of supersymmetric particles. These
heavy neutrinos mix with the light neutrino states implying a
reduction of the couplings of the light neutrinos to the $Z$-boson
and, thus, also a reduction of the invisible width of the
$Z$-boson. Taking the data from ref.~\cite{Nakamura:2010zzi} this can
be translated into the following condition on the $3 \times 3$ sub-block
$U^\nu_{ij}$, $i,j\le 3$, of the neutrino mixing matrix:
\begin{equation}
\left|1 - \sum_{ij=1, i \le j}^3 \left| 
 \sum_{k=1}^3 U^\nu_{ik} U^{\nu,*}_{jk} \right|^2 \right| < 0.009
\end{equation}
at the 3-$\sigma$ level. We have checked that all our benchmark points
fullfill this condition.

The main decay modes of the heavy neutrinos are\footnote{
For related discussions  see e.g.\ 
\cite{delAguila:2007em,AguilarSaavedra:2009ik,Das:2012ii} and references
therein.}
\begin{eqnarray}
\nu_j &\to& W^\pm l^\mp \\
\nu_j &\to& Z \nu_i \\
\nu_j &\to& h_k \nu_i 
\end{eqnarray}
where $j\ge 4$, $i\le 3$, $k=1,2$ and $l=e,\mu,\tau$, provided they
are kinemtically allowed. If there is no kinematical suppression we
find in general the branching ratios scale like $BR(\nu_j \to W^\pm
l^\mp):BR(\nu_j \to Z \nu_i):BR(\nu_j \to h_k \nu_i) \simeq 0.5 : 0.25: 0.25$   where we have summed over the light
Higgs bosons, the light neutrinos and leptons, respectively.  We
stress that these states are quasi-Dirac neutrinos and, thus, for six
heavy neutrinos at LHC the existence of up to three new particles
could be established.  Note, that the final states containing a
$W$-boson allow for a direct mass measurement. 

Beside the above decay modes, also decays into SUSY particle are
possible if kinematics allow for it. For example we find that for
BLRSP4 the decay into $\tilde \nu_{1,2,3} \tilde \chi^0_1$ are
possible and have branching ratios of about 3 per-cent. In 
scenarios like BLRSP3, BLRSP4 and BLRSP5 
the main production of the heavy neutrinos is via the $Z'$
and, thus, a high luminosity will be required to observe such final
states.

\subsection{SUSY cascade decays}
\label{sec:susycascades}

In this section we point out several features which distinguish the
BLR model from the usual MSSM. For the sake of preparing the ground,
let us first summarize the main features of the MSSM relevant for the
LHC, focusing for the time being on scenarios where the gluino is
heavier than the squarks: (i) The gluino decays dominantly into
squarks and quarks. (ii) L-squarks and L-sleptons decay dominantly
into the chargino and the neutralino which are mainly
$SU(2)_L$-gauginos. Apart from kinematical effects the branching ratio
for decays into the charged wino divided by the branching ratio into
the neutral wino is about 2:1. (iii) R-squarks and R-sleptons decay
dominantly into the bino-like neutralino with a branching ratio often
quite close to 100 per-cent.  (iv) In case of third generation
sfermions also decays into higgsinos are important.

In the BLR model one has two main new features: (i) there are
additional neutralinos and (ii) the sneutrino sector is enlarged as
well. The latter implies that sneutrino LSPs are possible consistent
with all astrophysical constraints and direct dark matter searches
\cite{Gopalakrishna:2006kr,Arina:2007tm,Thomas:2007bu,Cerdeno:2008ep,%
Bandyopadhyay:2011qm,Belanger:2011rs,Dumont:2012ee}. This
feature is for example realised in study points BLRSP1 and
BLRSP3\footnote{We note for completeness, that the relic abundance is
actually somewhat too large in this point but can easily be adjusted
by changing for example in BLRSP1 $\tan\beta_R$ from 1.05 to 1.0475
without changing the collider features of BLRSP1.}.

Let us start the discussion with BLRSP1. In this point the four
lighter neutralinos are the usual MSSM neutralinos with the standard
hierarchy. The fifth state corresponds to the additional
$U(1)$-gaugino, which we call ${\tilde B}_\perp$, whereas the two
additional states are the additional higgsinos. Note that the lighest
neutralino is not stable anymore but decays into final states
containing all nine neutrinos as well as the three lightest
sneutrinos. Of the latter ones the second lightest is so long lived
that it will lead to a displaced vertex in a typical collider detector.
The third sneutrino decays dominatly via three-body
decays into $l^+ l^- \tilde \nu_i$ and $\nu_k \nu_l \tilde \nu_i$ with
$i=1,2$ and $k,l=1,2,3$.  As discussed in section
\ref{sec:heavyneutrinos} the heavy neutrinos decay dominantly into
$W$-bosons and charged leptons, thus the decays of the lightest
neutralino are {\em not} invisible.

${\tilde B}_\perp$ appears for example in the decays of $\tilde d_R$
and $\tilde s_R$ with branching ratios BR($q \tilde \chi^0_1) \simeq
0.8$ and BR($q \tilde \chi^0_5) \simeq 0.2$. For completeness
we remark that the decays of $\tilde u_R$ and $\tilde c_R$ into
$\tilde \chi^0_5$ is suppressed as the corresponding coupling is
supressed as are the couplings of $Z'$ to $u$-type quarks in this
model.  $\tilde \chi^0_5$ decays dominantly into sleptons and
sneutrinos.  Combining all the above together one gets a much richer
structure for the decays of the R-squarks, e.g.\ the following decay
chains:
\begin{eqnarray}
\label{eq:RsquarktoZ}
\tilde q_R &\to& q \tilde \chi^0_1 \to q \nu_{k} \tilde \nu_1
              \to q \nu_j Z  \tilde \nu_1 \\
\tilde q_R &\to& q \tilde \chi^0_1 \to q \nu_{k} \tilde \nu_1
              \to q l^\pm W^\mp  \tilde \nu_1 \\
\tilde q_R &\to& q \tilde \chi^0_1 \to q \nu_{k} \tilde \nu_3
              \to q l^\pm W^\mp  \tilde l'{}^+ l'{}^- \nu_1 \\
\tilde q_R &\to& q \tilde \chi^0_5 \to q l^\pm \tilde l_i^\mp
               \to q l^\pm l^\mp   \tilde  \chi^0_1 
             \to q l^\pm l^\mp  \nu_{k}  \tilde  \nu_1 
             \to  q l^\pm l^\mp l'{}^\pm W^\mp     \tilde  \nu_1   
\label{eq:RsquarktoZ2}               
\end{eqnarray}
with $k\in\{4,5,6,7,8,9\}$ and $j\in \{1,2,3\}$.  Of course, several
other combinations are possible as well. 

From equations~(\ref{eq:RsquarktoZ}) to (\ref{eq:RsquarktoZ2}) 
one sees immediatly that the standard
signature of R-squarks, namely jet and missing energy, 
is only realized
in a few cases in this study point, e.g.\ if in
eq.~(\ref{eq:RsquarktoZ}) the $Z$ decays into neutrinos.
Interestingly, the chain via $\tilde \chi^0_5$ into sleptons leads to a
characteristic edge in the invariant mass of the lepton which can be
used to determine the corresponding masses once combined with
information from other decay chains. Also in the study points BLRSP2
and BLRSP5 $\tilde d_R$ and $\tilde s_R$ decay into heavy neutralinos,
which contain sizable content of the extra $U(1)$ gaugino, with a
sizable branching ratio. However, there the situation is somewhat less
involved as in these study points the lightest bino-like neutralino is
the LSP.

Another interesting feature is, that $\tilde \chi^0_5$ decays also
into the heavier sneutrinos which themselves decay into the LSP plus
$h_1$. Similarely $h_1$ can be produced in the decays of the heavy
neutrinos implying that this state can be produced with sizable
rate in SUSY cascade decays. However, as the corresponding
final states are quite complicated a dedicated Monte Carlo study 
will be necessary to decide if this is indeed a discovery channel
for $h_1$.

From the point of view of SUSY cascade decays BLRSP2 looks essentially
like a standard MSSM point. Inspection of the spectrum shows  that
$\tilde \chi^0_2$ is essentially a higgsino corresponding to the
extended $U(1)$ sector but it shows hardly up in the cascade decays.
Its main production channel is via an $s$-channel $Z'$ but even in
this case the corresponding cross section is so low that it will
not be dedected at the LHC even with an integrated luminosity of
300 fb$^{-1}$. Another interesting feature shows up in the
decays of $\tilde \chi^0_3$ which is mainly the neutral wino and gets
copiously produced in the decays of the  L-squarks: it decays with
about 77 (15) per-cent into $h_1$ ($h_2$), implying that the
cascade decays are an important source of Higgs bosons.

In case of BLRSP3 one has sneutrino LSPs like in BLRSP1 but with a
different hierarchy in the spectrum, as the three lightest sleptons
are lighter then the lighest neutralino. Therefore the $\tilde
\chi^0_1$ has also sizable decay rates into charged sleptons which sum
to about 30 per-cent. The sleptons decay then further into $W^- \tilde
\nu_{1,2,3}$ and $\nu_{2,3}$ via 3-body decays into $f
\bar{f}$-pairs. The latter, however, are rather soft due to the small
mass difference. In addition we have the decay channel into a light
neutrino and one of the heavier sneutrinos which themselves decay into
a lighter  sneutrino and either one of the Higgs boson or the
$Z$-boson.  Putting again all these decays together one obtains for the
$\tilde \chi^0_1$ decays
\begin{eqnarray}
\tilde \chi^0_1 &\to& l^\pm \tilde l^\mp \to l^\pm W^\mp  \tilde \nu_1 \\
\tilde \chi^0_1 &\to& l^\pm \tilde l^\mp \to l^\pm W^\mp  \tilde \nu_{2,3}
 \to  l^\pm W^\mp f \bar{f} \tilde \nu_1 \\
\tilde \chi^0_1 &\to& \nu_j \tilde \nu_{2,3}
\to \nu_{1,2,3}  f \bar{f} \tilde \nu_1 \\
\tilde \chi^0_1 &\to& \nu_j \tilde \nu_1 \\
\label{eq:h1part1}
\tilde \chi^0_1 &\to& \nu_j \tilde \nu_k \to \nu_j h_{1,2} \tilde \nu_1 \\
\label{eq:h1part2}
\tilde \chi^0_1 &\to& \nu_j \tilde \nu_k \to \nu_j h_{1,2} f \bar{f}  \tilde \nu_1 
\end{eqnarray} 
with $j=1,2,3$ and $k=4,5,6$. 
This implies that the decays of the R-squarks show again a more
complicated structure compared to the usual CMSSM expectations.
Channels (\ref{eq:h1part1}) and  (\ref{eq:h1part2}) give $h_1$ 
in about 15 per-cent of the final states of $\tilde \chi^0_1$.
Moreover,  $\tilde \chi^0_2$ and $\tilde \chi^+_1$ decay dominantly
into sleptons and sneutrinos. Here a new feature is found for 
$\tilde \chi^+_1$, as also the following chains
\begin{eqnarray}
\tilde \chi^+_1 &\to& l^+ \tilde \nu_{5,6} \to l^+ Z \tilde \nu_{1} \\
\tilde \chi^+_1 &\to& l^+ \tilde \nu_{5,6} \to l^+ h_{1,2} \tilde \nu_{1}
\end{eqnarray}
gives rise to sharp edge structures. However, as the main final states
of $Z$ and $h_{1,2}$ are two jets, the feasability still needs to be 
investigated.

In BLRSP4 we have chosen $\mu_R = 260$~GeV in order to construct an
LSP which is essentially a $\tilde h_R$. Here, the $R$-sleptons are
lighter than $\tilde \chi^0_2$, which is essentially bino-like in this
point, giving rise to the following decay chain of the down-type
$R$-squarks
\begin{equation}
\tilde d_R \to d \tilde \chi^0_2 \to d l^\pm \tilde l^\mp \to
d l^\pm l^\mp\tilde \chi^0_1
\end{equation}
Nearly all cascade decays end in a $\tilde \chi^0_2$ or one of
the lighter sleptons. Due to the fact, that in this particular
case the additional sneutrino states are hardly produced, 
it might
be difficult to disthinguish it from the NMSSM, at least as long as 
the $Z'$ is not discovered. The heavier $L$-sleptons do not show
up in the cascade decays of squarks and gluinos but can be
produced via the $Z'$ as discussed in section \ref{tab:ZpBRs}.

BLRSP5 is similar to BLRSP1 but compatible with pure GUT conditions,
e.g.\ $\mu_R$ and $m_{A_R}$ are not input in this case put derived
quantities.  To fullfill the tadpole equations we have to choose $Y_s
= 0.3$ and $\tan\beta_R = 1.03$ if we want a relatively low $m_0 =
500$~GeV while $M_{1/2} = 850$~GeV. The choice of $Y_s$ leads
automatically to large masses for the heavy neutrinos such that the
lightest Higgs can not decay into those states.  As in BLRSP1 the
down-type $R$-squarks decay not only into $\tilde \chi^0_1$ but also
into $\tilde \chi^0_6$ with a branching ratio of about 13
per-cent. For completeness, we note that here $\tilde \chi^0_3 \simeq
\tilde h_R$. However this state gets hardly produced in any of the
SUSY decays or via the $Z'$. Therefore, it is likely that LHC will
miss it and also at a linear collider such as ILC or CLIC it
will be difficult to study, due to the small production
cross section.

\section{Conclusions}

We have studied the minimal supersymmetric $U(1)_{B-L}\times U(1)_R$ 
extension of the standard model. The model is minimal in the sense 
that the extended gauge symmetry is broken with the minimal number 
of Higgs fields. In the matter sector the model contains (three copies 
of) a superfield ${\hat \nu}^c$, to cancel anomalies. Adding three 
singlet superfields ${\hat S}$ allows to generate small neutrino masses 
with an inverse seesaw mechanism.

The phenomenology of the model differs from the MSSM in a number of 
interesting aspects. We have foccused on the Higgs phenomenology 
and discussed changes in SUSY spectra and decays with respect to 
the MSSM. The model is less constrained then the CMSSM from the 
possible measurement of a Higgs with a mass of the order of 125 GeV. 
If the hints found in LHC data \cite{ATLAS:2012ae,Chatrchyan:2012tx} 
is indeed correct our model predicts two relatively light states 
should exist, with the second $h^0$ corresponding (mostly) to the 
lightest of the ``right'' Higgses, added to break the extended 
gauge group.

It is interesting, as we have discussed, that very often a right 
sneutrino is found to be the LSP. This will affect all constraints 
on CMSSM parameter space derived from constraints on the dark 
matter abundance. In fact, if the right sneutrino is indeed the 
LSP in our model, no constraint on any CMSSM parameters can be 
derived from DM constraints.

The model has new D-terms in all scalar mass matrices, which can 
lead to sizeable changes in the SUSY spectra, of potential 
phenomenological interest. We have discussed a few benchmark points, 
covering a number of features which could allow to distinguish the 
model from the CMSSM. Obviously this includes the discovery
of a $Z'$ at the LHC where we have shown that the current
bounds from LHC data depend on the details of the particle spectrum.
Also the cascade decays of supersymmetric particles can be
significantly more involved than in the usual CMSSM as the
additional neutralinos, neutrinos and sneutrinos lead to 
enhancement of the multiplicities in the final states.
This implies that the existing limits on the CMSSM parameter space get
modified as standard final states have reduced branching ratios and at
the same time additional final states are present. In case that the
mBLR model is indeed realized these new cascade decays will offer
additional kinematical information on the particle spectrum.

\section*{Acknowledgements}

W.P.~thanks the IFIC for hospitality during an extended stay.
M.H. and L.R. acknowledge support from the Spanish MICINN grants 
FPA2011-22975, MULTIDARK CSD2009-00064 and 
by the Generalitat Valenciana grant Prometeo/2009/091 and the
EU~Network grant UNILHC PITN-GA-2009-237920. L.R thanks the 
Instituto Superior T\'ecnico, Lisbon, for kind hospitality 
during an extended stay. 
W.P.\ has been supported by the Alexander von Humboldt foundation
and in part by the DFG, project no.\ PO-1337/2-1 and the Helmholtz
Alliance ``Physics at the Terascale''.

\begin{appendix}
\section{Appendix}

\subsection{Mass matrices}
Here we list the tree-level mass matrices of the model not given in the 
main text.

\begin{itemize} 
\item {\bf Mass matrix for Down-Squarks}, Basis: $\left(\tilde{d}_{L}, \tilde{d}_{R}\right)$ 
\begin{equation} 
m^2_{\tilde{d}} = \left( 
\begin{array}{cc}
m^2_{LL} &\frac{1}{\sqrt{2}} \Big(v_d T_{d}^{\dagger}  - v_u \mu Y_{d}^{\dagger} \Big) \\ 
\frac{1}{\sqrt{2}}  \Big(v_d T_d  - v_u Y_d \mu^* \Big) &m^2_{RR}\end{array} 
\right) 
\end{equation} 
\begin{align} 
m^2_{LL} &= m_q^2 + \frac{v_{d}^{2}}{2} Y_{d}^{\dagger}  Y_d
-\frac{1}{24}  \Big(
  \Big(g_{BL}^{2} + g_{BLR}^2 - g_{BLR}g_R -g_{BL} g_{RBL} \Big)
      \Big(v_{\chi_R}^{2}- v_{\bar{\chi}_R}^{2} \Big) \nonumber \\ 
& \hspace{4cm}
   +\Big(3 g_{L}^{2} +g_{BL} g_{RBL} +g_{BLR}  g_R \Big)
   \Big(  v_{d}^{2} - v_{u}^{2}\Big) \Big) {\bf 1}\\ 
m^2_{RR} &= m_d^2 + \frac{v_{d}^{2}}{2} Y_{d}  Y_d^{\dagger}
\nonumber \\ 
& +\frac{1}{24}  \Big(
  \Big(g_{BL}^{2} + g_{BLR}^2 - 4 (g_{BLR}g_R +g_{BL} g_{RBL})
   +3 (g_{R}^{2} + g_{RBL}^{2})\Big)
      \Big(v_{\chi_R}^{2}- v_{\bar{\chi}_R}^{2} \Big) \nonumber \\ 
& \hspace{1cm}
   +\Big(g_{BL} g_{RBL} +g_{BLR}  g_R - 3 (g_{R}^{2} + g_{RBL}^{2}) \Big)
   \Big(  v_{d}^{2} - v_{u}^{2}\Big) \Big) {\bf 1}
\end{align}

\item {\bf Mass matrix for Up-Squarks}, Basis: $\left(\tilde{u}_{L}, \tilde{u}_{R}\right)$
\begin{equation} 
m^2_{\tilde{u}} = \left( 
\begin{array}{cc}
m^2_{LL} &\frac{1}{\sqrt{2}} \Big(- v_d \mu Y_{u}^{\dagger}  + v_u T_{u}^{\dagger} \Big) \\ 
\frac{1}{\sqrt{2}}  \Big(- v_d Y_u \mu^*  + v_u T_u \Big) &m^2_{RR}\end{array} 
\right) 
\end{equation} 
\begin{align} 
m^2_{LL} &= m_q^2 + \frac{v_{u}^{2}}{2} Y_{u}^{\dagger}  Y_u
-\frac{1}{24}  \Big(
  \Big(g_{BL}^{2} + g_{BLR}^2 - g_{BLR}g_R -g_{BL} g_{RBL} \Big)
      \Big(v_{\chi_R}^{2}- v_{\bar{\chi}_R}^{2} \Big) \nonumber \\ 
& \hspace{4cm}
   +\Big(3 g_{L}^{2} -g_{BL} g_{RBL} -g_{BLR}  g_R \Big)
   \Big(  v_{u}^{2} - v_{d}^{2}\Big) \Big) {\bf 1}\\ 
m^2_{RR} &= m_u^2 + \frac{v_{u}^{2}}{2} Y_{u}  Y_u^{\dagger}
\nonumber \\ 
& +\frac{1}{24}  \Big(
  \Big(g_{BL}^{2} + g_{BLR}^2 + 2 (g_{BLR}g_R +g_{BL} g_{RBL})
   -3 (g_{R}^{2} + g_{RBL}^{2})\Big)
      \Big(v_{\chi_R}^{2}- v_{\bar{\chi}_R}^{2} \Big) \nonumber \\ 
& \hspace{1cm}
   +\Big(g_{BL} g_{RBL} +g_{BLR}  g_R + 3 (g_{R}^{2} + g_{RBL}^{2}) \Big)
   \Big(  v_{d}^{2} - v_{u}^{2}\Big) \Big) {\bf 1}
\end{align}

\item {\bf Mass of the Charged Higgs boson}:
One obtains the same expression as  in the MSSM:
\begin{equation} 
m^2_{H^+} = B_\mu \left( \tan\beta+\cot\beta\right) + m_W^2
\end{equation} 

\item {\bf Mass matrix for Charginos}, Basis: $\left(\tilde{W}^-, \tilde{H}_d^-\right), \left(\tilde{W}^+, \tilde{H}_u^+\right)$
\begin{equation} 
m_{\tilde{\chi}^-} = \left( 
\begin{array}{cc}
M_2 &\frac{1}{\sqrt{2}} g_L v_u \\ 
\frac{1}{\sqrt{2}} g_L v_d  &\mu\end{array} 
\right) 
\end{equation} 


\end{itemize} 

\subsection{Calculation of the mass spectrum}
\label{sec:LoopCorrections}

We are going to present now the basic steps to calculate the mass spectrum. 
As starting point we use electroweak precision data to get the gauge 
and Yukawa couplings: the SM-like Yukawa couplings are
calculated from the fermion masses and the one-loop relations of
ref.~\cite{Pierce:1996zz} which have been adjusted to our model. Similarly, 
also the standard model gauge couplings are calculated by the same procedure
presented in ref.~\cite{Pierce:1996zz}, but again, including all new contributions of 
the mode under consideration. Since the entire RGE running is performed in the basis
$SU(3)_C \times SU(2)_L \times U(1)_R \times U(1)_{B-L}$, the value of the
GUT normalized $g_{BL}$ and $g_R$
are matched to the GUT normalized hypercharge coupling $g_Y$ by
\begin{align}
 g_R =& c_{RY}\, gY \, , \\
 g_{BL} = & \frac{5 g_{BLR} g_{RBL} g_R  - \sqrt{6} g_{RBL} g_Y^2 + \sqrt{(3 g_{BLR}^2 - 2 \sqrt{6} g_{BLR} g_R + 2 g_R^2)(5(g_R^2+g_{RBL}^2 - 3 g_Y^2)g_Y^2}}{5 g_R^2 - 3 g_Y^2} \, . 
\end{align}
This is nothing else then an inversion of the well known relation 
between the gauge couplings for $U(1)_R \times U(1)_{B-L} \to U(1)_Y$ 
including the off-diagonal gauge couplings given in eq.~(\ref{eq:gY_from_gR_gBL}). 
We are using the $SO(10)$ GUT normalization
of $\sqrt{\frac{3}{5}}$ for $U(1)_Y$ and  $\sqrt{\frac{3}{2}}$ for $U(1)_{B-L}$.
To get the correct values of $c_{RY}$ as well as $g_{RBL}$ and $g_{BLR}$ an 
iterative procedure is used: $c_{RY}$ is calculated as ratio of the $g_Y$
and $g_{BL}$ when running down from the GUT scale and applying
\begin{equation}
\label{eq:gY_from_gR_gBL}
 g_Y = \sqrt{\frac{5(g_{BL} g_R - g_{BLR} g_{RBL})^2}{3(g_{BL}^2 + g_{BLR}) + 2 (g_R^2 + g_{RBL}^2) - 2 \sqrt{6}(g_R g_{BLR} + g_{BL} g_{RBL}) }}
\end{equation}

When the gauge and Yukawa couplings are derived, the RGEs are 
then evaluated up to the GUT scale where the corresponding boundary 
conditions of eqs.~(\ref{eq:msugra}), (\ref{eq:msugra2}) and (\ref{eq:GUToffdiagonal})
 are applied. Afterwards a RGE running of the full set of parameters to the 
SUSY scale is performed. We use always 2-loop RGEs which include the 
full effect of kinetic mixing \cite{Martin:1993zk,Fonseca:2011vn}.

The running parameters are then used to calculate the tree level mass spectrum. 
However, it is well known that the one-loop corrections can be very important for particular
particles and have to be taken into account. The best known example is the light MSSM Higgs
boson which get shifted by up to 50\% per-cent in case of heavy stops. Similar effects can be
expected in the extended Higgs sector especially since these can be very light at tree-level. 
Similarly, the gauginos arising in an extended gauge sector can be potentially light and receive
important corrections at one-loop \cite{O'Leary:2011yq}. To take these and all other possible 
effects into account we use a complete one-loop correction of the entire mass spectrum. Our 
procedure to calculate the one-loop masses is based on the method proposed in Ref.\cite{Pierce:1996zz}:
first, all running $\overline{\text{DR}}$ parameters are calculated at the SUSY scale and the SUSY masses
at tree-level are derived. The EW vevs $v_d$ and $v_u$ are afterwards re-calculated using the one-loop corrected
$Z$ mass and demanding
\begin{equation}
\label{eq:conditionMZ}
 m_Z^2 + \delta m_Z^2 = \frac{(g_{BL}^2 g_L^2 +  g_{BL}^2 g_R^2 + g_L^2 g_R^2) v^2}
{4 (g_{BL}^2 + g_R^2)} (v_d^2 + v_u^2)
\end{equation}
in addition with the running value of $\tan\beta$. Note that $\delta m_Z^2$ as well as all other
self-energies include the corrections originated by all particles present in the mBLR. These
calculations are performed in $\overline{\text{DR}}$ scheme and 't Hooft gauge. Also the
complete dependence on the external momenta are taken into account. The re-calculated vevs
are afterwards used to solve the tree-level tadpole equations again and to re-calculate the 
tree-level mass spectrum as well as all vertices entering
the one-loop corrections. Using these vertices and masses, the one-loop corrections  $\delta t_i$ to the
tadpole equations are derived and we use as renormalization condition
\begin{equation}
t_i - \delta t_i = 0 \, .
\end{equation}
These one-loop corrected tadpole equations are solved with respect to the same parameter as 
at tree level resulting in new parameters $\mu^{(1)}, B^{(1)}_\mu, \mu^{(1)}_R$, 
$B^{(1)}_{\mu_R}$ respectively $\mu^{(1)}, B^{(1)}_\mu$, $m^{2,(1)}_{\chi_R}$, $m^{2,(1)}_{{\bar\chi}_R}$.
The final step is to calculate all self-energies for different particles and to use those to get the one-loop
corrected mass spectrum. 
\begin{enumerate}
 \item {\bf Real scalars}:  for a real scalar $\phi$, the one-loop corrections  are included by calculating
 the real part of the poles of the corresponding propagator matrices \cite{Pierce:1996zz}
\begin{equation}
\mathrm{Det}\left[ p^2_i \mathbf{1} - m^2_{\phi,1L}(p^2) \right] = 0,
\label{eq:propagator}
\end{equation}
where
\begin{equation}
 m^2_{\phi,1L}(p^2) = \tilde{m}^{2}_{\phi,T} -  \Pi_{\phi}(p^2) .
\end{equation}
Equation (\ref{eq:propagator}) has to be solved for each
eigenvalue $p^2=m^2_i$ which can be achieved in an iterative
procedure. This has to be done also for charged scalars as well as the fermions. 
Note, $\tilde{m}^2_T$ is the tree-level mass matrix
but for the parameters fixed by the tadpole equations the one-loop
corrected values $X^{(1)}$ are used.

 \item {\bf Complex scalars}: for a complex scalar $\eta$ field we use at one-loop level
\begin{equation}
	m^{2,\eta}_{1L}(p^2_i) = \tilde{m}^{2,\eta}_{T} - \Pi_{\eta}(p^2_i) ,
\end{equation}
While in case of sfermions $\tilde{m}^{2,\eta}_{T}$ agrees exactly with the tree-level mass matrix, 
for charged Higgs bosons $\mu^{(1)}$ and $B^{(1)}_\mu$ or $m_{H_d}^{(1)}$ and $m_{H_d}^{(1)}$  has 
to be used depending on the set of parameters the tadpole equations are solved for.

 \item {\bf Majorana fermions}: the one-loop mass matrix of a Majorana $\chi$ fermion is related
to the tree-level mass matrix by
\begin{eqnarray}
M^{\chi}_{1L} (p^2_i) &=& M^{\chi}_T - 
\frac{1}{2} \bigg[ \Sigma^0_S(p^2_i) + \Sigma^{0,T}_S(p^2_i)
 + \left(\Sigma^{0,T}_L(p^2_i)+   \Sigma^0_R(p^2_i)\right) M^{\chi}_T
 \nonumber \\
&& \hspace{16mm}
+ M^{\tilde\chi^0}_T \left(\Sigma^{0,T}_R(p^2_i) +  \Sigma^0_L(p^2_i) \right)
 \bigg] ,
\end{eqnarray}
where we have denoted the wave-function corrections by $\Sigma^{0}_R$,
 $\Sigma^{0}_L$ and the direct one-loop contribution to the mass by 
$\Sigma^{0}_S$.
 
\item {\bf Dirac fermions}: for a Dirac fermion $\Psi$ one has to add the self-energies as
\begin{eqnarray}
M^{\Psi}_{1L}(p^2_i) =  M^{\Psi}_T - \Sigma^+_S(p^2_i)
 - \Sigma^+_R(p^2_i) M^{\Psi}_T - M^{\Psi}_T \Sigma^+_L(p^2_i) .
\end{eqnarray}
\end{enumerate}
Note, this procedure agrees with the method implemented in {\tt SPheno 3.1.10} to calculate the
loop masses in the MSSM as well as with the code produced by {\tt SARAH 3.0.39} or later. 
However, there are small differences to earlier versions of {\tt SPheno}
as well as other spectrum calculators: the MSSM equivalent of condition eq~(\ref{eq:conditionMZ}) is
often solved in an iterative way using the one-loop corrected parameters from the tadpole equations
to calculate $\delta m_Z^2$ until $m_Z^2 + \delta m_Z^2$ has converged. In this context also 
$\mu^{(1)}$ and $B_\mu^{(1)}$ are used in the vertices entering the one-loop corrections. However, 
these steps mix tree- and one-loop level and break therefore gauge invariance: when we tried this
approach the relation between Goldstone and gauge bosons mass is violated. However, the numerical 
differences in case of the MSSM turned out to be rather small. 

As example we give the necessary formulae to calculate the one-loop corrections to the tadpole equations and
the scalar Higgs masses in appendix~\ref{app:1LoopHiggs}.

\subsection{1-loop corrections of the Higgs sector}
\label{app:1LoopHiggs}
As discussed in section~\ref{sec:LoopCorrections} we have calculated the entire mass spectrum at one-loop. For that purpose it is necessary
to calculate all possible 1-loop diagrams for the one- and two-point functions. As example we here give the corresponding expressions for the one-loop corrections of the tadpoles as well as the self-energy for the scalar Higgs fields. For all other self-energies we refer to the output of \SARAH \footnote{To get the model files
of the mBLR which is not yet part of the public version of \SARAH please
send a mail to the authors.}.
The results are expressed via Passarino Veltman integrals \cite{Pierce:1996zz}. The basic integrals are
\begin{eqnarray}
A_0(m) &=& 16\pi^2Q^{4-n}\int{\frac{d^nq}{ i\,(2\pi)^n}}{\frac{1}{
q^2-m^2+i\varepsilon}} \thickspace ,\\
B_0(p, m_1, m_2) &=& 16\pi^2Q^{4-n}\int{\frac{d^nq}{ i\,(2\pi)^n}}
{\frac{1}{\biggl[q^2-m^2_1+i\varepsilon\biggr]\biggl[
(q-p)^2-m_2^2+i\varepsilon\biggr]}} \thickspace ,
\label{B0 def}
\end{eqnarray}
with the renormalization scale \(Q\).  All the other, 
necessary functions can be expressed by
$A_0$ and $B_0$. For instance,
\begin{equation}
 B_1(p, m_1,m_2) \ =\ {\frac{1}{2p^2}}\biggl[ A_0(m_2) - A_0(m_1) + (p^2
+m_1^2 -m_2^2) B_0(p, m_1, m_2)\biggr]~,
\end{equation}
and
\begin{align}
F_0(p,m_1,m_2) =& A_0(m_1)-2A_0(m_2)- (2p^2+2m^2_1-m^2_2)B_0(p,m_1,m_2)
\ , \\ 
G_0(p,m_1,m_2) =&
(p^2-m_1^2-m_2^2)B_0(p,m_1,m_2)-A_0(m_1)-A_0(m_2)
\end{align}
The numerical evalution of all loop-integrals is performed by \SPheno. With this conventions we can write the one-loop tadpoles as
{\allowdisplaybreaks
\begin{align} 
\delta t^{(1)}_{\sigma_{{i}}} = & \, +\frac{3}{2} {A_0\Big(m^2_{Z}\Big)} {\Gamma_{\sigma_{{i}},Z,Z}} +\frac{3}{2} {A_0\Big(m^2_{Z_R}\Big)} {\Gamma_{\sigma_{{i}},Z_R,Z_R}} +3 {A_0\Big(m^2_{W^-}\Big)} {\Gamma_{\sigma_{{i}},W^+,W^-}} \nonumber \\ 
 &+16 {A_0\Big(m^2_{\nu_{{1}}}\Big)} {\Gamma_{\sigma_{{i}},\nu_{{1}},\nu_{{1}}}} m^2_{\nu_{{1}}} - \sum_{a=1}^{2}{A_0\Big(m^2_{H^-_{{a}}}\Big)} {\Gamma_{\sigma_{{i}},H^+_{{a}},H^-_{{a}}}}  \nonumber \\ 
 &+4 \sum_{a=1}^{2}{A_0\Big(m^2_{\tilde{\chi}^-_{{a}}}\Big)} {\Gamma^L_{\sigma_{{i}},\tilde{\chi}^+_{{a}},\tilde{\chi}^-_{{a}}}} m^2_{\tilde{\chi}^-_{{a}}}  +12 \sum_{a=1}^{3}{A_0\Big(m^2_{d_{{a}}}\Big)} {\Gamma^L_{\sigma_{{i}},\bar{d}_{{a}},d_{{a}}}} m^2_{d_{{a}}}  \nonumber \\ 
 &+4 \sum_{a=1}^{3}{A_0\Big(m^2_{e_{{a}}}\Big)} {\Gamma^L_{\sigma_{{i}},\bar{e}_{{a}},e_{{a}}}} m^2_{e_{{a}}}  +12 \sum_{a=1}^{3}{A_0\Big(m^2_{u_{{a}}}\Big)} {\Gamma^L_{\sigma_{{i}},\bar{u}_{{a}},u_{{a}}}} m^2_{u_{{a}}}  \nonumber \\ 
 &-\frac{1}{2} \sum_{a=1}^{4}{A_0\Big(m^2_{A_{0,{a}}}\Big)} {\Gamma_{\sigma_{{i}},A_{0,{a}},A_{0,{a}}}}  -\frac{1}{2} \sum_{a=1}^{4}{A_0\Big(m^2_{h_{{a}}}\Big)} {\Gamma_{\sigma_{{i}},h_{{a}},h_{{a}}}}  -3 \sum_{a=1}^{6}{A_0\Big(m^2_{\tilde{d}_{{a}}}\Big)} {\Gamma_{\sigma_{{i}},\tilde{d}^*_{{a}},\tilde{d}_{{a}}}}  \nonumber \\ 
 &- \sum_{a=1}^{6}{A_0\Big(m^2_{\tilde{e}_{{a}}}\Big)} {\Gamma_{\sigma_{{i}},\tilde{e}^*_{{a}},\tilde{e}_{{a}}}}  -3 \sum_{a=1}^{6}{A_0\Big(m^2_{\tilde{u}_{{a}}}\Big)} {\Gamma_{\sigma_{{i}},\tilde{u}^*_{{a}},\tilde{u}_{{a}}}}  \nonumber \\ 
 &+2 \sum_{a=1}^{7}{A_0\Big(m^2_{\tilde{\chi}^0_{{a}}}\Big)} {\Gamma^L_{\sigma_{{i}},\tilde{\chi}^0_{{a}},\tilde{\chi}^0_{{a}}}} m^2_{\tilde{\chi}^0_{{a}}}  - \sum_{a=1}^{9}{A_0\Big(m^2_{\tilde{\nu}_{{a}}}\Big)} {\Gamma_{\sigma_{{i}},\tilde{\nu}^*_{{a}},\tilde{\nu}_{{a}}}}  \nonumber \\ 
 &+2 \sum_{a=1}^{9}{A_0\Big(m^2_{\nu_{{a}}}\Big)} {\Gamma_{\sigma_{{i}},\nu_{{a}},\nu_{{a}}}} m^2_{\nu_{{a}}}   
\end{align} 
with $\sigma_i = \left(\sigma_d, \sigma_u, \sigma_R, \bar{\sigma}_R\right)^T_i$. $\Gamma_{xyz}$ denotes the vertex of the three particles {$x$, $y$ $z$}, while  $\Gamma_{wxyz}$ will be used for four-point interactions. For chiral couplings we use $\Gamma^L$ as coefficient of the left and $\Gamma^R$ as coefficient of the right polarization operator. For instance, $\Gamma_{\sigma_d, Z, Z}$ is the coupling of a pure down-type Higgs to a $Z$ boson  while $\Gamma^L_{\sigma_{R},\tilde{\chi}^0_{2},\tilde{\chi}^0_{2}}$ corresponds to the left-chiral part of the interaction of a $R$-Higgs to a neutralino of the second generation. The expressions for all vertices can be obtained with \SARAH. \\
Using these conventions the self-energy for the scalar Higgs fields reads 
\begin{align} 
\Pi_{\sigma_i,\sigma_j}(p^2) &=\frac{7}{4} {B_0\Big(p^{2},m^2_{Z},m^2_{Z}\Big)} {\Gamma^*_{\sigma_{{j}},Z,Z}} {\Gamma_{\sigma_{{i}},Z,Z}} \nonumber \\ 
 &+\frac{7}{2} {B_0\Big(p^{2},m^2_{Z},m^2_{Z_R}\Big)} {\Gamma^*_{\sigma_{{j}},Z_R,Z}} {\Gamma_{\sigma_{{i}},Z_R,Z}} +\frac{7}{4} {B_0\Big(p^{2},m^2_{Z_R},m^2_{Z_R}\Big)} {\Gamma^*_{\sigma_{{j}},Z_R,Z_R}} {\Gamma_{\sigma_{{i}},Z_R,Z_R}} \nonumber \\ 
 &+\frac{7}{2} {B_0\Big(p^{2},m^2_{W^-},m^2_{W^-}\Big)} {\Gamma^*_{\sigma_{{j}},W^+,W^-}} {\Gamma_{\sigma_{{i}},W^+,W^-}} +2 {A_0\Big(m^2_{Z}\Big)} {\Gamma_{\sigma_{{i}},\sigma_{{i}},Z,Z}} +2 {A_0\Big(m^2_{Z_R}\Big)} {\Gamma_{\sigma_{{i}},\sigma_{{i}},Z_R,Z_R}} \nonumber \\ 
 &+4 {A_0\Big(m^2_{W^-}\Big)} {\Gamma_{\sigma_{{i}},\sigma_{{i}},W^+,W^-}} - \sum_{a=1}^{2}{A_0\Big(m^2_{H^-_{{a}}}\Big)} {\Gamma_{\sigma_{{i}},\sigma_{{i}},H^+_{{a}},H^-_{{a}}}}  \nonumber \\ 
 &+\sum_{a=1}^{2}\sum_{b=1}^{2}{B_0\Big(p^{2},m^2_{H^-_{{a}}},m^2_{H^-_{{b}}}\Big)} {\Gamma^*_{\sigma_{{j}},H^+_{{a}},H^-_{{b}}}} {\Gamma_{\sigma_{{i}},H^+_{{a}},H^-_{{b}}}} \nonumber \\ 
 &-2 \sum_{a=1}^{2}m_{\tilde{\chi}^+_{{a}}} \sum_{b=1}^{2}{B_0\Big(p^{2},m^2_{\tilde{\chi}^-_{{a}}},m^2_{\tilde{\chi}^-_{{b}}}\Big)} m_{\tilde{\chi}^-_{{b}}} \Big({\Gamma^{L*}_{\sigma_{{j}},\tilde{\chi}^+_{{a}},\tilde{\chi}^-_{{b}}}} {\Gamma^R_{\sigma_{{i}},\tilde{\chi}^+_{{a}},\tilde{\chi}^-_{{b}}}}  + {\Gamma^{R*}_{\sigma_{{j}},\tilde{\chi}^+_{{a}},\tilde{\chi}^-_{{b}}}} {\Gamma^L_{\sigma_{{i}},\tilde{\chi}^+_{{a}},\tilde{\chi}^-_{{b}}}} \Big)  \nonumber \\ 
 &+\sum_{a=1}^{2}\sum_{b=1}^{2}{G_0\Big(p^{2},m^2_{\tilde{\chi}^-_{{a}}},m^2_{\tilde{\chi}^-_{{b}}}\Big)} \Big({\Gamma^{L*}_{\sigma_{{j}},\tilde{\chi}^+_{{a}},\tilde{\chi}^-_{{b}}}} {\Gamma^L_{\sigma_{{i}},\tilde{\chi}^+_{{a}},\tilde{\chi}^-_{{b}}}}  + {\Gamma^{R*}_{\sigma_{{j}},\tilde{\chi}^+_{{a}},\tilde{\chi}^-_{{b}}}} {\Gamma^R_{\sigma_{{i}},\tilde{\chi}^+_{{a}},\tilde{\chi}^-_{{b}}}} \Big)\nonumber \\ 
 &-6 \sum_{a=1}^{3}m_{\bar{d}_{{a}}} \sum_{b=1}^{3}{B_0\Big(p^{2},m^2_{d_{{a}}},m^2_{d_{{b}}}\Big)} m_{d_{{b}}} \Big({\Gamma^{L*}_{\sigma_{{j}},\bar{d}_{{a}},d_{{b}}}} {\Gamma^R_{\sigma_{{i}},\bar{d}_{{a}},d_{{b}}}}  + {\Gamma^{R*}_{\sigma_{{j}},\bar{d}_{{a}},d_{{b}}}} {\Gamma^L_{\sigma_{{i}},\bar{d}_{{a}},d_{{b}}}} \Big)  \nonumber \\ 
 &+3 \sum_{a=1}^{3}\sum_{b=1}^{3}{G_0\Big(p^{2},m^2_{d_{{a}}},m^2_{d_{{b}}}\Big)} \Big({\Gamma^{L*}_{\sigma_{{j}},\bar{d}_{{a}},d_{{b}}}} {\Gamma^L_{\sigma_{{i}},\bar{d}_{{a}},d_{{b}}}}  + {\Gamma^{R*}_{\sigma_{{j}},\bar{d}_{{a}},d_{{b}}}} {\Gamma^R_{\sigma_{{i}},\bar{d}_{{a}},d_{{b}}}} \Big) \nonumber \\ 
 &-2 \sum_{a=1}^{3}m_{\bar{e}_{{a}}} \sum_{b=1}^{3}{B_0\Big(p^{2},m^2_{e_{{a}}},m^2_{e_{{b}}}\Big)} m_{e_{{b}}} \Big({\Gamma^{L*}_{\sigma_{{j}},\bar{e}_{{a}},e_{{b}}}} {\Gamma^R_{\sigma_{{i}},\bar{e}_{{a}},e_{{b}}}}  + {\Gamma^{R*}_{\sigma_{{j}},\bar{e}_{{a}},e_{{b}}}} {\Gamma^L_{\sigma_{{i}},\bar{e}_{{a}},e_{{b}}}} \Big)  \nonumber \\ 
 &+\sum_{a=1}^{3}\sum_{b=1}^{3}{G_0\Big(p^{2},m^2_{e_{{a}}},m^2_{e_{{b}}}\Big)} \Big({\Gamma^{L*}_{\sigma_{{j}},\bar{e}_{{a}},e_{{b}}}} {\Gamma^L_{\sigma_{{i}},\bar{e}_{{a}},e_{{b}}}}  + {\Gamma^{R*}_{\sigma_{{j}},\bar{e}_{{a}},e_{{b}}}} {\Gamma^R_{\sigma_{{i}},\bar{e}_{{a}},e_{{b}}}} \Big)\nonumber \\ 
 &-6 \sum_{a=1}^{3}m_{\bar{u}_{{a}}} \sum_{b=1}^{3}{B_0\Big(p^{2},m^2_{u_{{a}}},m^2_{u_{{b}}}\Big)} m_{u_{{b}}} \Big({\Gamma^{L*}_{\sigma_{{j}},\bar{u}_{{a}},u_{{b}}}} {\Gamma^R_{\sigma_{{i}},\bar{u}_{{a}},u_{{b}}}}  + {\Gamma^{R*}_{\sigma_{{j}},\bar{u}_{{a}},u_{{b}}}} {\Gamma^L_{\sigma_{{i}},\bar{u}_{{a}},u_{{b}}}} \Big)  \nonumber \\ 
 &+3 \sum_{a=1}^{3}\sum_{b=1}^{3}{G_0\Big(p^{2},m^2_{u_{{a}}},m^2_{u_{{b}}}\Big)} \Big({\Gamma^{L*}_{\sigma_{{j}},\bar{u}_{{a}},u_{{b}}}} {\Gamma^L_{\sigma_{{i}},\bar{u}_{{a}},u_{{b}}}}  + {\Gamma^{R*}_{\sigma_{{j}},\bar{u}_{{a}},u_{{b}}}} {\Gamma^R_{\sigma_{{i}},\bar{u}_{{a}},u_{{b}}}} \Big) \nonumber \\ 
 &-\frac{1}{2} \sum_{a=1}^{4}{A_0\Big(m^2_{A_{0,{a}}}\Big)} {\Gamma_{\sigma_{{i}},\sigma_{{i}},A_{0,{a}},A_{0,{a}}}}  -\frac{1}{2} \sum_{a=1}^{4}{A_0\Big(m^2_{h_{{a}}}\Big)} {\Gamma_{\sigma_{{i}},\sigma_{{i}},h_{{a}},h_{{a}}}}  \nonumber \\ 
 &+\frac{1}{2} \sum_{a=1}^{4}\sum_{b=1}^{4}{B_0\Big(p^{2},m^2_{A_{0,{a}}},m^2_{A_{0,{b}}}\Big)} {\Gamma^*_{\sigma_{{j}},A_{0,{a}},A_{0,{b}}}} {\Gamma_{\sigma_{{i}},A_{0,{a}},A_{0,{b}}}}  \nonumber \\ 
 &+\frac{1}{2} \sum_{a=1}^{4}\sum_{b=1}^{4}{B_0\Big(p^{2},m^2_{h_{{a}}},m^2_{h_{{b}}}\Big)} {\Gamma^*_{\sigma_{{j}},h_{{a}},h_{{b}}}} {\Gamma_{\sigma_{{i}},h_{{a}},h_{{b}}}}  -3 \sum_{a=1}^{6}{A_0\Big(m^2_{\tilde{d}_{{a}}}\Big)} {\Gamma_{\sigma_{{i}},\sigma_{{i}},\tilde{d}^*_{{a}},\tilde{d}_{{a}}}}  \nonumber \\ 
 &- \sum_{a=1}^{6}{A_0\Big(m^2_{\tilde{e}_{{a}}}\Big)} {\Gamma_{\sigma_{{i}},\sigma_{{i}},\tilde{e}^*_{{a}},\tilde{e}_{{a}}}}  -3 \sum_{a=1}^{6}{A_0\Big(m^2_{\tilde{u}_{{a}}}\Big)} {\Gamma_{\sigma_{{i}},\sigma_{{i}},\tilde{u}^*_{{a}},\tilde{u}_{{a}}}}  \nonumber \\ 
 &+3 \sum_{a=1}^{6}\sum_{b=1}^{6}{B_0\Big(p^{2},m^2_{\tilde{d}_{{a}}},m^2_{\tilde{d}_{{b}}}\Big)} {\Gamma^*_{\sigma_{{j}},\tilde{d}^*_{{a}},\tilde{d}_{{b}}}} {\Gamma_{\sigma_{{i}},\tilde{d}^*_{{a}},\tilde{d}_{{b}}}}  +\sum_{a=1}^{6}\sum_{b=1}^{6}{B_0\Big(p^{2},m^2_{\tilde{e}_{{a}}},m^2_{\tilde{e}_{{b}}}\Big)} {\Gamma^*_{\sigma_{{j}},\tilde{e}^*_{{a}},\tilde{e}_{{b}}}} {\Gamma_{\sigma_{{i}},\tilde{e}^*_{{a}},\tilde{e}_{{b}}}} \nonumber \\ 
 &+3 \sum_{a=1}^{6}\sum_{b=1}^{6}{B_0\Big(p^{2},m^2_{\tilde{u}_{{a}}},m^2_{\tilde{u}_{{b}}}\Big)} {\Gamma^*_{\sigma_{{j}},\tilde{u}^*_{{a}},\tilde{u}_{{b}}}} {\Gamma_{\sigma_{{i}},\tilde{u}^*_{{a}},\tilde{u}_{{b}}}}  \nonumber \\ 
 &- \sum_{a=1}^{7}m_{\tilde{\chi}^0_{{a}}} \sum_{b=1}^{7}{B_0\Big(p^{2},m^2_{\tilde{\chi}^0_{{a}}},m^2_{\tilde{\chi}^0_{{b}}}\Big)} m_{\tilde{\chi}^0_{{b}}} \Big({\Gamma^{L*}_{\sigma_{{j}},\tilde{\chi}^0_{{a}},\tilde{\chi}^0_{{b}}}} {\Gamma^R_{\sigma_{{i}},\tilde{\chi}^0_{{a}},\tilde{\chi}^0_{{b}}}}  + {\Gamma^{R*}_{\sigma_{{j}},\tilde{\chi}^0_{{a}},\tilde{\chi}^0_{{b}}}} {\Gamma^L_{\sigma_{{i}},\tilde{\chi}^0_{{a}},\tilde{\chi}^0_{{b}}}} \Big)  \nonumber \\ 
 &+\frac{1}{2} \sum_{a=1}^{7}\sum_{b=1}^{7}{G_0\Big(p^{2},m^2_{\tilde{\chi}^0_{{a}}},m^2_{\tilde{\chi}^0_{{b}}}\Big)} \Big({\Gamma^{L*}_{\sigma_{{j}},\tilde{\chi}^0_{{a}},\tilde{\chi}^0_{{b}}}} {\Gamma^L_{\sigma_{{i}},\tilde{\chi}^0_{{a}},\tilde{\chi}^0_{{b}}}}  + {\Gamma^{R*}_{\sigma_{{j}},\tilde{\chi}^0_{{a}},\tilde{\chi}^0_{{b}}}} {\Gamma^R_{\sigma_{{i}},\tilde{\chi}^0_{{a}},\tilde{\chi}^0_{{b}}}} \Big) \nonumber \\ 
 &- \sum_{a=1}^{9}{A_0\Big(m^2_{\tilde{\nu}_{{a}}}\Big)} {\Gamma_{\sigma_{{i}},\sigma_{{i}},\tilde{\nu}^*_{{a}},\tilde{\nu}_{{a}}}}  +\sum_{a=1}^{9}\sum_{b=1}^{9}{B_0\Big(p^{2},m^2_{\tilde{\nu}_{{a}}},m^2_{\tilde{\nu}_{{b}}}\Big)} {\Gamma^*_{\sigma_{{j}},\tilde{\nu}^*_{{a}},\tilde{\nu}_{{b}}}} {\Gamma_{\sigma_{{i}},\tilde{\nu}^*_{{a}},\tilde{\nu}_{{b}}}} \nonumber \\ 
 &- \sum_{a=1}^{9}m_{\nu_{{a}}} \sum_{b=1}^{9}{B_0\Big(p^{2},m^2_{\nu_{{a}}},m^2_{\nu_{{b}}}\Big)} m_{\nu_{{b}}} \Big({\Gamma^{L*}_{\sigma_{{j}},\nu_{{a}},\nu_{{b}}}} {\Gamma^R_{\sigma_{{i}},\nu_{{a}},\nu_{{b}}}}  + {\Gamma^{R*}_{\sigma_{{j}},\nu_{{a}},\nu_{{b}}}} {\Gamma^L_{\sigma_{{i}},\nu_{{a}},\nu_{{b}}}} \Big)  \nonumber \\ 
 &+\frac{1}{2} \sum_{a=1}^{9}\sum_{b=1}^{9}{G_0\Big(p^{2},m^2_{\nu_{{a}}},m^2_{\nu_{{b}}}\Big)} \Big({\Gamma^{L*}_{\sigma_{{j}},\nu_{{a}},\nu_{{b}}}} {\Gamma^L_{\sigma_{{i}},\nu_{{a}},\nu_{{b}}}}  + {\Gamma^{R*}_{\sigma_{{j}},\nu_{{a}},\nu_{{b}}}} {\Gamma^R_{\sigma_{{i}},\nu_{{a}},\nu_{{b}}}} \Big) \nonumber \\ 
 &+2 \sum_{b=1}^{2}{\Gamma^*_{\sigma_{{j}},W^+,H^-_{{b}}}} {\Gamma_{\sigma_{{i}},W^+,H^-_{{b}}}} {F_0\Big(p^{2},m^2_{H^-_{{b}}},m^2_{W^-}\Big)}  +\sum_{b=1}^{4}{\Gamma^*_{\sigma_{{j}},\gamma,A_{0,{b}}}} {\Gamma_{\sigma_{{i}},\gamma,A_{0,{b}}}} {F_0\Big(p^{2},m^2_{A_{0,{b}}},0\Big)} \nonumber \\ 
 &+\sum_{b=1}^{4}{\Gamma^*_{\sigma_{{j}},Z,A_{0,{b}}}} {\Gamma_{\sigma_{{i}},Z,A_{0,{b}}}} {F_0\Big(p^{2},m^2_{A_{0,{b}}},m^2_{Z}\Big)} +\sum_{b=1}^{4}{\Gamma^*_{\sigma_{{j}},Z_R,A_{0,{b}}}} {\Gamma_{\sigma_{{i}},Z_R,A_{0,{b}}}} {F_0\Big(p^{2},m^2_{A_{0,{b}}},m^2_{Z_R}\Big)}  
\end{align} 
}

\subsection{RGEs}
 
\label{app:rges}
The calculation of the renormalization group equations performed by \SARAH is
 based on the generic
expression of  \cite{Martin:1993zk}. In addition, the results of
 \cite{Fonseca:2011vn} are used to
include the effect of kinetic mixing. \\
The $\beta$ functions for the parameters of a general superpotential written as 
\begin{equation}
 W (\phi) = \frac{1}{2}{\mu}^{ij}\phi_i\phi_j + \frac{1}{6}Y^{ijk}
\phi_i\phi_j\phi_k
\end{equation}
can be easily obtained from the shown results for the anomalous dimensions by
 using the relations \cite{West:1984dg,Jones:1984cx}
\begin{eqnarray}
 \beta_Y^{ijk} &= & Y^{p(ij} {\gamma_p}^{k)} \thickspace, \\
 \beta_{\mu}^{ij} &= & \mu^{p(i} {\gamma_p}^{j)} \thickspace .
\end{eqnarray}
For the results of the other parameters as well as for the two-loop results
 which we skip here because of their length we suggest to use
the function {\tt CalcRGEs[]} of \SARAH.

\subsubsection{Anomalous dimensions}
{\allowdisplaybreaks \begin{align} 
\gamma_{\hat{q}}^{(1)} & =  
\frac{1}{12} \Big(12 \Big({Y_{d}^{\dagger}  Y_d} + {Y_{u}^{\dagger}  Y_u}\Big) - \Big(18 g_{L}^{2}  + 32 g_{s}^{2}  + g_{BL}^{2} + g_{BLR}^{2}\Big){\bf 1} \Big)\\ 
\gamma_{\hat{l}}^{(1)} & =  
-\frac{3}{4} \Big(2 g_{L}^{2}  + g_{BL}^{2} + g_{BLR}^{2}\Big){\bf 1}  + {Y_{e}^{\dagger}  Y_e} + {Y_{v}^{\dagger}  Y_v}\\ 
\gamma_{\hat{H}_d}^{(1)} & =  
\frac{1}{2} \Big(2 \mbox{Tr}\Big({Y_e  Y_{e}^{\dagger}}\Big)  -3 g_{L}^{2}  + 6 \mbox{Tr}\Big({Y_d  Y_{d}^{\dagger}}\Big)  - g_{R}^{2}  - g_{RBL}^{2} \Big)\\ 
\gamma_{\hat{H}_u}^{(1)} & =  
\frac{1}{2} \Big(2 \mbox{Tr}\Big({Y_v  Y_{v}^{\dagger}}\Big)  -3 g_{L}^{2}  + 6 \mbox{Tr}\Big({Y_u  Y_{u}^{\dagger}}\Big)  - g_{R}^{2}  - g_{RBL}^{2} \Big)\\ 
\gamma_{\hat{\chi}_R}^{(1)} & =  
\frac{1}{4} \Big(-2 g_{R}^{2}  -2 g_{RBL}^{2}  + 2 \sqrt{6} g_{BL} g_{RBL}  + 2 \sqrt{6} g_{BLR} g_R  -3 g_{BL}^{2}  -3 g_{BLR}^{2}  + 4 \mbox{Tr}\Big({Y_s  Y_{s}^{\dagger}}\Big) \Big)\\ 
\gamma_{\hat{\bar{\chi}}_R}^{(1)} & =  
\frac{1}{4} \Big(-2 \Big(g_{R}^{2} + g_{RBL}^{2}\Big) + 2 \sqrt{6} g_{BL} g_{RBL}  + 2 \sqrt{6} g_{BLR} g_R  -3 g_{BL}^{2}  -3 g_{BLR}^{2} \Big)\\ 
\gamma_{\hat{S}}^{(1)} & =  
{Y_{s}^{\dagger}  Y_s}\\ 
\gamma_{\hat{u}}^{(1)} & =  
\frac{1}{12} \Big(24 {Y_u^*  Y_{u}^{T}}  - \Big(2 \sqrt{6} g_{BL} g_{RBL}  + 2 \sqrt{6} g_{BLR} g_R  + 32 g_{s}^{2}  + 6 g_{R}^{2}  + 6 g_{RBL}^{2}  + g_{BL}^{2} + g_{BLR}^{2}\Big){\bf 1} \Big)\\ 
\gamma_{\hat{d}}^{(1)} & =  
\frac{1}{12} \Big(24 {Y_d^*  Y_{d}^{T}}  - \Big(-2 \sqrt{6} g_{BL} g_{RBL}  -2 \sqrt{6} g_{BLR} g_R  + 32 g_{s}^{2}  + 6 g_{R}^{2}  + 6 g_{RBL}^{2}  + g_{BL}^{2} + g_{BLR}^{2}\Big){\bf 1} \Big)\\ 
\gamma_{\hat{\nu}}^{(1)} & =  
\frac{1}{4} \Big(- \Big(2 \Big(g_{R}^{2} + g_{RBL}^{2}\Big) -2 \sqrt{6} g_{BL} g_{RBL}  -2 \sqrt{6} g_{BLR} g_R  + 3 g_{BL}^{2}  + 3 g_{BLR}^{2} \Big){\bf 1}  \nonumber \\
 &  \hspace{1cm} + 4 \Big(2 {Y_v^*  Y_{v}^{T}}  + {Y_s^*  Y_{s}^{T}}\Big)\Big)\\ 
\gamma_{\hat{e}}^{(1)} & =  
\frac{1}{4} \Big(- \Big(2 \Big(g_{R}^{2} + g_{RBL}^{2}\Big) + 2 \sqrt{6} g_{BL} g_{RBL}  + 2 \sqrt{6} g_{BLR} g_R  + 3 g_{BL}^{2}  + 3 g_{BLR}^{2} \Big){\bf 1}  + 8 {Y_e^*  Y_{e}^{T}} \Big)
\end{align} } 

\subsubsection{Gauge Couplings}
{\allowdisplaybreaks  \begin{align} 
\beta_{g_{BL}}^{(1)} & =  
\frac{1}{4} \Big(27 g_{BL}^{3}  -2 \sqrt{6} g_{BL}^{2} g_{RBL}  + g_{BL} \Big(27 g_{BLR}^{2}  + 30 g_{RBL}^{2}  - \sqrt{6} g_{BLR} g_R \Big) \nonumber \\
 &  \hspace{1cm} + g_{BLR} \Big(30 g_R  - \sqrt{6} g_{BLR} \Big)g_{RBL} \Big)\\ 
\beta_{g_R}^{(1)} & =  
\frac{1}{4} \Big(27 g_{BL} g_{BLR} g_{RBL}  + 27 g_{BLR}^{2} g_R  -2 \sqrt{6} g_{BLR} g_{R}^{2}  + 30 g_{R}^{3}  + 30 g_R g_{RBL}^{2} \nonumber \\
 & \hspace{1cm}  - \sqrt{6} g_{BL} g_R g_{RBL}  - \sqrt{6} g_{BLR} g_{RBL}^{2} \Big)\\ 
\beta_{g_{BLR}}^{(1)} & =  
\frac{1}{4} \Big(g_{BL}^{2} \Big(27 g_{BLR}  - \sqrt{6} g_R \Big) + g_{BL} \Big(30 g_R g_{RBL}  - \sqrt{6} g_{BLR} g_{RBL} \Big) \nonumber \\
 & \hspace{1cm} + g_{BLR} \Big(27 g_{BLR}^{2}  -2 \sqrt{6} g_{BLR} g_R  + 30 g_{R}^{2} \Big)\Big)\\ 
\beta_{g_{RBL}}^{(1)} & =  
\frac{1}{4} \Big(27 g_{BL}^{2} g_{RBL}  + 30 g_{RBL} \Big(g_{R}^{2} + g_{RBL}^{2}\Big) + g_{BL} \Big(27 g_{BLR} g_R \nonumber \\ & \hspace{1cm} - \sqrt{6} \Big(2 g_{RBL}^{2}  + g_{R}^{2}\Big)\Big) - \sqrt{6} g_{BLR} g_R g_{RBL} \Big) \\
\beta_{g_L}^{(1)} & =  
g_{L}^{3}\\ 
\beta_{g_s}^{(1)} & =  
-3 g_{s}^{3} 
\end{align} }

\subsubsection{Gaugino Mass Parameters}
{\allowdisplaybreaks  \begin{align} 
\beta_{M_{BL}}^{(1)} & =  
\frac{1}{2} \Big(27 g_{BL}^{2} M_{BL}  - g_{BL} \Big(-27 g_{BLR} M_{B R}  + 2 \sqrt{6} g_{RBL} M_{BL}  + \sqrt{6} g_R M_{B R} \Big) \nonumber \\
 &  \hspace{1cm} + g_{RBL} \Big(30 g_{RBL} M_{BL}  + 30 g_R M_{B R}  - \sqrt{6} g_{BLR} M_{B R} \Big)\Big)\\ 
\beta_{M_R}^{(1)} & =  
\frac{1}{2} \Big(27 g_{BLR}^{2} M_R  + 30 g_R \Big(g_{RBL} M_{B R}  + g_R M_R \Big) + g_{BL} \Big(27 g_{BLR}  - \sqrt{6} g_R \Big)M_{B R} \nonumber \\
 &  \hspace{1cm}  - \sqrt{6} g_{BLR} \Big(2 g_R M_R  + g_{RBL} M_{B R} \Big)\Big)\\ 
\beta_{M_{B R}}^{(1)} & =  
\frac{1}{4} \Big(27 g_{BL}^{2} M_{B R} +27 g_{BLR}^{2} M_{B R} - \sqrt{6} g_{BLR} \Big(2 g_R M_{B R}  + g_{RBL} \Big(M_{BL} + M_R\Big)\Big)+30 \Big(g_{R}^{2} M_{B R}  \nonumber \\
 &  \hspace{1cm} + g_{RBL}^{2} M_{B R}  + g_R g_{RBL} \Big(M_{BL} + M_R\Big)\Big)+g_{BL} \Big(27 g_{BLR} \Big(M_{BL} + M_R\Big) \nonumber \\
 &  \hspace{1cm} - \sqrt{6} \Big(2 g_{RBL} M_{B R}  + g_R \Big(M_{BL} + M_R\Big)\Big)\Big)\Big) \\ 
\beta_{M_2}^{(1)} & =  
2 g_{L}^{2} M_2 \\ 
\beta_{M_3}^{(1)} & =  
-6 g_{s}^{2} M_3 
\end{align}} 

\end{appendix}

\end{document}